\documentclass[12pt]{article}

\usepackage{amsbsy,amsmath,amsthm,latexsym,amssymb}
\usepackage{graphicx,stmaryrd,sectsty}
\usepackage{setspace}
\usepackage[export]{adjustbox} % LW
\newcommand{\bc}{\begin{center}}
\newcommand{\ec}{\end{center}}
\newcommand{\bfr}{\begin{flushright}}
\newcommand{\efr}{\end{flushright}}

\newcommand{\no}{\noindent}
\newcommand{\be}{\begin{enumerate}}
\newcommand{\ee}{\end{enumerate}}
\newcommand{\bi}{\begin{itemize}}
\newcommand{\ei}{\end{itemize}}
\newcommand{\bd}{\begin{description}}
\newcommand{\ed}{\end{description}}
\newcommand{\beq}{\begin{equation}}
\newcommand{\eeq}{\end{equation}}
\newcommand{\bea}{\begin{eqnarray}}
\newcommand{\eea}{\end{eqnarray}}

\newcommand{\bfi}{\begin{figure}}
\newcommand{\efi}{\end{figure}}
\newcommand{\bay}{\begin{array}{l}}
\newcommand{\eay}{\end{array}}

\newcommand{\cref}[1]{(\ref{#1})}   %to make cross reference easy.

\sectionfont{\normalsize}
\subsectionfont{\small}

\usepackage[usenames, dvipsnames]{color} % LW
\definecolor{NU}{RGB}{82,0,99} % LW
\usepackage{color}
\usepackage{tabularx}
\usepackage{float}
\usepackage[section]{placeins}
\begin{document}
%\baselineskip 10.7pt
%% BEGIN SEGIM PAGE
%\baselineskip 10.7pt
\begin{titlepage}
\clearpage\thispagestyle{empty}
%\noindent {\footnotesize {{\em
%\hfill To be submitted to Cement and Concrete Composites} }} \\
\noindent
\hrulefill
\begin{figure}[h!]
\centering
\includegraphics[width=2 in]{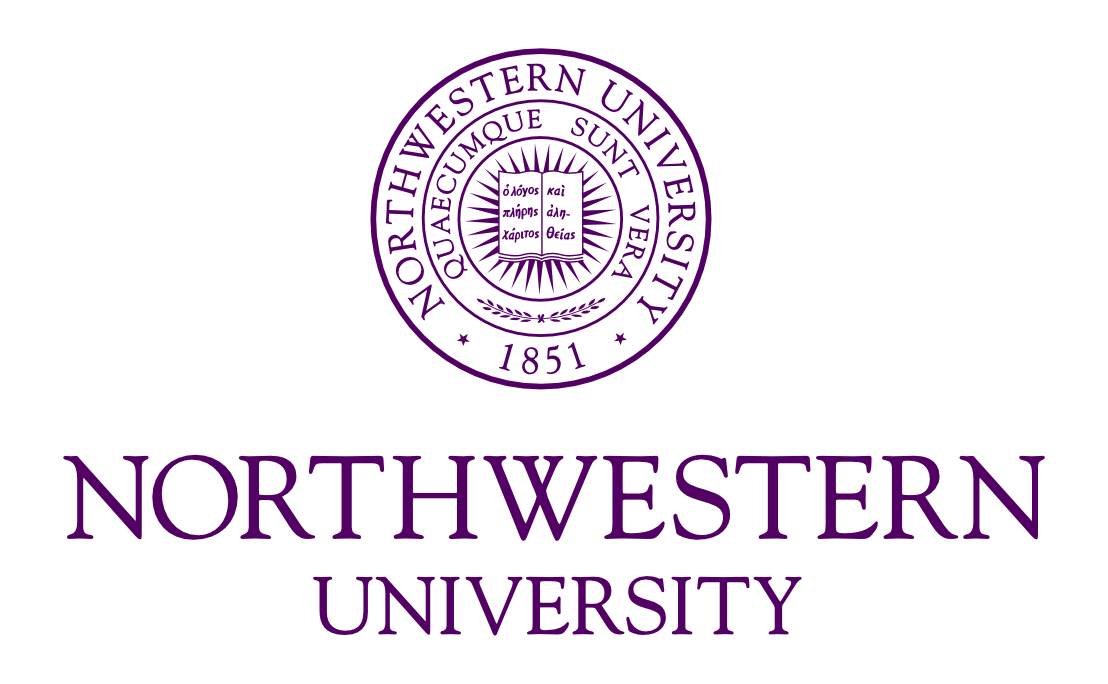}
\end{figure}
%{\color{NU} some text}
\begin{center}
{\color{NU}{
{\bf Center for Sustainable Engineering of Geological and Infrastructure Materials (SEGIM)} \\ [0.1in]
Department of Civil and Environmental Engineering \\ [0.1in]
McCormick School of Engineering and Applied Science \\ [0.1in]
Evanston, Illinois 60208, USA
}
}
\end{center} %\vskip 5mm
\hrulefill \\ \vskip 2mm
\vskip 0.5in
\begin{center}
{\large {\bf AGE-DEPENDENT SIZE EFFECT AND FRACTURE CHARACTERISTICS OF ULTRA HIGH PERFORMANCE CONCRETE}}\\[0.5in]
{\large {\sc Lin Wan, Roman Wendner, Gianluca Cusatis}}\\[0.75in]
{\sf \bf SEGIM INTERNAL REPORT No. 16-08/465A}\\[0.75in]
\end{center}
\noindent {\footnotesize {{\em Submitted to Cement and Concrete Composites  \hfill September 2016} }}
\end{titlepage}

\newpage
\clearpage \pagestyle{plain} \setcounter{page}{1}
%% END SEGIM PAGE

\begin{center}
{\large {\bf Age-dependent Size Effect and Fracture Characteristics of Ultra High Performance Concrete}}
\\[9mm]
%{\large
{\bf  {\bf  By \\
Lin Wan \footnote{Ph.D., Researcher, Department of Civil and Environmental Engineering, Northwestern University, 2145 Sheridan Rd. Evanston IL, 60208 USA. E-mail: lin.wan@u.northwestern.edu},
Roman Wendner$^*$ \footnote{Corresponding Author: Director Christian Doppler Laboratory LiCRoFast, Department of Civil Engineering and Natural Hazards, University of Natural Resources and Life Sciences (BOKU) Vienna. E-mail: roman.wendner@boku.ac.at},
Gianluca Cusatis \footnote{Associate Professor, Department of Civil and Environmental Engineering, Northwestern University, 2145 Sheridan Rd. Evanston IL, 60208 USA. E-mail: g-cusatis@northwestern.edu, Phone: (847)-491-4027} %Member ASCE, %
}  }
%\end{center}  \vskip 6mm

\vspace{30mm}

A Paper submitted for publication in

Cement and Concrete Composites 

%\today
September 8, 2016

Corresponding Author

Roman Wendner

Director - Christian Doppler Laboratory LiCRoFast

Department of Civil Engineering and Natural Hazards

University of Natural Resources and Life Sciences (BOKU)

Peter Jordan Strasse 82, A-1190

Vienna, Austria

Tel: +43 1 47654 5252

Fax: +43 1 47654 5299

Email: roman.wendner@boku.ac.at

Website: http://www.baunat.boku.ac.at/cd-labor

\end{center}

\newpage

\no {\bf   Abstract:}\\ {\sf

This paper presents an investigation of the age-dependent size effect and fracture characteristics of an ultra high performance concrete (UHPC). The study is based on a unique set of experimental data connecting aging tests for two curing protocols of one size and scaled size effect tests of one age. Both aging and size effect studies are performed on notched three point bending tests. Experimental data is augmented by state of the art simulations employing a recently developed discrete element based early-age computational framework. The framework is constructed by coupling a hygro-thermo-chemical (HTC) model and the Lattice Discrete Particle Model (LDPM) through a set of aging functions. The HTC component allows taking into account variable curing conditions and predicts the maturity of concrete. The mechanical component, LDPM, simulates the failure behavior of concrete at the length scale of major heterogeneities. After careful calibration and validation the mesoscale HTC-LDPM model is uniquely posed to perform predictive simulations. The ultimate flexural strengths from experiments and simulations are analyzed by the cohesive size effect curve (CSEC) method, and the classical size effect law (SEL). The fracture energies obtained by LDPM, CSEC, SEL, and cohesive crack analyses are compared and an aging formulation for fracture properties is proposed. Based on experiments, simulations, and size effect analyses, the age-dependence of size effect and the robustness of analytical size effect  methods are evaluated. 

keywords: UHPC; aging; size effect; cohesive crack analysis; fracture energy; tensile characteristic length

\section{Introduction}\no

Ultra high performance concretes (UHPCs) are cementitious composites characterized by high compressive strength, low water-binder ratio, and optimized gradation curve. In many cases thermal activation, fiber reinforcement and superplasticizers are employed to increase strength, enhance ductility and ensure workability. UHPC became commercially available in the beginning of the 21st century and has been utilized in the construction industry, especially for bridge applications and tall buildings, around the world, across North America, Europe, and Asia. While more and more UHPCs are developed and utilized in the construction industry, what is lacking in the available literature is a model capable of describing the evolution of fracture characteristics of UHPCs at early age and beyond. This is crucial in terms of structural design, project planning, and building optimization. The performance assessment of existing structures is gaining in importance as well and requires an accurate reproduction of the initial construction stages in which the long-term behavior is anchored \cite{Strauss13,Wendner14}. In an earlier paper the authors developed a computational framework for the analysis of aging UHPC structures \cite{WanUHPCI}. Such a framework is also quintessential for the scientific investigation of time-dependent processes such as creep, shrinkage \cite{WanConcreep2015b,Hubler15,Wendner15}, and ASR \cite{Alnaggar13} which are affected by spatial gradients in material properties. 
In this paper, the authors discuss in depth the evolution of fracture properties by the example of a chosen UHPC, the choice of a suitable parameter for the formulation of fracture related aging laws, and the age-dependence of size effect. The computational analysis is based on an experimental size effect test campaign and compared to established analytical formulations. 

It is well known that the strength of cementitious concretes increases rapidly at early age. However, the chemical and physical mechanisms behind this phenomenon are complex and consist of multiple dimensions. The cross-effects between hydration reaction, temperature and humidity evolution, and member deformation involve complex chemo-physical mechanisms that operate over a broad range of length and time scales. Notably, evolution laws for maturing concrete based on Arrhenius-type time acceleration concepts are widely supported by a good agreement with experimental data \cite{Byfors1980, Regourd1980, Ulm1995}. 

A review of relevant research is given in \cite{WanUHPCI} and summarized below. 
Ulm and Coussy \cite{Ulm1995} studied the thermo-chemo-mechanical coupling of concrete at early age with a formulation based upon thermodynamics of open porous media composed of a skeleton and several fluid phases saturating the porous space. It accounts explicitly for the hydration of cement by considering the thermodynamic imbalance between the chemical constituents in the constitutive model at the macrolevel, however neglecting the effects from stress and temperature evolutions. Afterwards they extended the thermo-chemo-mechanical cross effects characterizing the autogeneous shrinkage, hydration heat and strength growth, within the framework of chemoplasticity \cite{Ulm1998}. Cervera et al. \cite{Cervera1999} applied the reactive porous media theory and introduced a novel aging model which accounts for the effect of curing temperature evolution featuring the aging degree as an internal variable. They suggested that the evolution of the compressive and tensile strengths and elastic moduli can be predicted in terms of the aging degree \cite{Cervera2000, Cervera2000II}. The model considers the short-term mechanical behavior based on the continuum damage mechanics theory and the long-term mechanical behavior based upon the microprestress-solidification theory \cite{Cervera1999II}. Bernard, Ulm and Lemarchand \cite{Bernard2003} developed a multi scale micromechanics-hydration model to predict the aging elasticity of cement-based materials starting at the nano level of the C-S-H matrix. Lackner and Mang \cite{Lackner2004} proposed a 3-D material model for the simulation of early-age cracking of concrete based on the Rankine criterion formulated in the framework of multi surface chemoplasticity. Gawin, Pesavento, and Schrefler \cite{Gawin2006, Gawin2006II} proposed a solidification-type early-age model and extended it to account for coupled hygro-thermo-chemo-mechanical phenomena, which was already applied to practical problems \cite{Sciume2013, Cervenka2014}.

Di Luzio and Cusatis \cite{DiLuzio2009I,DiLuzio2009II} formulated, calibrated, and validated a hygro-thermo-chemical (HTC) model suitable for the analysis of moisture transport and heat transfer for standard as well as high performance concrete. In this study, classical macroscopic mass and energy conservation laws were formulated in terms of humidity and temperature as primary variables and by taking into account explicitly various chemical reactions including cement hydration, silica fume reaction, and silicate polymerization  \cite{DiLuzio2009I}. Furthermore, Di Luzio and Cusatis \cite{DiLuzio2013}, amalgamated the microplane model %\cite{DiLuzio2007, Cusatis2008I, Cusatis2008II} 
and the microprestress-solidification theory. This unified model takes into account all the most significant aspects of concrete behavior, such as creep, shrinkage, thermal deformation, and cracking starting from the initial stages of curing up to several years of age. 

While continuum mechanics and finite element solvers are broadly utilized for mechanical analyses of concrete structures at the macroscopic level, the Lattice Discrete Particle Model (LDPM) \cite{Cusatis2011a,Cusatis2011b,WanMarsCon} provides additional insights into the failure behavior of concrete at smaller length scales. LDPM simulates concrete at the length scale of coarse aggregate pieces (mesoscale) and is formulated within the framework of discrete models, which enables capturing the salient aspects of material heterogeneity while keeping the computational cost manageable \cite{Cusatis2011a}. 

The HTC model and LDPM are selected as basis for the early age mechanical model formulated in the authors' previous work \cite{WanUHPCI}. The hygro-thermo-chemical (HTC) model is coupled with the Lattice Discrete Particle Model (LDPM) by aging functions formulated in terms of aging degree. The proposed computational framework can accurately simulate the development of the internal structure of concrete (C-S-H reactions) and the corresponding effects on the mechanical properties. The details of the computational framework can be found in Wan et al. \cite{WanUHPCI} and are summarized in Appendix~A.

Many infrastructure materials are quasi-brittle, including concrete. This fact is well known in the mechanics and material sciences community. Yet, it took decades for the concrete engineering societies all over the world to slowly accept the quasi-brittle nature of (ultra high performance) concretes and, as a consequence, the existence of the size effect on structural strength. Most recently, the discussion on updated shear provisions for the next generation of design codes in Europe and the US was centered on the size effect \cite{Yu2016}. The main reason is the difficulty to test size effect of structural scale components in a laboratory, although SE is an essential aspect for the transfer of laboratory scale test results to real structures. This situation is further complicated by the fact that the fracture properties of concrete evolve over time due to chemical reactions that continue for decades. In the literature, only a limited number of studies on the age-dependence of fracture characteristics are available \cite{Gettu, Kim, Ostergaard}. Furthermore, computational frameworks capable of comprehensively capturing and predicting the age-dependent fracture characteristics are scarce, especially for UHPCs. Although size effect has long been known in the civil engineering community, its age dependence has not been studied in the available literature. With the recently developed aging framework HTC-LDPM \cite{WanUHPCI}, it becomes now possible to investigate the age-dependence of size effect through calibrated simulations and validated predictions. During the formulation of the framework, namely the aging functions to compute tensile strength, tensile characteristic length, and fracture energy, major insights regarding the evolution of fracture characteristics were gained, which are extensively discussed in the following sections.

Fully cured UHPC beams with scaled geometries are utilized in three-point-bending tests for the size-effect study. Test data of beams of one size but on various ages are available from previous studies \cite{WanUHPCI} and they serve for the calibration and validation of the aging framework. After confirming the model's ability to correctly predict the size effect at a given age as well as the aging effect of a given size, predictions are carried out for all early ages and scaled specimen sizes. The nominal flexural strengths obtained from experiments as well as predictive simulations are then analyzed by various methods including the size effect law (SEL) \cite{Bazant1984} and the cohesive size effect curve (CSEC) method \cite{Schauffert2009}. The tensile strength and tensile characteristic length can be directly obtained through data fitting by SEL and CSEC. Afterwards the initial and total fracture energy are computed and further analyzed in terms of aging. The stress field orthogonal to the crack and the associated dissipated energy for various sizes and ages, determined from simulations, are discussed. Lastly, the magnitude of size effect is evaluated in terms of concrete aging. 
   
\section{Literature Review on Size Effect}

The mechanical size effect of quasi-brittle materials, e.g. concrete, rock, ceramics, etc., has been broadly described in the literature as the dependence of the structural strength on the structural size. More specifically, the structural strength, a normalized measure of the load-carrying capacity of the structure, decreases as the structural size increases. In the case of beam specimens, tested in three-point-bending (TPB), the structural flexural nominal stress $\sigma_N$ can be defined as: 
\beq
\sigma_N=\frac{3PS}{2BD^2}
\label{stress}
\eeq
where $P$ = load, $S$ = specimen span, $B$ = specimen thickness, and $D$ = specimen depth \cite{Bazant1998}. 

Based on the cohesive crack model, with a linear softening law, the size effect for mode~$I$ fracture can be described by the following equation \cite{Planas1997,Cedolin2008}: 
\beq
\left(\frac{f'_t}{\sigma_{Nu}}\right)^2 = \Phi\left(\frac{D}{\ell_{ch}}\right)
\label{SizeEffectFracture}
\eeq
where $f'_t$ = tensile strength, $\sigma_{Nu}$ = nominal strength associated with the peak load $P_u$, $D$ = size of structural member, and $\ell_{ch}$ is Hillerborg's characteristic length: $\ell_{ch}=EG_F/f_t^{'2}$, $E=$ Young's modulus, and $G_F$ = total fracture energy (energy required to create a unit area of stress-free crack). The approximation of $\Phi$ can be obtained by carrying out numerical simulations of cohesive crack propagation in geometrically similar structures. Following Cusatis and Schauffert \cite{Schauffert2009}, size effect relationships represented symbolically by Eq.~\ref{SizeEffectFracture}, and all other similar relationships involving normalized ultimate nominal stress as a function of normalized structural size, will be termed \emph{cohesive size effect curves} and abbreviated \textquotedblleft CSEC\textquotedblright. 

On the other hand, with no reference to the cohesive crack model, the size effect law (SEL), first formulated by Ba\v{z}ant in 1984 \cite{Bazant1984}, can be derived on the basis of equivalent linear elastic fracture mechanics (LEFM). The LEFM crack initiation condition is written with reference to an \textquotedblleft effective crack length\textquotedblright as $G(\alpha_0+c_F/D) = \sigma_{Nu}^2 D g(\alpha_0+c_F/D)/E = G_F$, where $G(\alpha)$ = energy release rate, $\alpha_0=a_0/D$ is the initial dimensionless notch depth, $a_0$ = initial notch depth, $g(\alpha)$ = dimensionless energy release rate, and $c_F$ = effective fracture process zone length, assumed to be a material property \cite{Bazant1998,Schauffert2009}. By approximating $g(\alpha_0 + c_F/D)$ with its Taylor series expansion at $\alpha_0$ and retaining only up to the linear term of the expansion, the classical form of Ba\v{z}ant's SEL can be obtained:
\beq
\sigma_{Nu} = \sqrt{\frac{EG_F}{g'_0 c_F + g_0D}}
\label{nominal}
\eeq
where $g_0$ is the dimensionless energy release rate and $g'_0$ is its derivative for $\alpha=\alpha_0$. By introducing Hillerborg's characteristic length, $\ell_{ch}$, Ba\v{z}ant's SEL, can be recast into the following form: 
\beq
\left(\frac{f'_t}{\sigma_{Nu}}\right)^2 = g_0\frac{D}{\ell_{ch}} + g'_0\frac{c_F}{\ell_{ch}}
\label{BSEL}
\eeq

As already proven by previous analytical and numerical studies \cite{Schauffert2009,Planas1986,Planas1992}, the SEL is equivalent to the asymptotic behavior of the CSEC, namely the CSEC tends asymptotically to a straight line for large sizes that corresponds to the SEL. For a nonlinear softening law and for realistic structural sizes, it can be shown that Eq.~\ref{SizeEffectFracture}$\sim$\ref{BSEL} still apply, if one considers the initial fracture energy $G_f$ characterizing the initial part of the softening law (see Fig.~\ref{fe}d) and the associated characteristic length $\ell_1=EG_f/f_t^{'2}$.

To facilitate the identification of $G_f$ and $\ell_1$ through the CSEC, an approximated analytical CSEC formula for TPB geometries was developed to match both the small-size (plastic limit) and large-size (asymptotic) behaviors \cite{Schauffert2009} with the form:

\beq
\frac{f_t^{'2}}{g'_0\sigma_{Nu}^2} = \frac{g_0D}{g'_0\ell_1} + \left(1+11\sqrt{\frac{g_0D}{g'_0\ell_1}}\right)\left(\beta_0+25\sqrt{\frac{g_0D}{g'_0\ell_1}}\right)^{-1}
\label{CSEC}
\eeq
where $\beta_0=9(1-\alpha_0)^4g'_0$. The associated SEL \cite{Schauffert2009} is:

\beq
\frac{f_t^{'2}}{g'_0\sigma_{Nu}^2} = \frac{g_0D}{g'_0\ell_1} + 0.44
\label{SEL}
\eeq
%Note the characteristic length $\ell_1$, distinguished from $\ell_{ch}$, corresponds to the initial fracture energy: $G_f=\ell_1f_t^{'2}/E$. A nonlinear data fitting by size effect experimental results with the CSEC and corresponding SEL, Eq.~\ref{CSEC}~\&~\ref{SEL}, provides estimates of the tensile strength, $f'_t$, and the characteristic length, $\ell_1$. 

\section{UHPC Size Effect Study}

\subsection{Mixture, constituents, and curing procedures}
The mixture proportions for the adopted UHPC mix design are reported in Table~\ref{mix}. The material composition consists of LaFarge Type~H cement, F-50 Ottawa sand, Sil-co-sil 75 silica flour, Elkem ES-900W silica fume, ADVA-190 Superplasticizer and tap water. The maximum particle size, 0.6 mm, is limited to that of silica sand, which is a foundry grade Ottawa sand  \cite{CorTuf2009}. In the author's previous studies \cite{WanUHPCI}, two curing protocols with and without hot water bath curing were explored. A first group of specimens was kept in the humidity room (HR) for 14 days, a second group, instead, was kept in the humidity room for 7~days after which it was placed in hot water bath (WB) at 85$^\circ$C for another 7~days. After 14~days HR and WB curing, the specimens of both groups were stored at room temperature at an average RH of 50\% until testing. 
The specimens for the experimental size effect study were cured following the WB curing procedures, while the predictive simulations were ran for both curing routines.

\begin{table}[h]
\caption{Constituents and mixing proportions of UHPC}\label{mix}
\centering
\begin{tabular}{l l c c}
\hline
Ingredient	&Type	&Proportion	&Weight per kg	\\
\hline
Cement 	&Lafarge Type H&1.0000	&0.3497	\\
Sand 	&F-50&0.9674	&0.3383	\\
Silica Flour  &	Sil-co-sil 75&0.2768	&0.0968	\\
Silica Fumes 	&Elkem ES-900W&	0.3890	&0.1360	\\
Superplasticizer&	ADVA-190&	0.0180&	0.0063\\
Water	&Tap Water&	0.2082&	0.0728	\\
\hline
\end{tabular}
\end{table}

\subsection{Experiments}

Three sets of beams were cast with a geometrical scaling factor of $\sqrt{5}$ for depth and span, but with a constant thickness of 25.4~mm (1~in). Their nominal dimensions are as follows: size~M, 25.4$\times$25.4$\times$127~mm (1$\times$1$\times$5~in); size~L, 56.8$\times$25.4$\times$254~mm ($\sqrt{5}$$\times$1$\times$10~in); size~XL, 127$\times$25.4$\times$558.8~mm (5$\times$1$\times$22~in). The depth to span ratio was kept at 1:4. After 120~days of average age, the specimens were machined to have 50\% notch depth, and were tested in a TPB configuration in crack mouth opening displacement (CMOD) control as shown in Fig.~\ref{exp}a. The actual dimensions measured on the day of testing are utilized for post-processing and are listed in Table~\ref{dima}. The testing results are shown in Fig.~\ref{exp}b, in which the nominal flexural stress is computed based on Eq.~\ref{stress}, and the nominal strain is calculated by $\epsilon_N=\delta/D$ based on the CMOD opening measurements $\delta$. The total fracture energy $G_F$, according to the work of fracture method, is computed as the area under the force-displacement curve divided by the ligament area. As one can see, the nominal strength decreases significantly with size and the total fracture energy is approximately constant with an average value of $G_F^{exp}$=63.3 J/m$^2$.

\begin{table}[h]
  \caption{Actual specimen dimensions for size effect tests}
  \label{dima}
  \centering
  \begin{tabular}{l l l l l c}
  \hline
  Size & length $L$ [mm] & Thickness $B$ [mm] & Depth $D$ [mm] & Test span $S$ [mm] & No. of specimens\\
  \hline
  M &	128.0$\pm$0.2\% &	26.3$\pm$1.9\% &	26.2$\pm$1.1\% &	101.6 & 8\\
  L &	254.1$\pm$0.4\% &	26.5$\pm$4.6\% &	58.0$\pm$0.7\% &	227.2 & 4\\
  XL &	 558.2$\pm$0.0\% &	26.5$\pm$4.3\% &	127.3$\pm$0.9\% &	508 & 3\\
  \hline
  \end{tabular}
\end{table}

\bfi [h] 
\centering
(a)  \includegraphics[height=2in,valign=c]{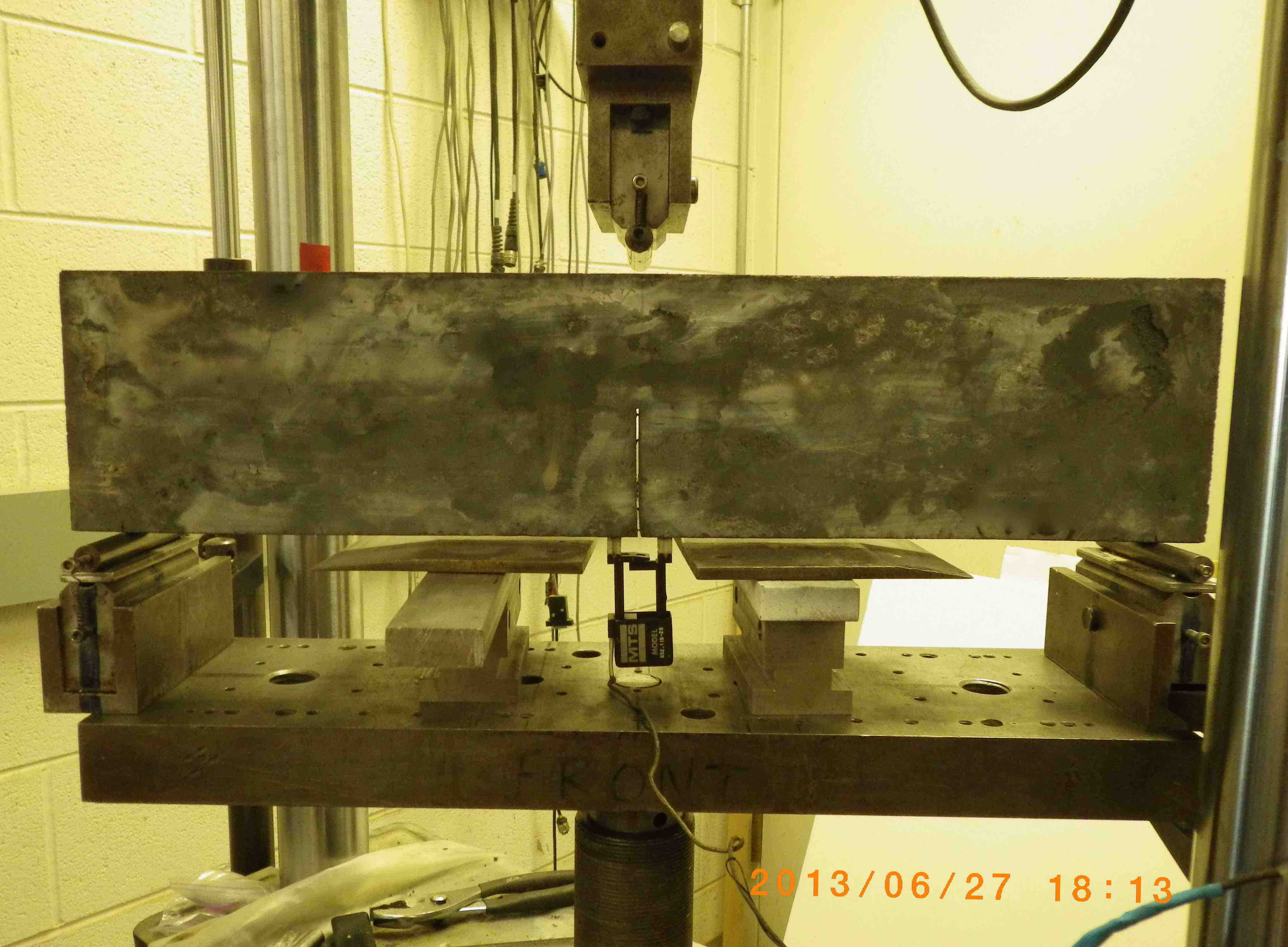} 
(b)    \includegraphics[height=3in,valign=c]{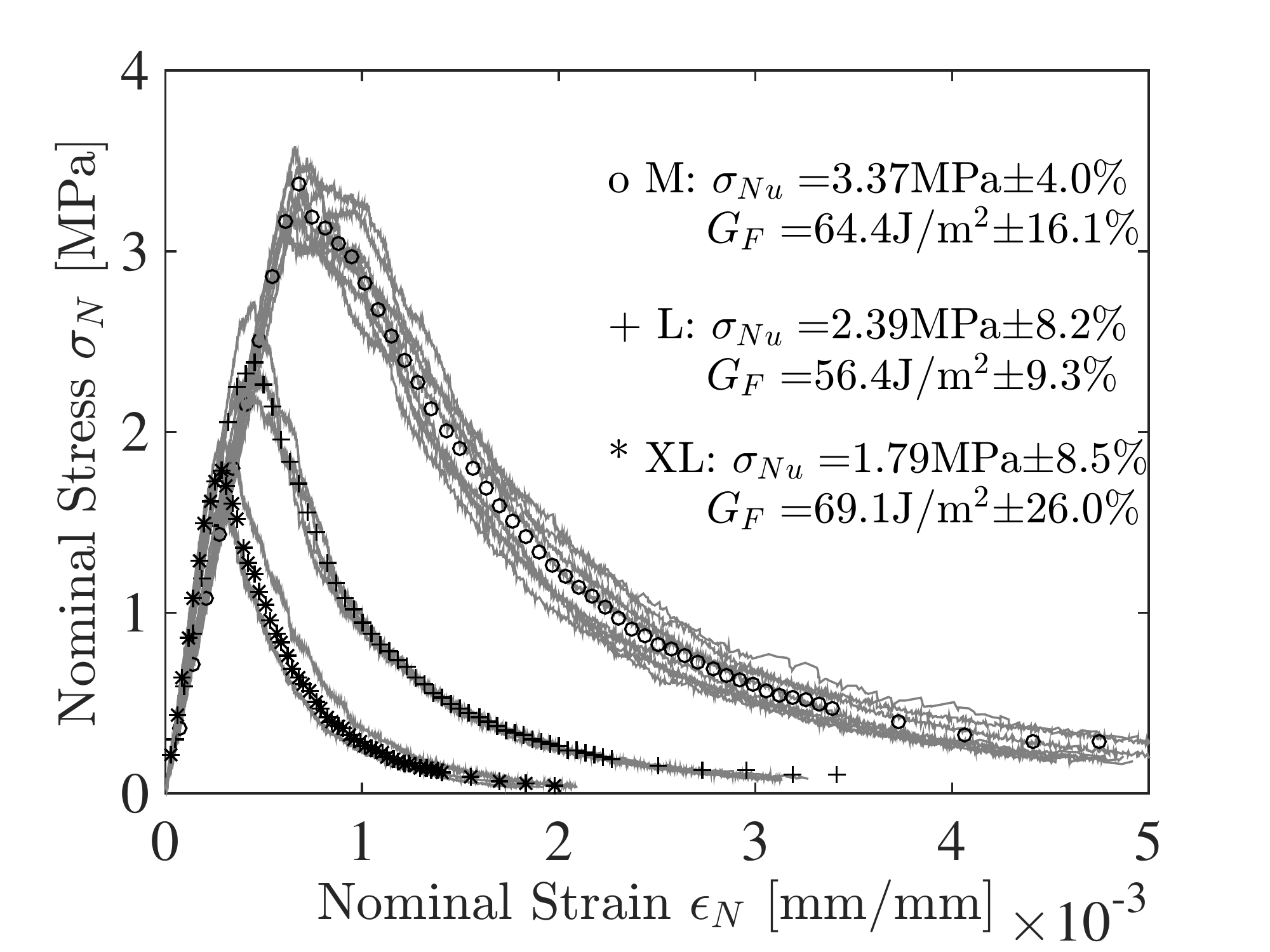} 
    \caption{Experimental UHPC size effect data: (a) test setup, (b) testing results: nominal stress-strain curves of three point bending tests}
    \label{exp}
\efi

\subsection{LDPM Simulations}
For the fracture analysis of the investigated UHPC, experimental data of one age for different sizes and aging data of one size for different ages are available. Numerical simulations are employed in order to (i) gain additional insights into the behavior of the available experimental data, and (ii) generate additional size-effect data for other ages. 
All numerical simulations are performed utilizing a recently formulated age-dependent mesoscale computational framework \cite{WanUHPCI, WanUHPCc, WanPhDNU, WanKEM2016}. It essentially links the hygro-thermo-chemical (HTC) model \cite{DiLuzio2009I, DiLuzio2009II, DiLuzio2013} and the Lattice Discrete Particle Model (LDPM) \cite{Cusatis2011a, Cusatis2011b} through a set of aging functions. The details of the computational framework are summarized in Appendix~A. Additional background is given in Wan et al. \cite{WanUHPCI} which also discusses the calibration of the HTC model and chemo-mechanical coupling through the aging functions which are defined in terms of aging degree $0 \leq \lambda \leq 1$. The aging degree is a suitable parameter to quantify the maturity of concretes arising from cement hydration and silica-fume reaction and takes into account the temperature evolution \cite{WanUHPCI}. The normal modulus $E_0$ is proportional to $\lambda$, strength parameters are assumed to be proportional to $\lambda^{n_a}$, and the tensile characteristic length $\ell_t \propto k_a (1-\lambda)$, where $n_a$ and $k_a$ are positive constants. It was found that $n_a = 2.33$ and $k_a = 22.2$ for the investigated UHPC. 
The asymptotic material properties at $\lambda = 1$ are given in Table~\ref{seldpm} together with the corresponding values for nominally 120-day old UHPC specimens, calculated using the aging functions. Parameters affecting the tensile fracturing behavior are listed in the first block. It must be observed here that the parameters reported in Table~\ref{seldpm} have been identified on the basis of the TPB test data of medium size only, along with data on unconfined compression \cite{WanUHPCI, Smith2014, WanConcreep2015}. Consequently, the simulations of large (L) and extra-large (XL) specimens must be regarded as pure predictions. Fig.~\ref{setup} shows the simulation setup as well as the obtained crack patterns. In order to save computational cost, the middle section of the specimen with a length equal to the specimen depth is simulated through the mesoscale model whereas the rest of the specimen is simulated with elastic finite elements. Continuity at the interfaces is enforced by means of a master-slave algorithm.

The LDPM simulation results as well as crack patterns for each size can be found in Fig.~\ref{simu}. Through the random particle placement a certain degree of spatial variability is introduced, resulting in differences in the macroscopic response. Consequently, at least three specimens with different particle placement were simulated and the results were averaged for each size to obtain a representative curve. The simulations have slightly higher stress values compared to the experimental data. This is most likely due to shrinkage cracking of the specimens tested experimentally, which is not considered in the current numerical analysis. The fracture energy from simulations are calculated as the area under the simulated force-displacement curve divided by the nominal ligament area. Note that the $G_F^{sim}$ values in Fig.~\ref{simu} are computed after truncating the simulated force-displacement at a point at which the simulated crack mouth opening values correspond to the respective experimental values at failure. $G_F^{sim}$ has an average value of 68.3 J/m$^2$, which is about 7.9\% higher than that of the experiments ($G_F^{exp}$=63.3 J/m$^2$). When the full force-displacement curve is considered, $G_F^{sim}$ has a value of about 100.6 J/m$^2$. Overall, the simulations can represent the UHPC behavior in a TPB test setup with high accuracy and capture very well the experimentally observed size effect for ``fully'' (aging degree of 0.95) cured specimens. The crack patterns from simulations also coincide with those of the TPB experiments. In addition, the capability of the framework to well reproduce the age-dependent response was shown in an earlier study \cite{WanUHPCI}. Thus, it can be assumed that the model can also predict well the age dependence of size effect.

\bfi [ht]
\centering
  \includegraphics[width=1in]{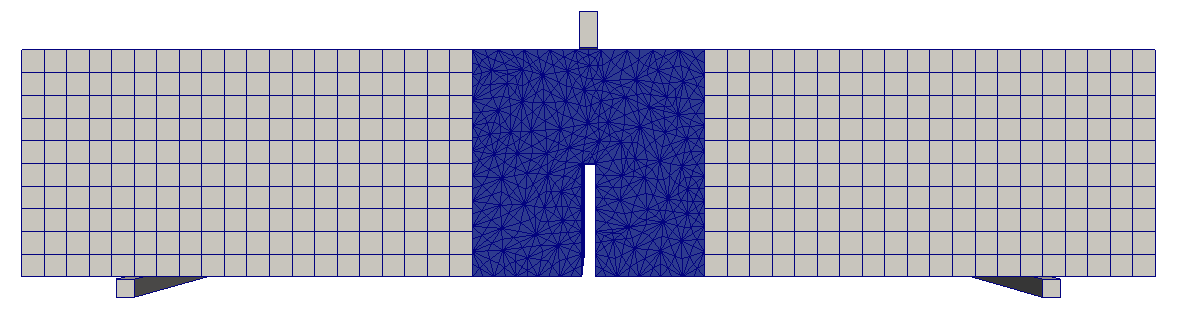} \\
   \includegraphics[width=2.2in]{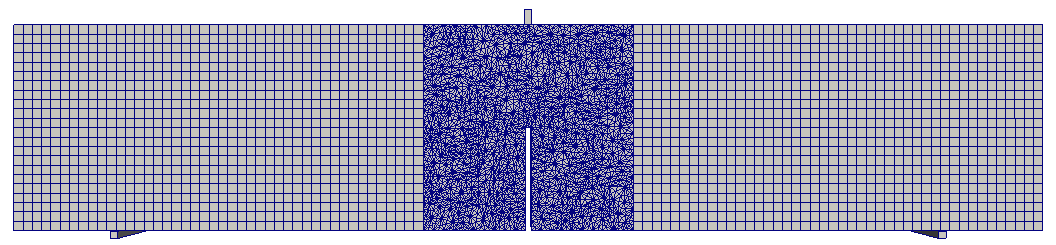} \\
    \includegraphics[width=5in]{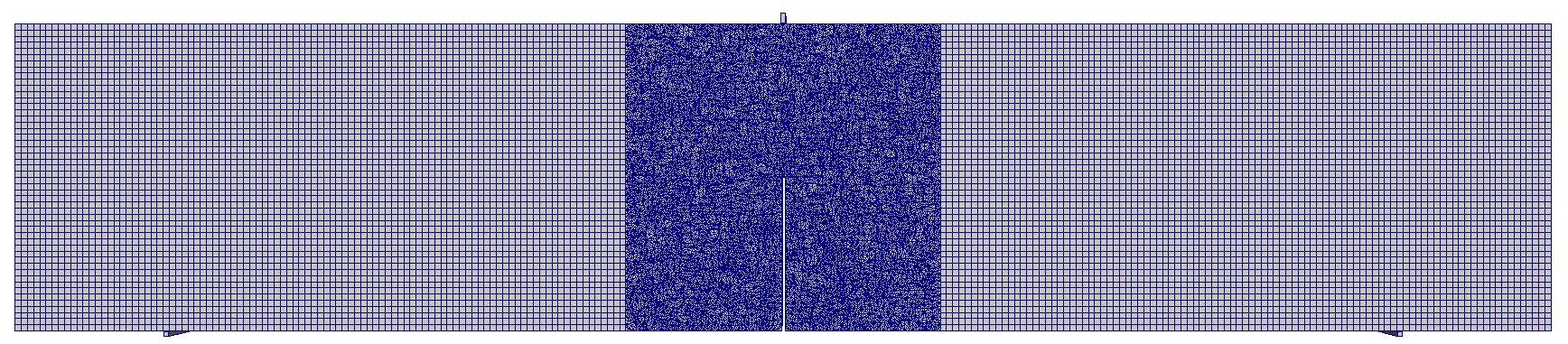} \\
    \caption{UHPC size effect LDPM simulation setup (top to bottom: M, L, XL)}
    \label{setup}
\efi    
 
\bfi [ht]
\centering
(a)    \includegraphics[height=2.5in,valign=t]{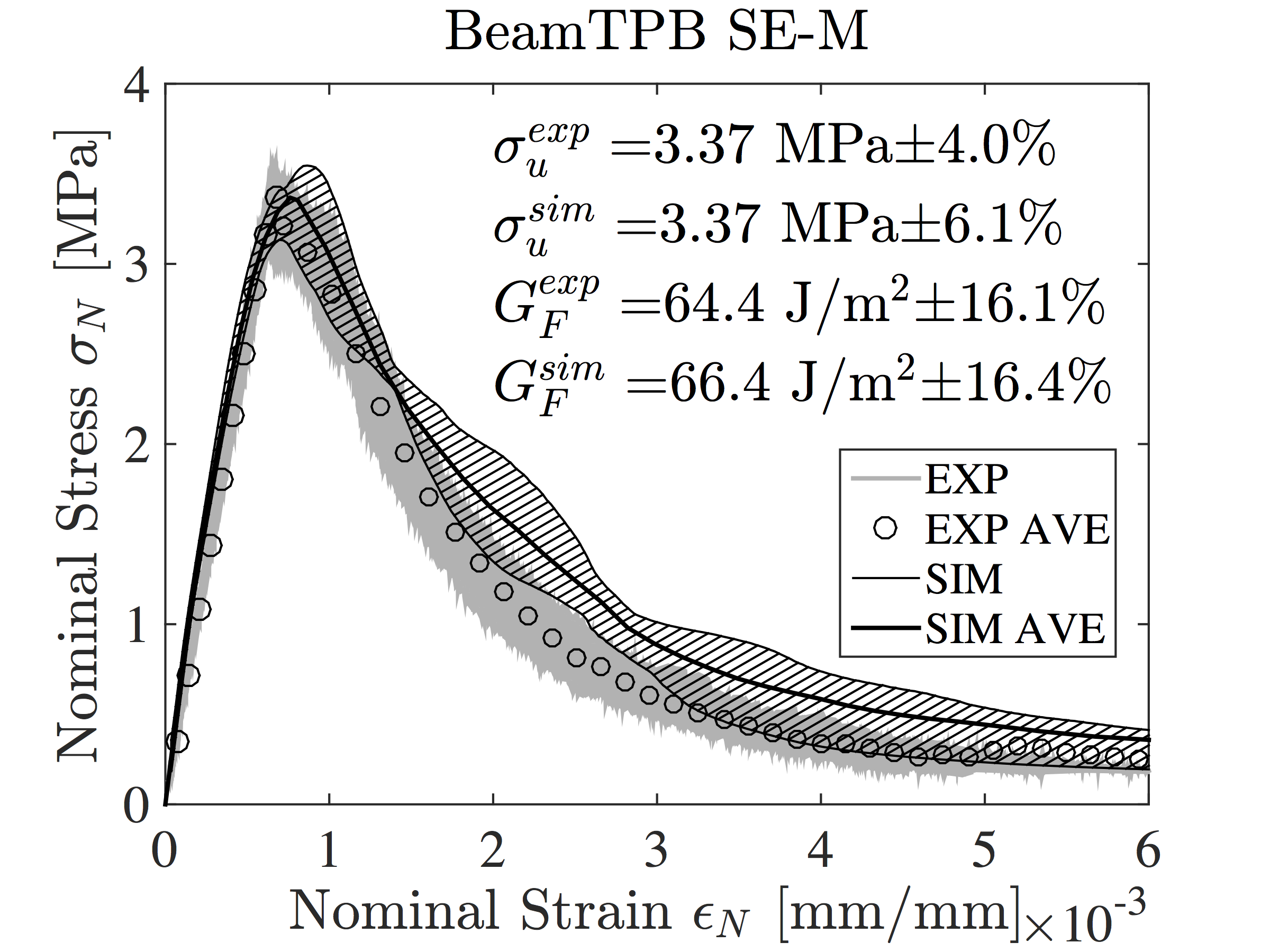} 
    \includegraphics[height=2.5in,valign=t]{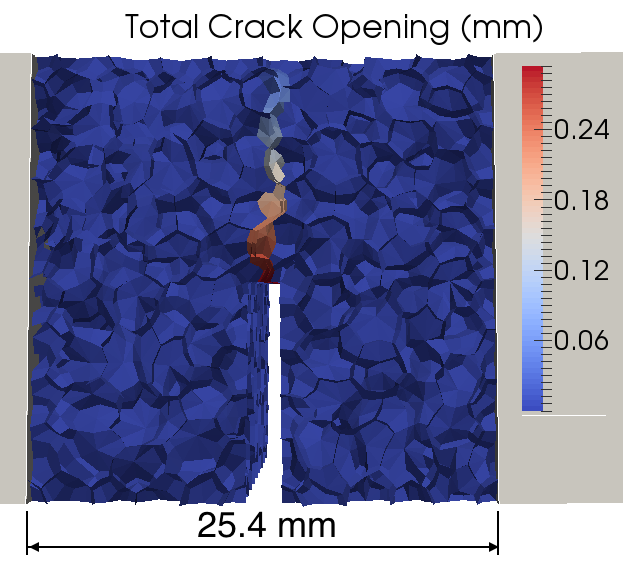}\\
(b)   \includegraphics[height=2.5in,valign=t]{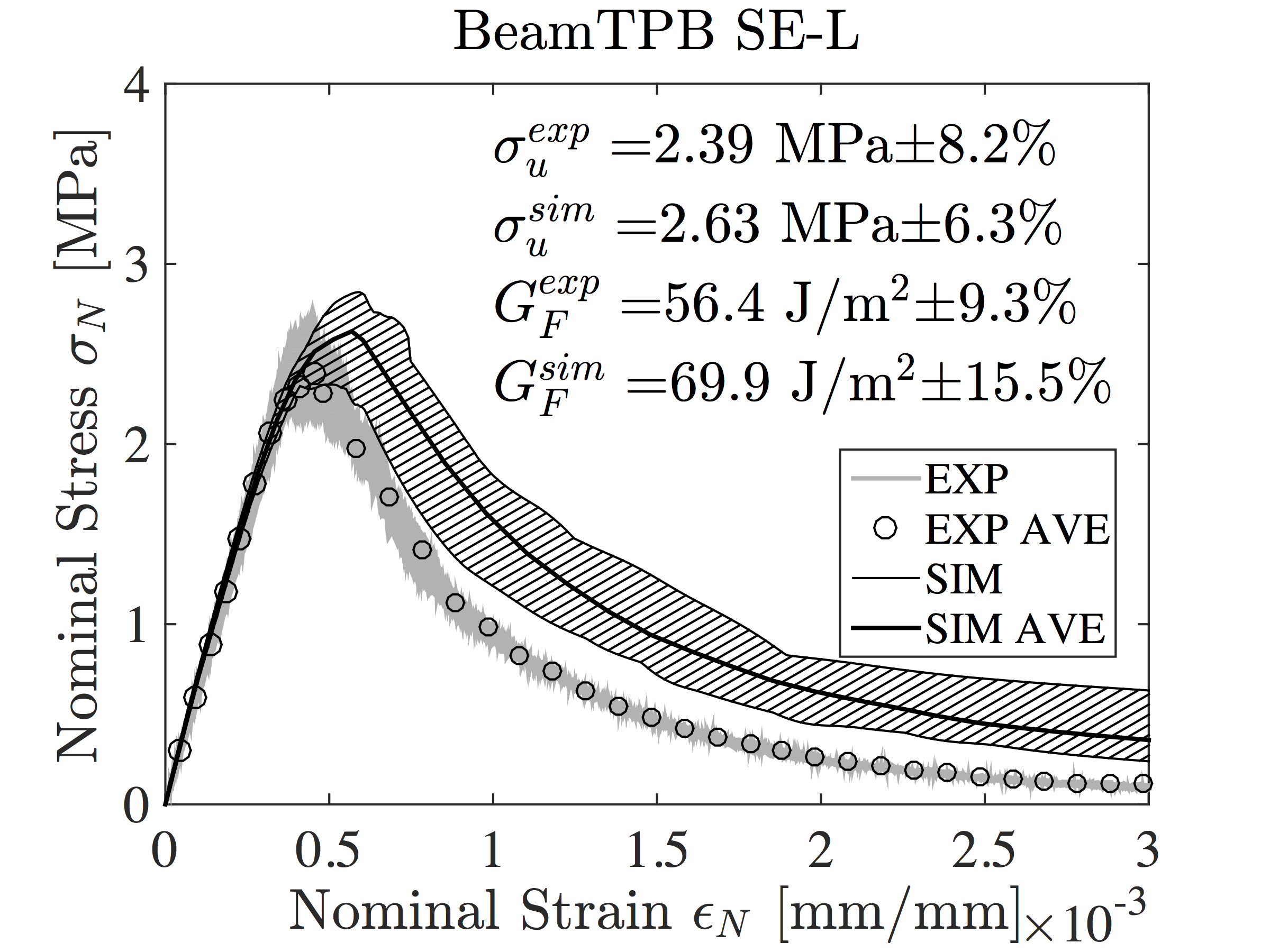} 
       \includegraphics[height=2.5in,valign=t]{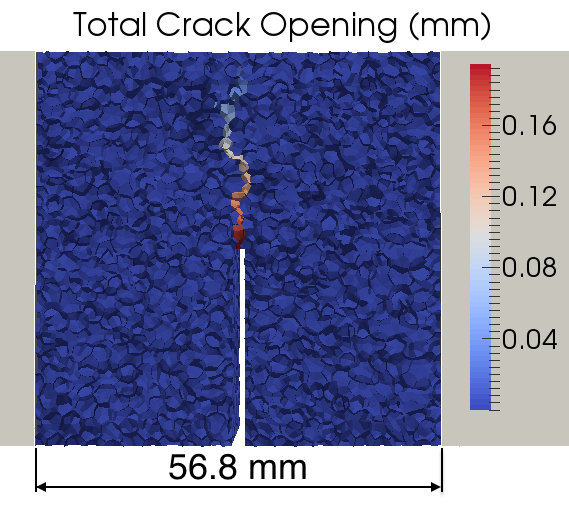}\\   
(c)    \includegraphics[height=2.5in,valign=t]{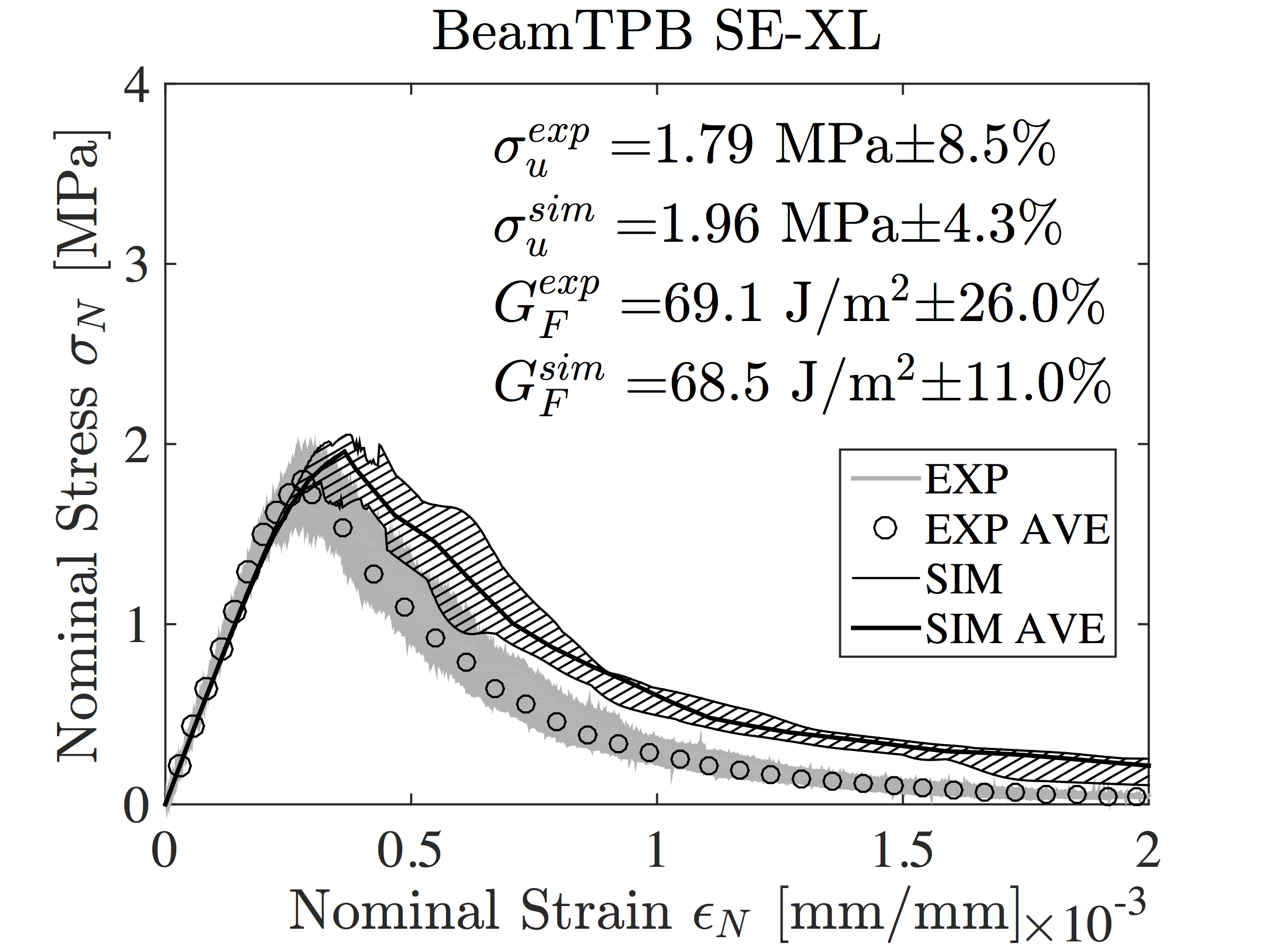} 
        \includegraphics[height=2.5in,valign=t]{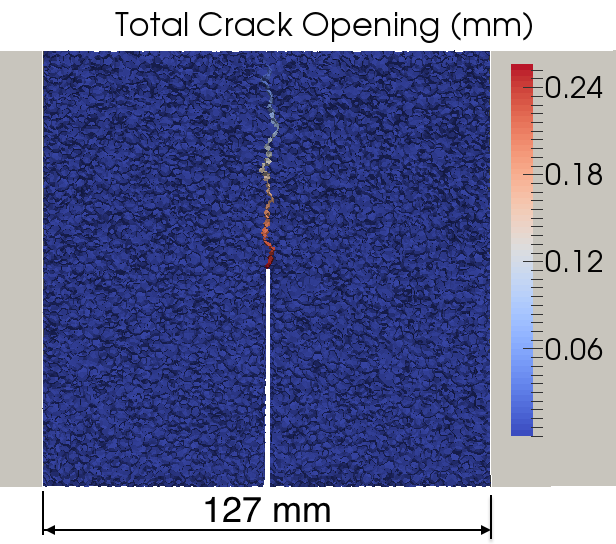}\\
    \caption{UHPC size effect experiments vs. LDPM simulations and simulated crack propagation for (a) size M, (b) size L, and (c) size XL}
    \label{simu}
\efi

\begin{table}[ht]
\caption{Size Effect - LDPM Parameters}\label{seldpm}
\centering
\begin{tabular}{l c c}
\hline\hline
Days & 120 (HR+WB) & Asymptote\\
\hline\hline
Ave. aging degree & 0.9546 & 1.0\\
$\pm$ standard deviation [-] & $\pm$0.010 & \\
\hline\hline
Normal modulus $E_0$ [MPa]  & 71595 & 75000\\
Tensile strength $\sigma_t$ [-] [MPa] & 11.9 & 13.3\\
Tensile characteristic length $\ell_{t}$ [mm] & 21.0 & 10.6\\
Fracture energy $G_t$ [J/m$^2$]$^*$  & 21.13 & 12.47\\
Softening exponent [-]  &0.28 & 0.28\\
\hline
Densification ratio $E_d/E_0$ [-]  &2.5 & 2.5\\
Alpha [-] & 0.25 & 0.25\\
Compressive yielding stress [MPa] & 449 & 500\\
Shear strength ratio [-]  &5.5 & 5.5\\
Initial hardening modulus ratio [-]  &0.36 & 0.36\\
Transitional strain ratio [-]  &4 & 4\\
Initial friction [-]  &0.0335 & 0.0335\\
Asymptotic friction [-]  &0 & 0\\
Transitional stress [MPa]  & 269 & 300\\
Volumetric deviatoric coupling [-]  &0 & 0\\
Deviatoric strain threshold ratio [-]  &1 & 1\\
Deviatoric damage parameter [-]  &5 & 5\\
\hline\hline
$^*$ Calculated as $G_t = \ell_{t} \sigma_t^2 /2E_0$
\end{tabular}
\end{table}

\subsection{Size Effect Analysis}

In this section CSEC \& SEL (Eq.~\ref{CSEC}\&\ref{SEL}) are fitted to the experimental and numerical nominal strengths in order to identify macroscopic fracture properties ($f'_t$, $\ell_1$, $G_f$) and assess the reliability of such identification procedure. For a single-edge cracked beam subjected to three-point bending, the expression for fracture analysis depends mildly on the shear force magnitude near the central cross-section, i.e., on the span-to-depth ratio \cite{Bazant1998}. For the case $S$/$D$=4 as in this study, the dimensionless energy release rate can be calculated, according to Pastor et al. \cite{Pastor1995, Bazant2009}, as $g(\alpha)=k(\alpha)^2$, $k(\alpha) = \sqrt{\alpha}p(\alpha)/((1+2*\alpha)(1-\alpha)^{1.5})$, and $p(\alpha) = 1.900 - \alpha [-0.089+0.603(1-\alpha)-0.441(1-\alpha)^2+1.223(1-\alpha)^3]$.

For 50\% notched beams, the initial relative notch depth $\alpha_0$ = 0.5, and one can obtain $g_0$ = 3.1415, $g'_0$ = 19.858. Both CSEC and SEL are fitted to the mean values of each size, with and without weighting inversely proportional to the sample variance. Due to the small number of sizes and the relatively constant scatter between sizes, the fits with and without weights correspond approximately. The fitted CSEC \& SEL curves for experimental and simulated ultimate strengths are presented in Fig.~\ref{fit}a\&b, respectively. 
The axes $X = g_0D/g'_0\ell_1$ and $Y = f_t^{'2}/g'_0\sigma_{Nu}^2$ are taken as in Eq.~\ref{CSEC}\&\ref{SEL}. The two parameters, tensile strength, $f'_t$, and tensile characteristic length, $\ell_1$, are obtained through the CSEC \& SEL data fitting. The initial fracture energy is calculated as $G_f=\ell_{1}f^{'2}_t/E$ using the respective fitted values ($f'_t$, $\ell_1$) of CSEC \& SEL. Note, in the figure the SEL lines are normalized based on CSEC fits for comparison reasons. 

As expected, CSEC and SEL fit the simulated SE data almost perfectly (R$^2>0.99$). In both cases, the CSEC fits the data slightly better than the SEL which represents the asymptotic shape of the CSEC for large sizes and, thus, can not fully capture the curvature of the size effect trend for small sizes.
The CSEC fitted tensile strength is 20\% lower than the SEL fitted value; The CSEC fitted tensile characteristic length is about 40\% higher than that of SEL; The calculated initial fracture energy for CSEC is roughly 4.8\% higher than that for SEL. 
When compared to the LDPM parameters $\sigma_t$ \& $\ell_t$ (see Table~\ref{seldpm}), the CSEC fitted $f'_t$ \& $\ell_1$ (see Fig.~\ref{fit}b) are 10.9\% lower and 17.6\% lower, respectively. In comparison, the SEL fitted values are 16.0\% higher and 54.3\% lower, respectively. Overall, the CSEC identification is closer to the LDPM parameters than the SEL results. The SEL tends to overestimate the tensile strength and largely underestimates the tensile characteristic length. However, the asymptotic behavior of CSEC and SEL tend to coincide with each other. 

In conclusion, both CSEC and SEL can provide accurate fits. However, the CSEC is more reliable than the SEL in terms of providing fracture related parameter values from data fitting, especially if data on smaller sizes is available for which the asymptotic shape of the SEL is not sufficient. 

\bfi [h]
\centering
(a) \includegraphics[height=2.5in,valign=t]{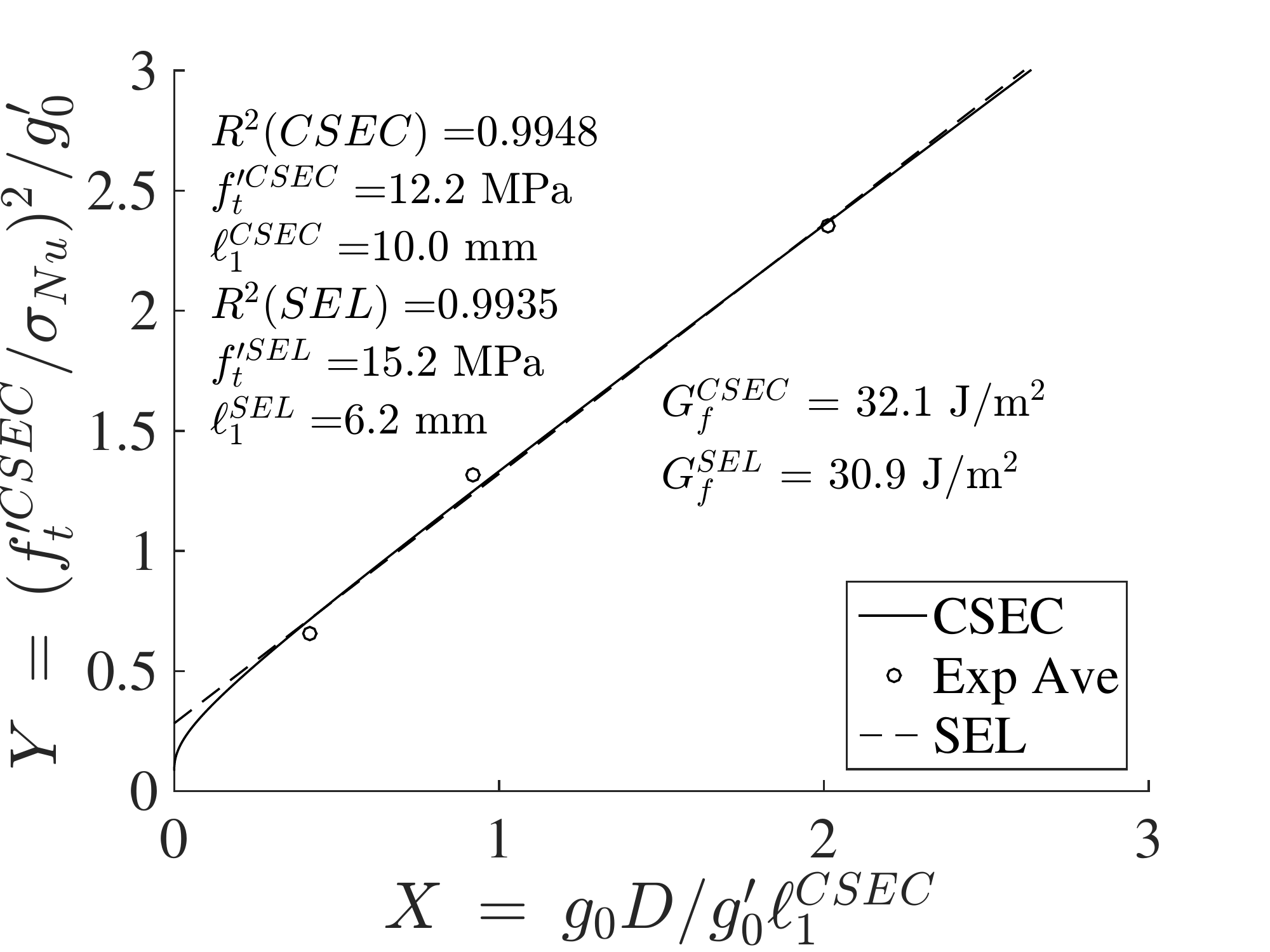} 
(b)  \includegraphics[height=2.5in,valign=t]{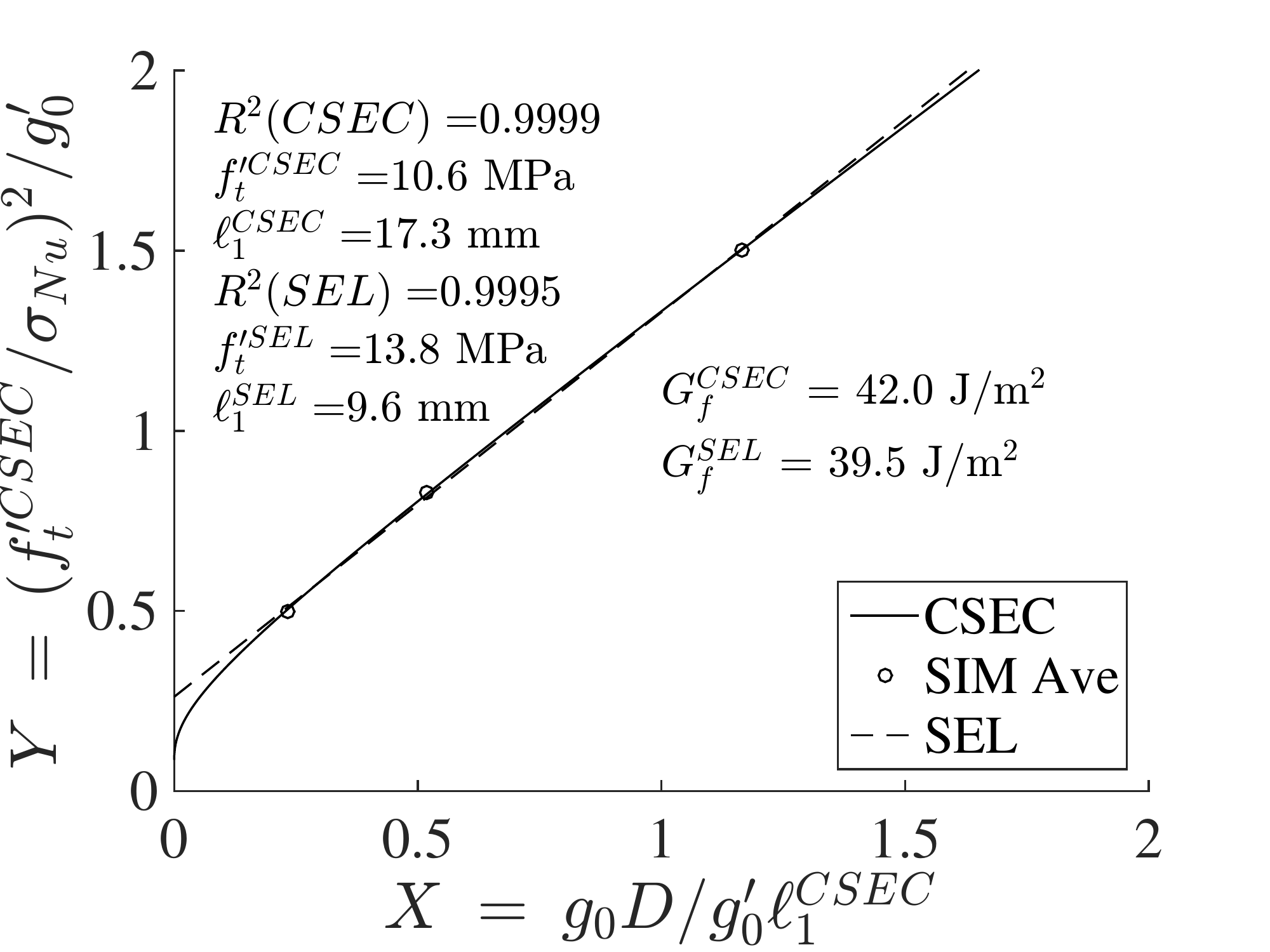}
 \caption{UHPC CSEC and SEL data fitting for (a) experimental, and (b) simulated ultimate strengths of size M, L, and XL}
 \label{fit}
 \efi 

\subsection{Sensitivity study of CSEC}

The use of size effect laws to obtain a unique set of fracture related material properties has been widely promoted, see e.g. \cite{Bazant2013}. In this section the sensitivity of such an identification strategy will be discussed by the example of the cohesive size effect curve (CSEC). The widely used SEL can be assumed to behave similarly for the available size range as the CSEC approaches the SEL asymptotically for large sizes. 
This analysis, furthermore, tries to shed light on the observed differences between the CSEC analysis for the experimental and simulated results. As shown in Fig.~\ref{simu}, the simulations are fairly close to the experimental data, representing well the actual crack propagation. The average nominal strengths $\sigma_N$ from simulations for size M, L and XL are 3.1\%, 9.5\% and 8.8\% higher than those of the experiments, respectively. 

As shown in Fig.~\ref{fit}, the simulated ultimate strengths are fitted by CSEC with R$^2$ value of 99.99\%. Similarly, the experimental data points are matched with R$^2$ = 99.48\%, indicating the excellent capability of CSEC to describe the size effect of concrete. However, the CSEC fitted parameters ($f'_t$, $\ell_1$, $G_f$) for the experimental results differ moderately from those for the simulated set of data. The CSEC fitted $f'_t$ for the experiments is 15\% higher than that for the simulations. The CSEC fitted $\ell_1$ for the experiments is 42\% lower compared to that for the simulations. The correspondingly calculated $G_f$ for experiments is about 24\% lower than that for the simulations, see Fig.~\ref{fit}. 

To study the sensitivity of the CSEC method, bootstrapping \cite{Efron1979} with 10,000 possible realizations for the ultimate flexural tensile strength, $\sigma_{Nu}$, was performed. Bootstrapping is a statistical resampling method 
by which inferences about the population can be made from limited sample data. For the purpose of this investigation it was assumed that the nominal strength of each size follows a normal distribution with known mean and variance. The specimen dimensions were considered deterministic and set equal to the nominal size. This is supported by the observed negligible scatter in the actual average dimensions, $D$, of lower than 1\%. In the Monte Carlo type analysis 10,000 uncorrelated possible realizations of ultimate nominal strength values for the three available sizes are generated and then fitted using the CSEC. Consequently, the mean value and variance of the original strength distributions are preserved and unbiased samples with size $n=$10,000 are generated. The resulting histograms of $f^\prime_t$ and $\ell_1$ can be found in Fig.~\ref{sens}, along with the correspondingly calculated initial fracture energy, $G_f=\ell_{1}f^{\prime 2}_t/E$. The obtained mean $f^\prime_t$ value from the experimental samples is 50\% higher than that for simulations while the mode (most likely value) almost coincide, see discussion below. The average $\ell_1$ and the average $G_f$ for the experimental samples are only 15\% lower and about 20\% lower, respectively, than those for the simulations. 

Overall, the parameter distributions identified from the simulated data exhibit low scatter, as expected, since the simulations do not suffer from measurement errors or unaccounted damage processes such as shrinkage and thermal shock. 
The identification from the experiments likely suffers from noise due to e.g. measurement errors that are sufficiently large compared to the parameters' sensitivities to cause divergence in several fits. The consequence are unrealistically high strength values (they qualify as outliers) which are balanced by very low (and without bounds even non-physical negative) tensile characteristic lengths. The fact that the computationally derived initial fracture energy $G_f$ follows a relatively smooth distribution while $f^\prime_t$ and $\ell_1$ suffer from outliers confirms the previous conclusion. In general, large scatter in the distribution indicates lower confidence in the identified parameters. Furthermore, it can be concluded that the CSEC is a relatively robust method to obtain initial fracture energy but much less so for tensile characteristic length and tensile strength, when noise from various sources is present in the data. 

It is important to note that the mean values obtained from multiple fits (here $f^\prime_t$ and $\ell_1$) to data (here $\sigma_N$ values) will not coincide with the results of a single fit to the experimental means, unless very specific conditions are met. The identified distributions of $f^\prime_t$ and $\ell_1$ clearly are asymmetric and skewed unlike the initial normal distribution of strength values, resulting in a relatively large deviation between the mode (most-likely value) and the mean value. Note that mean values by themselves are unsuitable statistical indicators for skewed distributions.
% we should plot as grey not mean +- sigma but a lower and a upper percentile  
The histograms of $f^\prime_t$ that are obtained for the experimental and the simulated data respectively have similar shapes and coinciding modes, in other words, they qualitatively agree with each other. Yet, the pure comparison of mean values indicates large differences. The reason for this difference is found in the skewness of the distribution which is further aggravated by the large number of outliers in the fitting results of the experimental data sets. 

Provided that the LDPM parameters are well calibrated and validated based on a comprehensive experimental campaign \cite{WanUHPCI}, the LDPM mesoscale tensile strength, $\sigma_t$ = 11.9 MPa, and the tensile characteristic length, $\ell_t$ = 21 mm, (see Table~\ref{seldpm}) can be utilized as reference to evaluate the CSEC fitted results. For the ``perfect'' SE data that was created by LDPM simulations, the CSEC analysis yields a tensile strength, $f^\prime_t$ = 11.4 MPa, and tensile characteristic length, $\ell_1$ = 21.5 mm, which are very close to the mesoscale LDPM parameter set.  

For the experimental dataset, however, this is not the case. The identified $\ell_1$ = 18.6 mm, is 11\% lower than $\ell_t$, and $f^\prime_t$ = 22.2 MPa with outliers almost exceeds $\sigma_t$ by a factor of 2. The likely reasons are, as discussed above, errors in the data or unaccounted phenomena. The resulting high sensitivity in obtaining $f_t^\prime$ from experimental samples can be overcome by introducing constraints during the fitting process. Introducing an upper limit of e.g. 40~MPa for tensile strength during fitting improves the results significantly. This is (almost) equivalent to discarding outliers with identified values higher than 40~MPa in the final analysis. In both cases, the average $f_t^\prime$ drops from 22.2~MPa to 16.9~MPa, with negligible change in the fitted average $\ell_1$ and $G_f$. 

In conclusion, the values of tensile strength and tensile characteristic length from CSEC analysis for the simulated SE data agree with those adopted in LDPM. In terms of sensitivity, obtaining $G_f$ by CSEC is fairly stable against (experimental) noise in the data. However, the CSEC method is very sensitive towards $f_t^\prime$ and in consequence $\ell_1$, which individually can not be trusted. The performance of the method can be improved by exerting upper and lower bounds according to the expected and physically meaningful range of the parameters, especially of $f_t^\prime$. 

\bfi [h]
\centering
(a) \includegraphics[height=2.5in,valign=t]{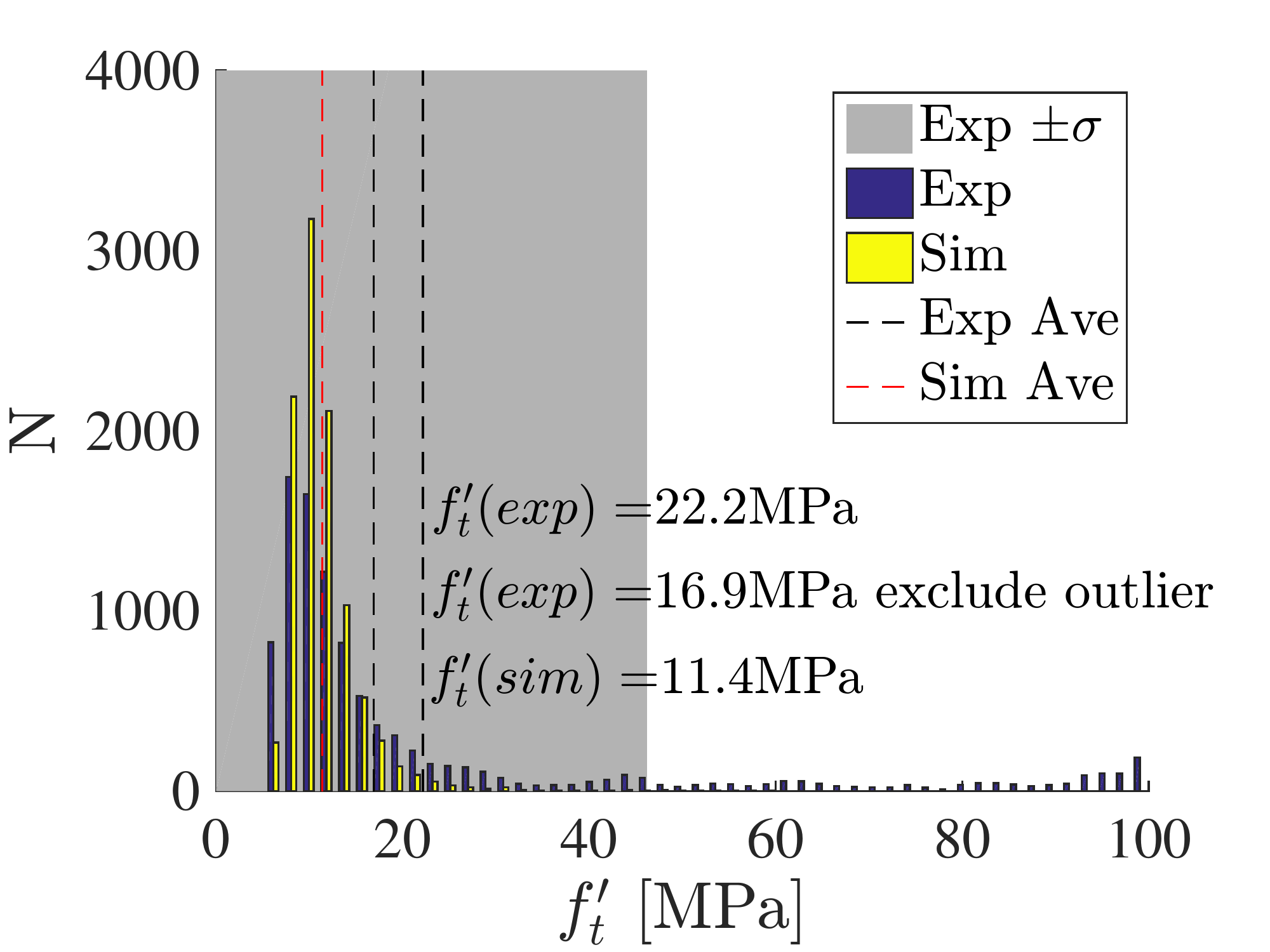}
(b) \includegraphics[height=2.5in,valign=t]{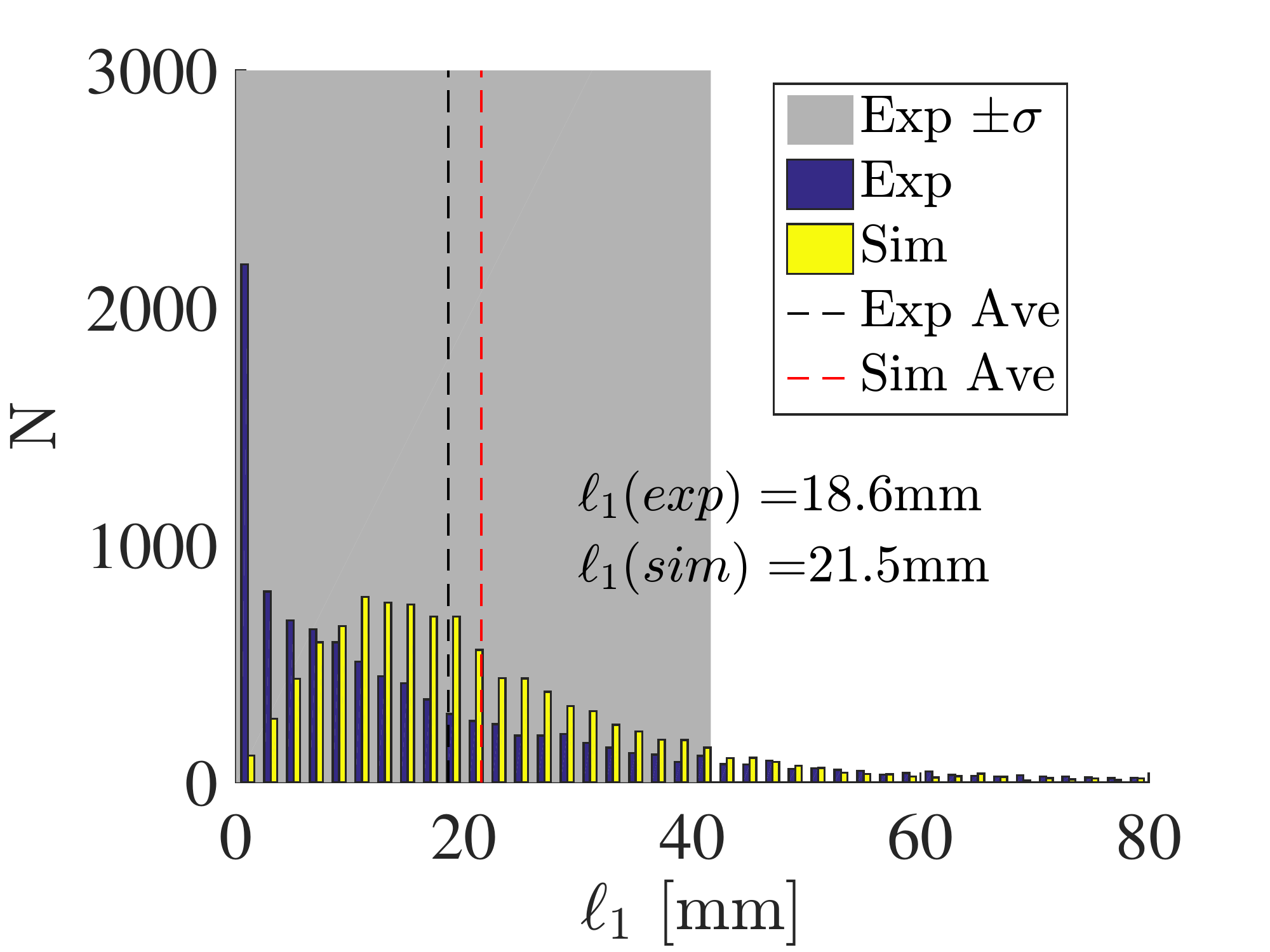}\\
(c) \includegraphics[height=2.5in,valign=t]{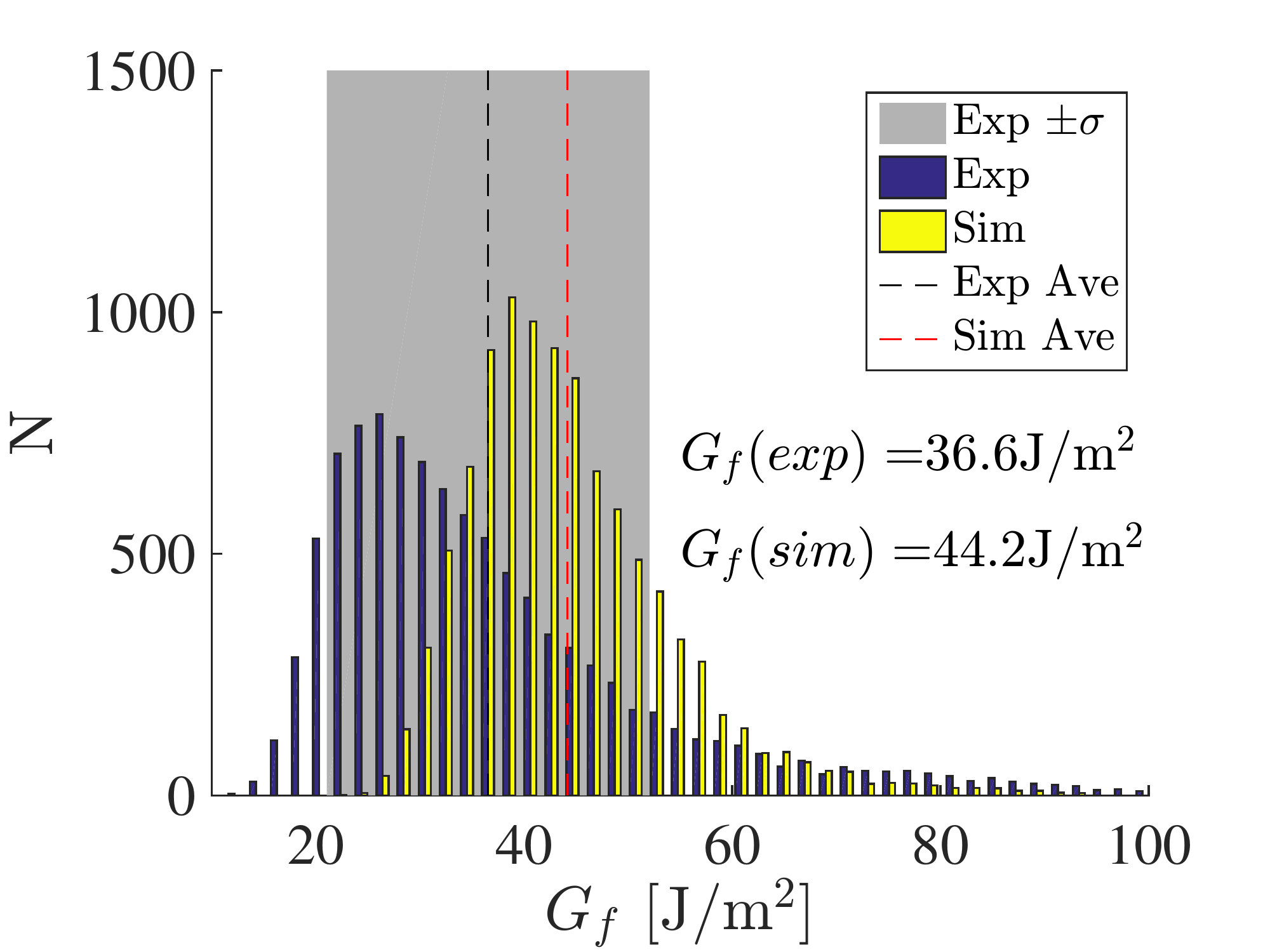}
\caption{Sensitivity study of CSEC based on experimental and simulated peak strength: (a) fitted tensile strength, (b) tensile characteristic length, and (c) initial fracture energy}
\label{sens}
\efi 

\section{HTC-LDPM Simulations and Predictions}

After calibration and validation, predictive size effect simulations for the UHPC (size M, L, XL, XXL) are carried out for the same early ages as investigated by Wan et al. \cite{WanUHPCI}: HR curing 3, 7, 14, 28 days, and HR+WB curing for 14 and 28 days. Size M, L, XL have the same nominal dimensions as described in the previous sections. The largest size XXL is geometrically upscaled from size XL by $\sqrt{5}$ on span and height while keeping the thickness constant. It has the nominal dimensions of 284$\times$25.4$\times$1420~mm (5$\sqrt{5}$$\times$1$\times$25$\sqrt{5}$~in) with a span of 1136~mm (20$\sqrt{5}$ in). No experimental data exists for this size. The simulated stress-strain curves for all sizes and ages are shown in Fig.~\ref{pred}, along with the fitted CSEC \& SEL curves for the corresponding early ages. At least five simulated specimens with different discrete particle placement were utilized for each size and age to ensure representative results. Size M has a relatively large scatter in the late post-peak phase. Size L shows a relatively large scatter in the ultimate peak stress. In spite of the scatter, the predictive size effect simulations by the HTC-LDPM computational framework show the expected clear trends, namely a decreasing nominal strength with size and an increasing strength with age. 

The strength-size relationship for each simulated age again can be analyzed by both the size effect models, CSEC \& SEL. The same procedures as discussed in the previous section of CSEC and SEL analysis (Eq.~\ref{CSEC}~\&~\ref{SEL}) are followed. The shape factor, $\alpha$, initial energy release rate, $g_0$, and its first order derivative, $g'_0$, remain the same. Tensile strength, $f'_t$, and characteristic length, $\ell_1$, are obtained from data fitting, which are then utilized to compute the initial fracture energy, $G_f$, for each age. This calculation requires the knowledge of the corresponding Young's modulus values for each age, which are identified from experimental size M test of that age \cite{WanUHPCI}, see Table~\ref{modulus}. The average correlation coefficient R$^2$ of the CSEC fits for the predicted ultimate strength on different ages is higher than 0.999. The SEL fits have a slightly lower, though also quite high, R$^2$ value of 0.995. 
Overall, the analytical CSEC fits can perfectly reproduce the numerically obtained size dependence of nominal strength at all ages. While the coefficients of determination indicate perfect fits for both size effect laws, qualitatively, the SEL is not able to capture the slight curvature that is present in the datasets of all ages if plotted in normalized space, see Fig.~\ref{pred}. For HR 3 days all data points satisfy $X<0.4$ and $Y<0.7$. With increasing aging degree the upper limit of the data range expands and exceeds for the most mature concrete of aging degree 0.954 already $X=1.2$ and $Y=1.5$.
The quality of fit is constant across all ages for CSEC but follows a clear trend for SEL: both visually and according to the coefficient of determination the best fit is obtained for the fully cured concrete of WB curing. As the data points move towards the plastic limit for younger concretes, see Fig.~\ref{pred}, the limitations of SEL become evident. The plastic limit is a size range for which the SEL only approximately applies, because the SEL represents the asymptotic shape of the CSEC for large sizes. 

In order to assess the mesoscale stress level and energy dissipation for the beams of different ages and sizes, the relevant data is extracted from the LDPM simulations. Stress along the ligament, obtained from the LDPM simulations as homogenized stress tensor of each tetrahedron for each simulated age and size, is plotted in Fig.~\ref{StressTensor}. The subfigures from left to right represent increasing sizes and from top to bottom increasing aging degree. On the abscissa the stress level is plotted, and on the ordinate the normalized ligament height, with value 0 being the initial notch tip and value 1 representing the top of the specimen. Investigated time-points during each simulated test include 50\% of peak in the linear elastic regime, at peak, 50\% post peak and at 20\% remaining strength in the post peak. 
Note, the LDPM simulations mimic the natural heterogeneity of concrete by randomly placed discrete particles that follow a Fuller distribution. As the true maximum aggregate size of 0.6~mm would be computationally prohibitive a coarse graining strategy was applied \cite{WanUHPCI}. All simulated specimens use the same minimum and maximum aggregate sizes of 2-4 mm. As a result, stress fields are neither continuous nor smooth and the spatial resolution is limited to that of the particle size. Thus, the larger the specimen size, the smoother the stress profile and the clearer the observed trends. The comparison of stress profiles of different ages and sizes reveals the following characteristics: The maximum stress of each age is size-independent. The maximum stress readings increase with age and are equal to the mesoscale tensile strength of LDPM for the corresponding ages. As the crack propagates, the corresponding section in tension travels upwards and causes a further concentration of compressive stresses in the top section (compressive zone decreases in height). The neutral axis at peak is located at about half of the ligament height for all sizes and ages.

Fig.~\ref{disenergy} presents the dissipated energy profiles along the ligament for all investigated sizes and ages. Similar to the stress profiles in Fig.~\ref{StressTensor}, the dissipated energy profiles are plotted at 50\% pre peak, at peak, 50\% post peak and at 20\% remaining strength. The dissipated energy is determined for horizontal stripes of 4~mm height by numerical integration along the LDPM section of the beam. The plotted values are dissipated energies per unit area and correspond to the fracture energy for a fully softened traction free crack segment. For each size, the maximum observed dissipated energy increases in terms of aging degree until humidity curing for 28 days and drops again for the case of 28 day hot water bath curing. This agrees with the age-dependent fracture energy obtained from experiments, calibrated and validated HTC-LDPM simulations, as well as CSEC and SEL analyses, which all conclude that the fracture energy can reach a peak in terms of aging degree. In general, the larger the size, the higher the total accumulated dissipated energy, which can be estimated as the area under the curves in Fig.~\ref{disenergy}. However, the maximum dissipated per unit area is roughly the same for different sizes if the fracture process zone fully develops. 
For size XL and XXL, the dissipated energy after peak reaches the maximum value for about 50\% of the ligament, indicating that a fracture process zone fully develops and travels. For size L, the maximum dissipated energy is observed close to notch tip where it reaches roughly the same level as for sizes XL \& XXL. This means that, for size L, a fracture process zone barely develops. For size M, the maximum dissipated energy per unit area is lower than the values observed for sizes L, XL \& XXL, indicating that the fracture process zone can not fully develop due to its limited size. 

\begin{table}[ht]
\caption{UHPC Young's modulus evolution at early age}\label{modulus}
\centering
\begin{tabular}{l cccccc}
\hline
Age  & HR 3d & HR 7d  & HR 14d & HR 28d & WB 14/28 & WB 120d \\
\hline
$E$ [MPa] & 27089 &	34267 &	37669 &	39323 &	46302 &	46326\\
\hline
\end{tabular}
\end{table}

\bfi [ht]
\centering
(a)   \includegraphics[height=1.8in,valign=t]{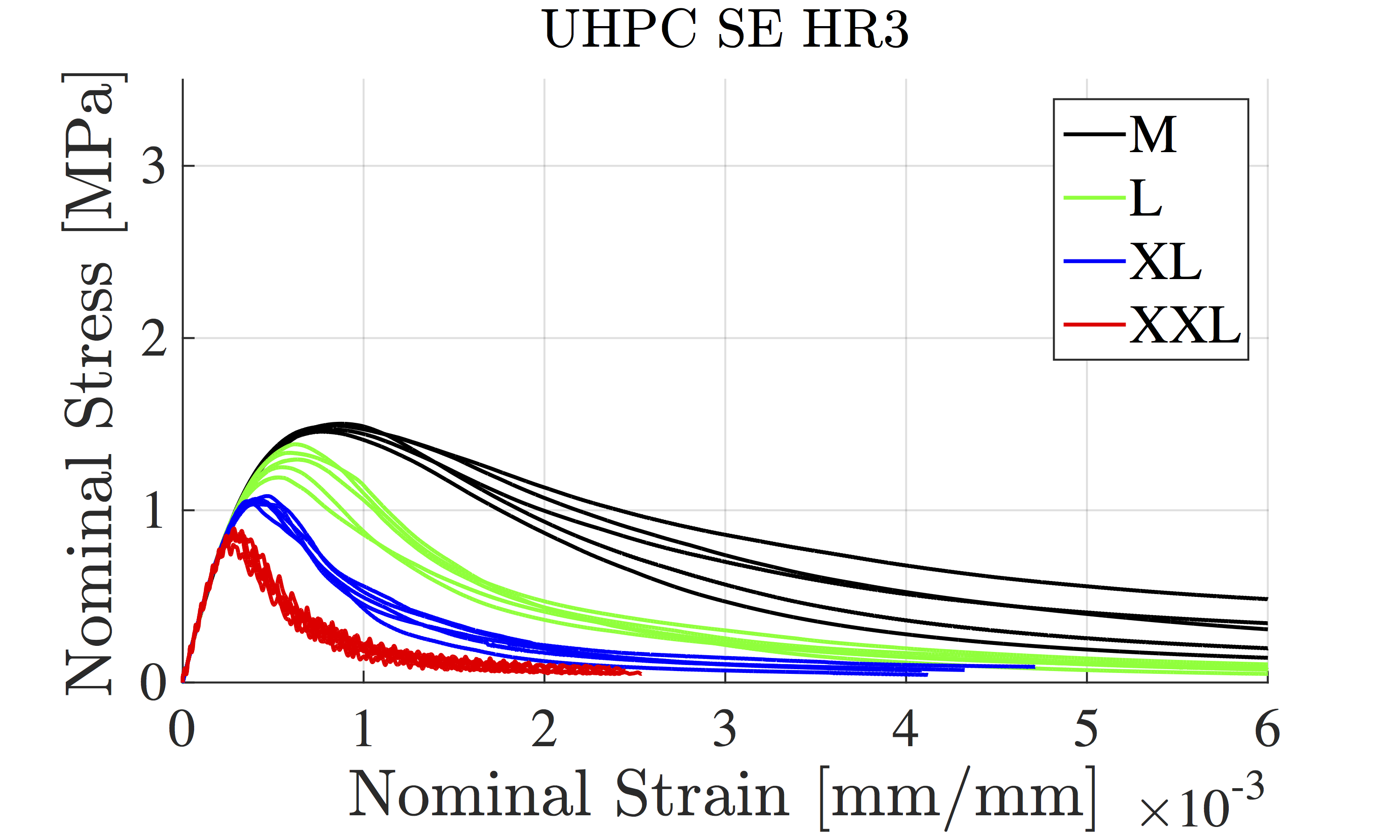}   
    \includegraphics[height=1.8in,valign=t]{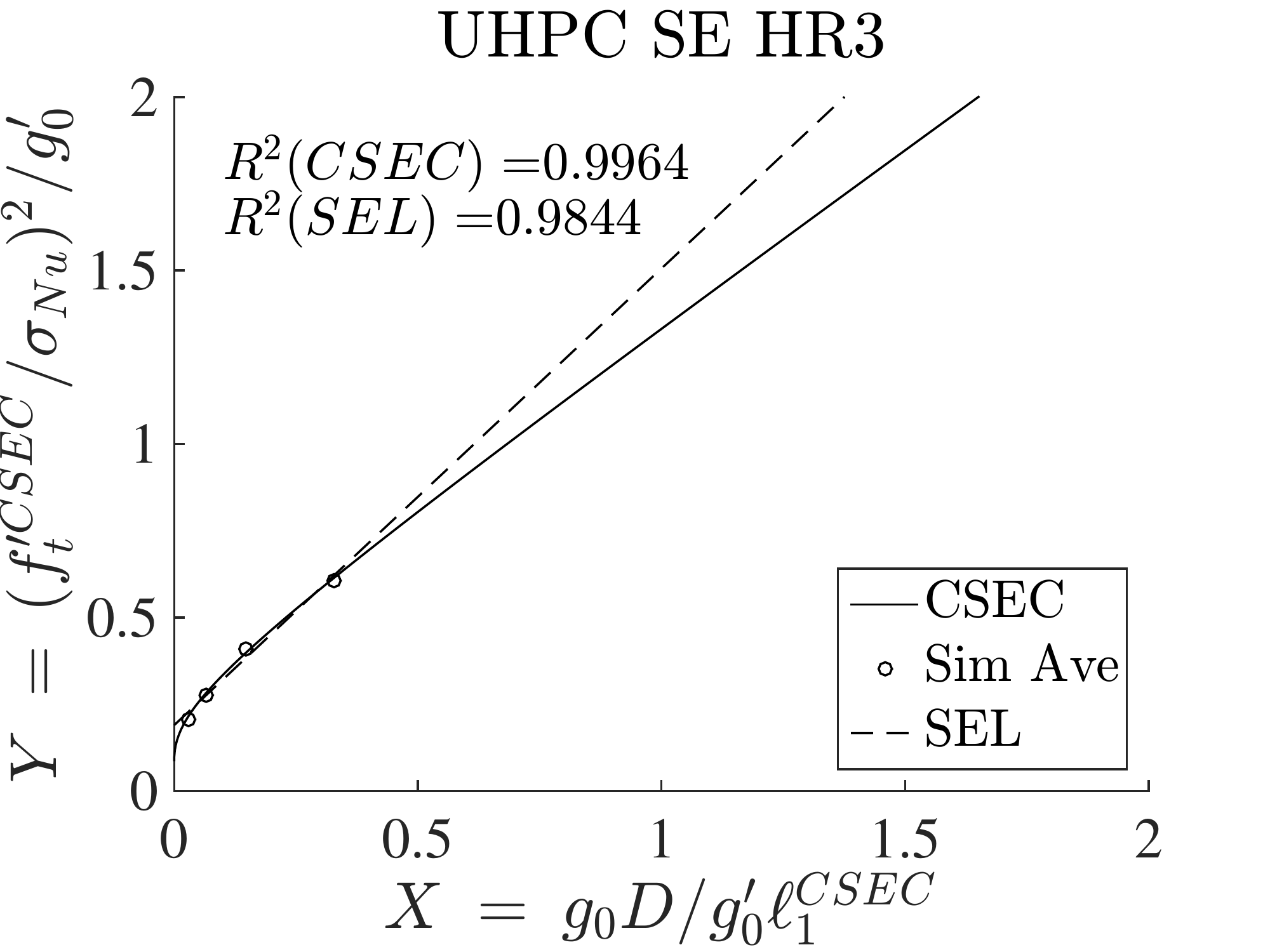}\\ 
(b)   \includegraphics[height=1.8in,valign=t]{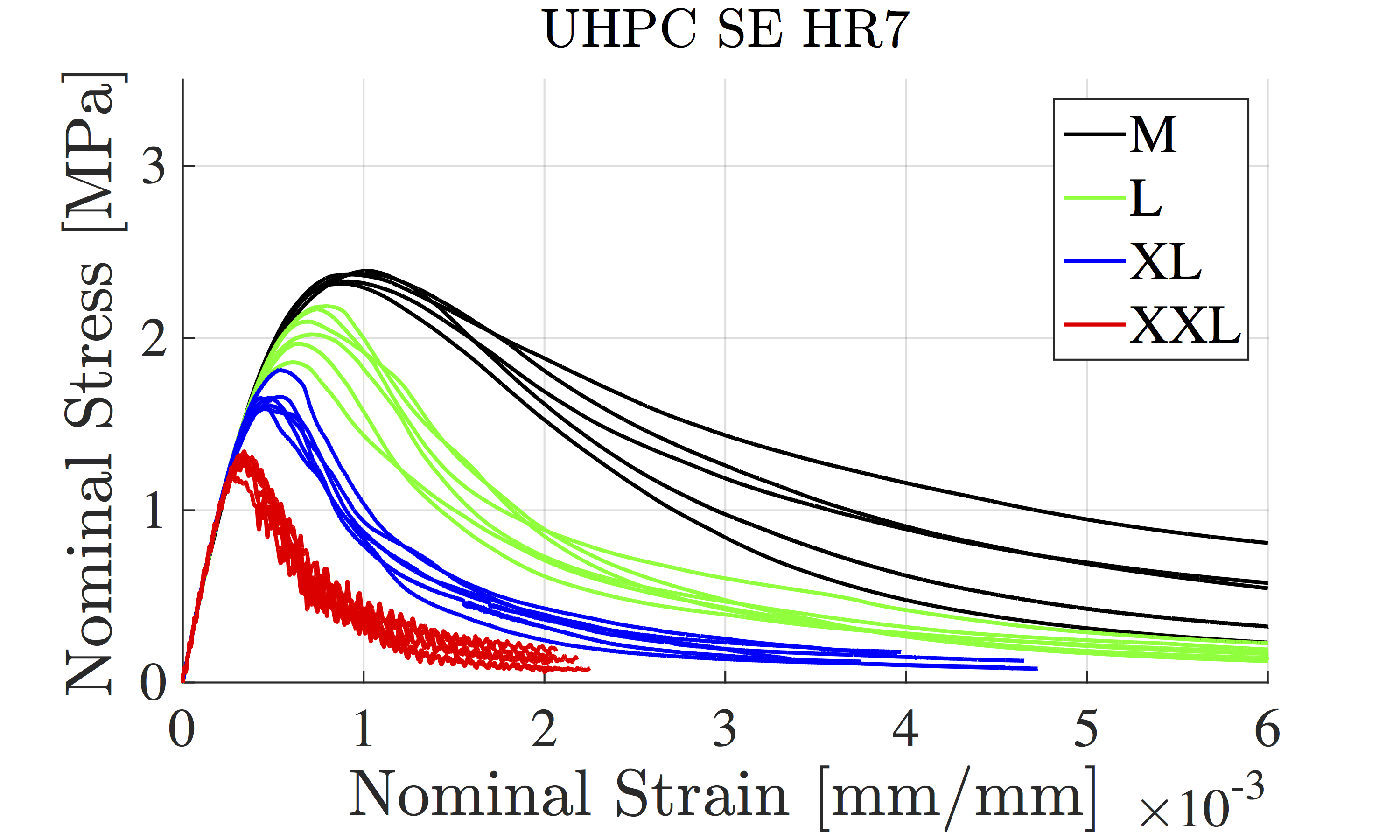}   
    \includegraphics[height=1.8in,valign=t]{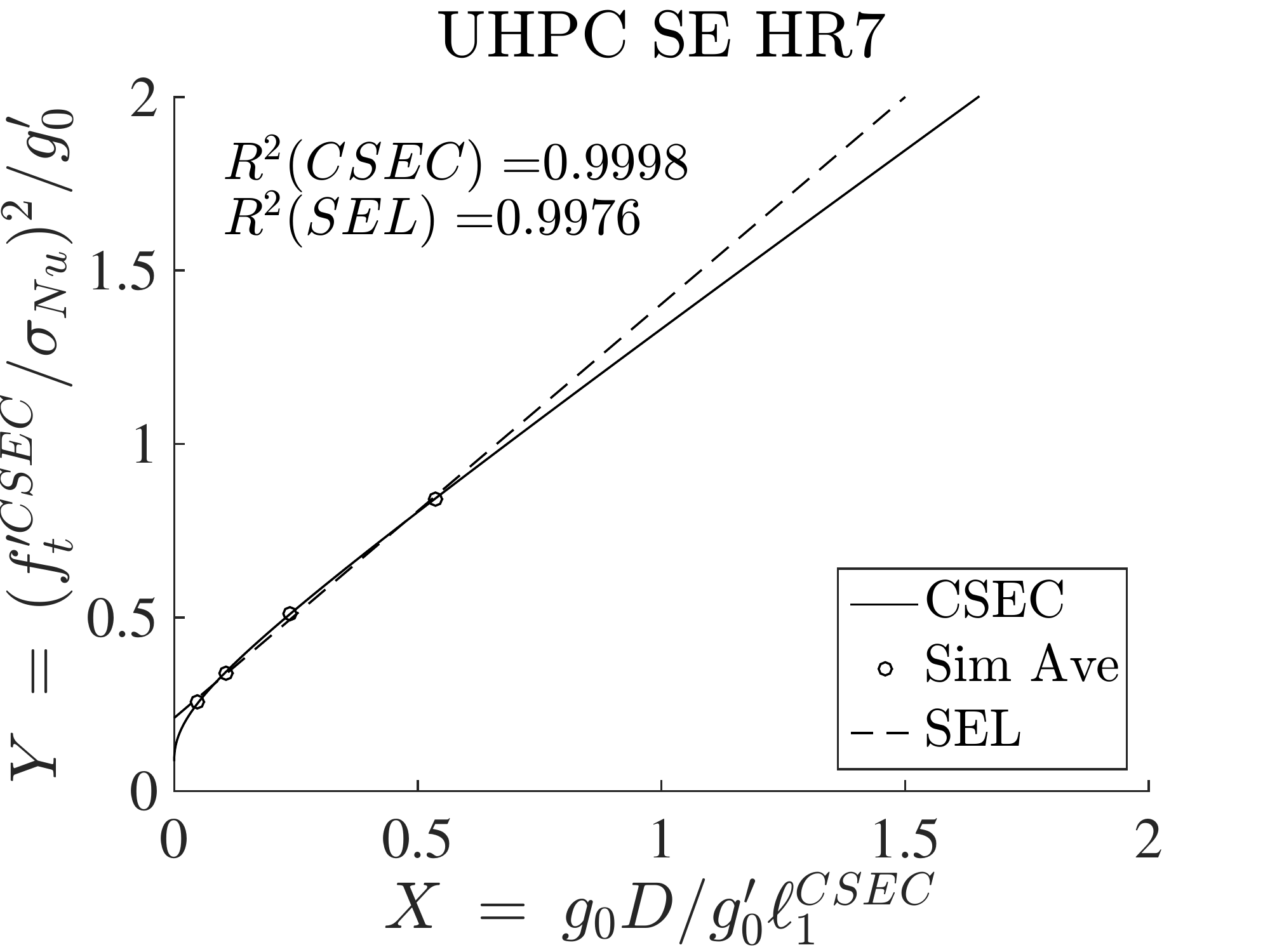} \\
(c)   \includegraphics[height=1.8in,valign=t]{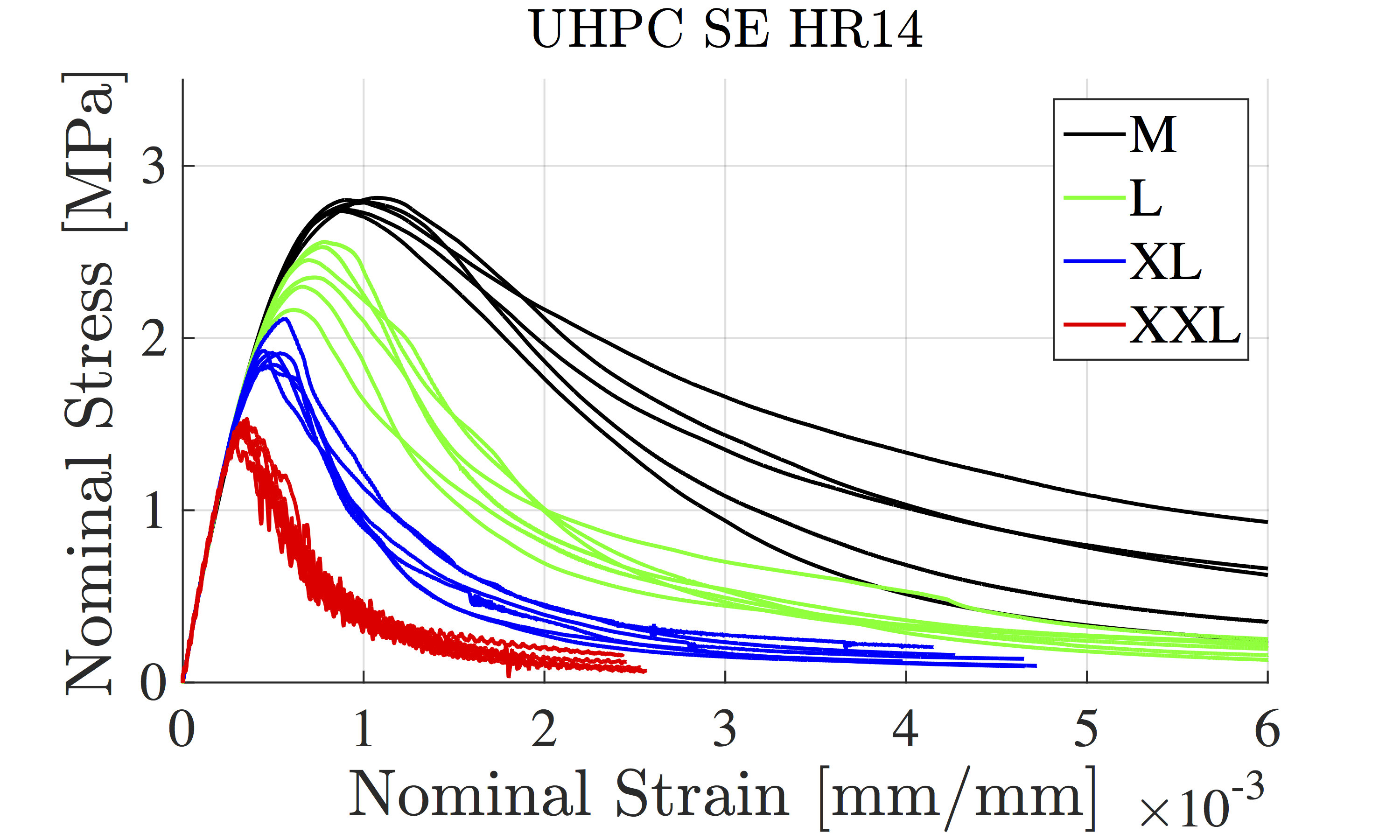}   
    \includegraphics[height=1.8in,valign=t]{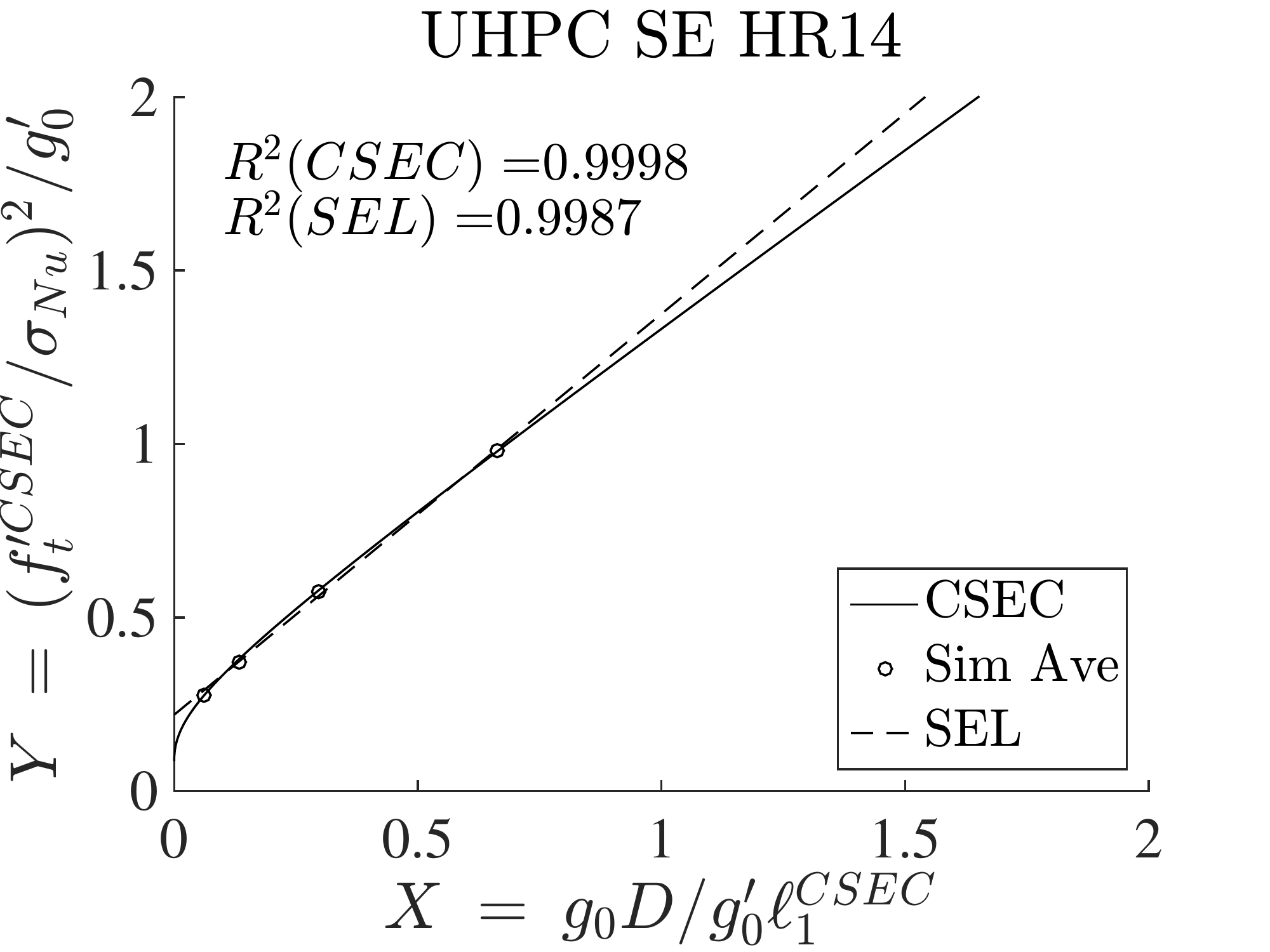} \\
(d)   \includegraphics[height=1.8in,valign=t]{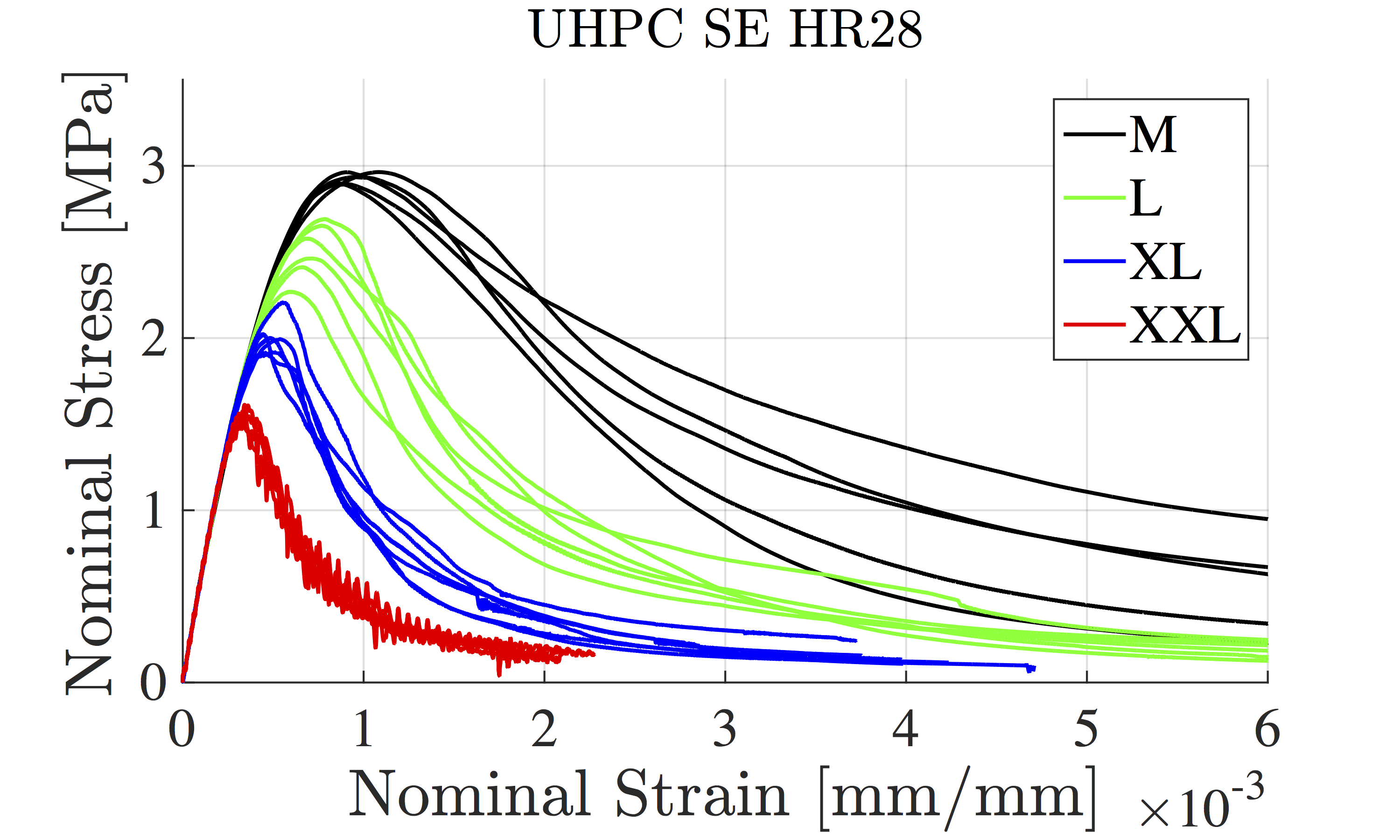}   
    \includegraphics[height=1.8in,valign=t]{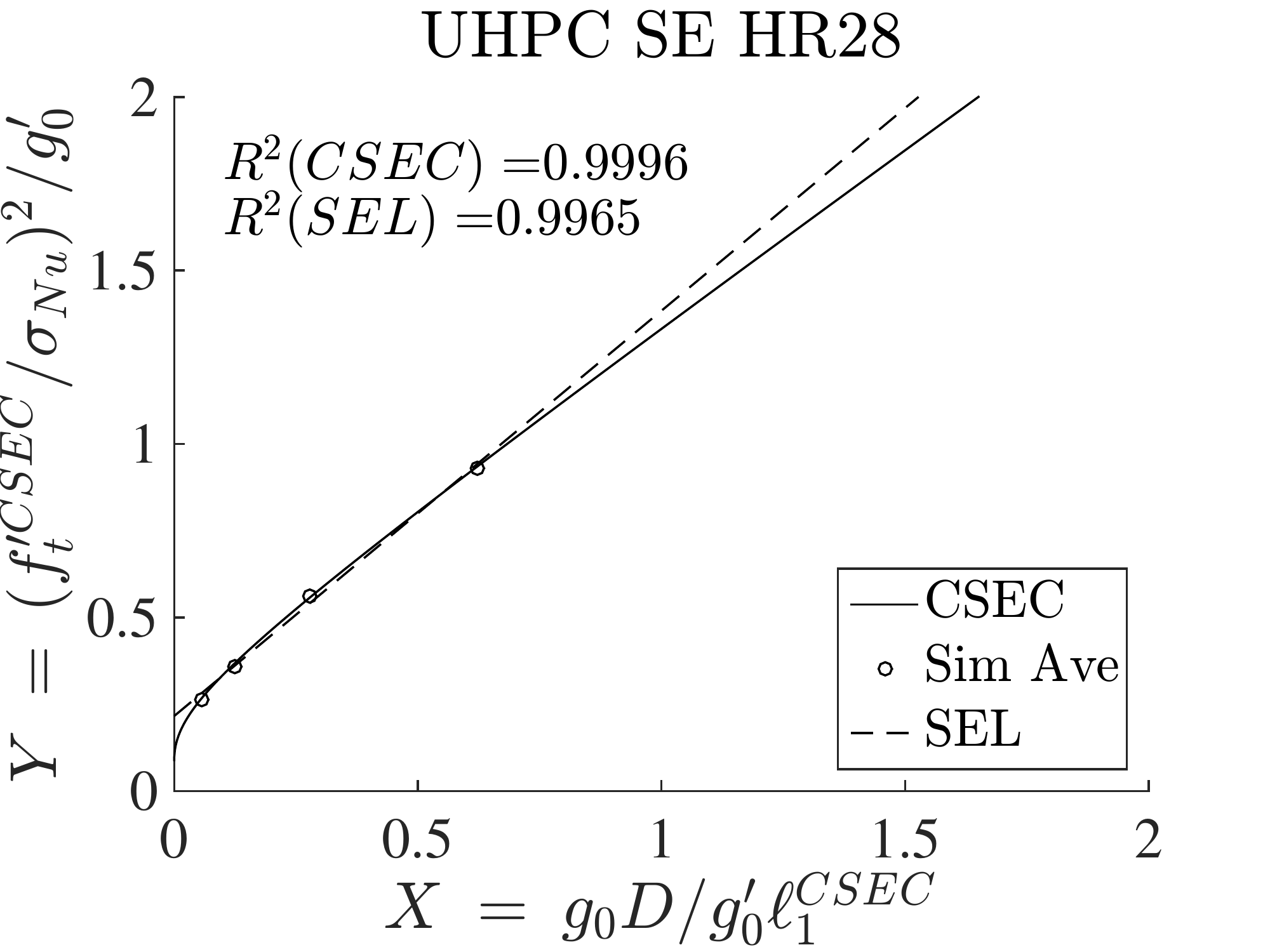} \\
(e)   \includegraphics[height=1.8in,valign=t]{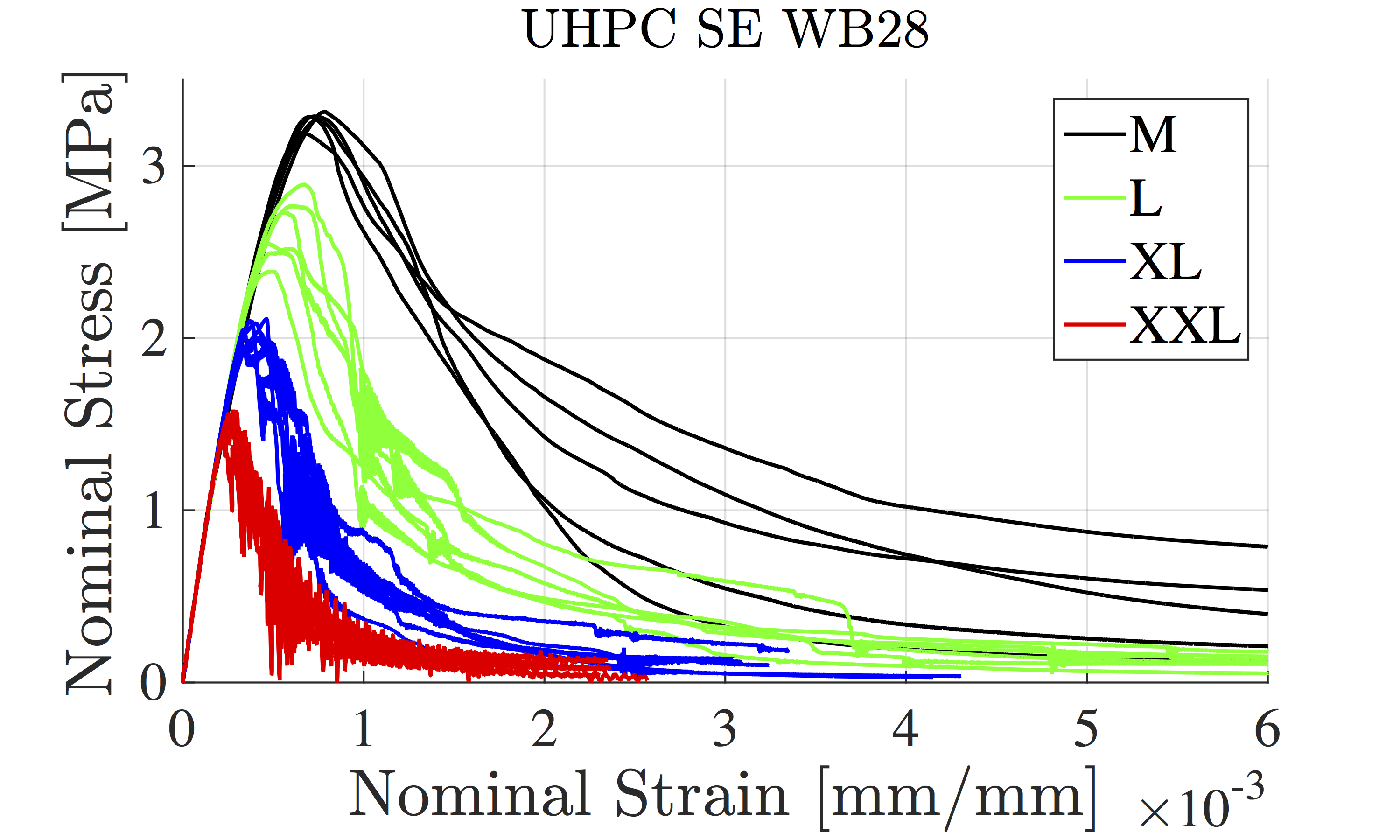}   
    \includegraphics[height=1.8in,valign=t]{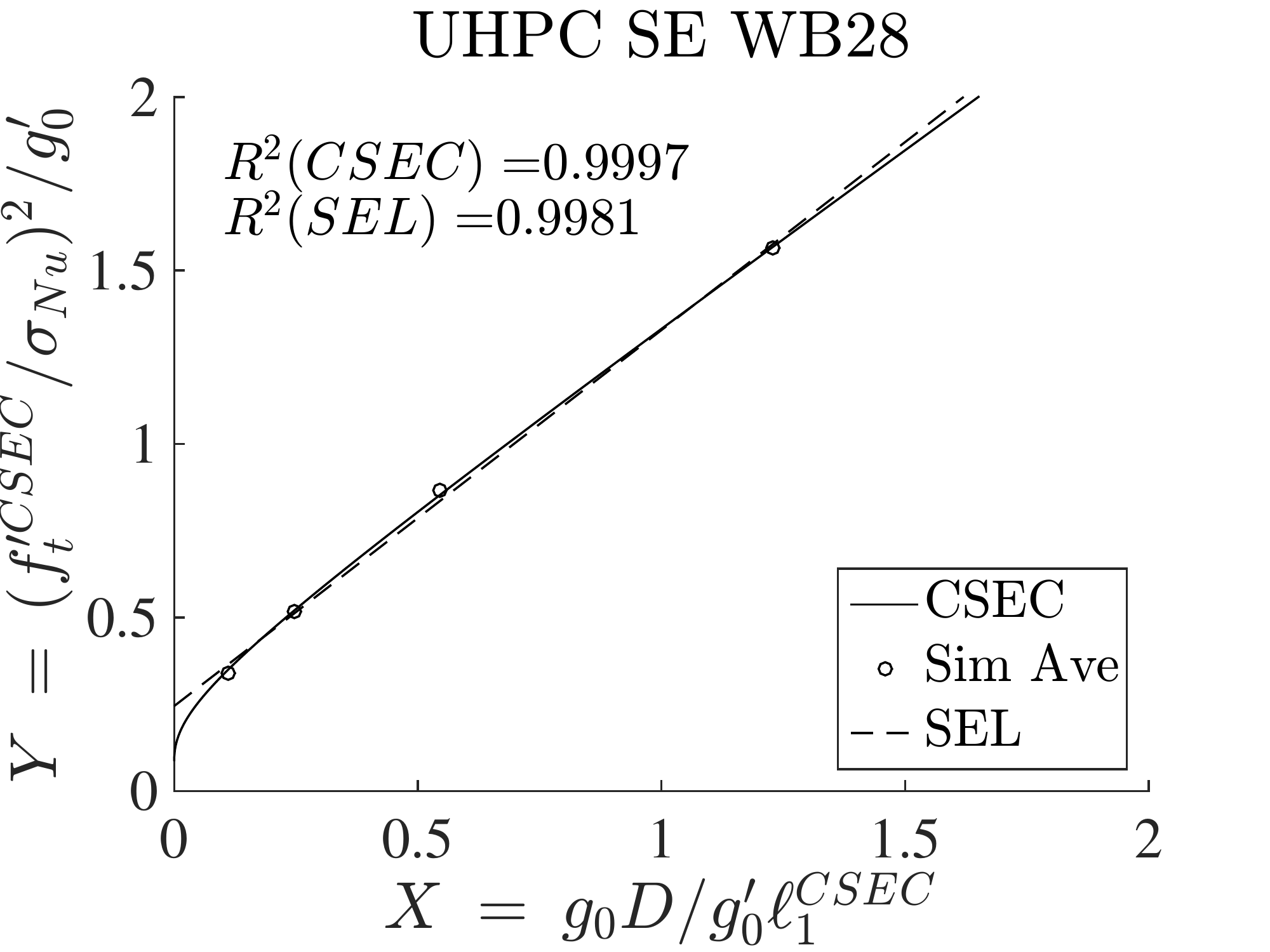} 
    \caption{HTC-LDPM size effect predictions and CSEC \& SEL analyses for (a) humidity room (HR) curing for 3 days, (b) HR 7 days, (c) HR 14 days, (d) HR 28 days, and (e) hot water bath (WB) curing for 28 days. SEL curves are normalized based on CSEC fitting}
    \label{pred}
\efi

\bfi [ht]
\centering
  \includegraphics[height=1.35in]{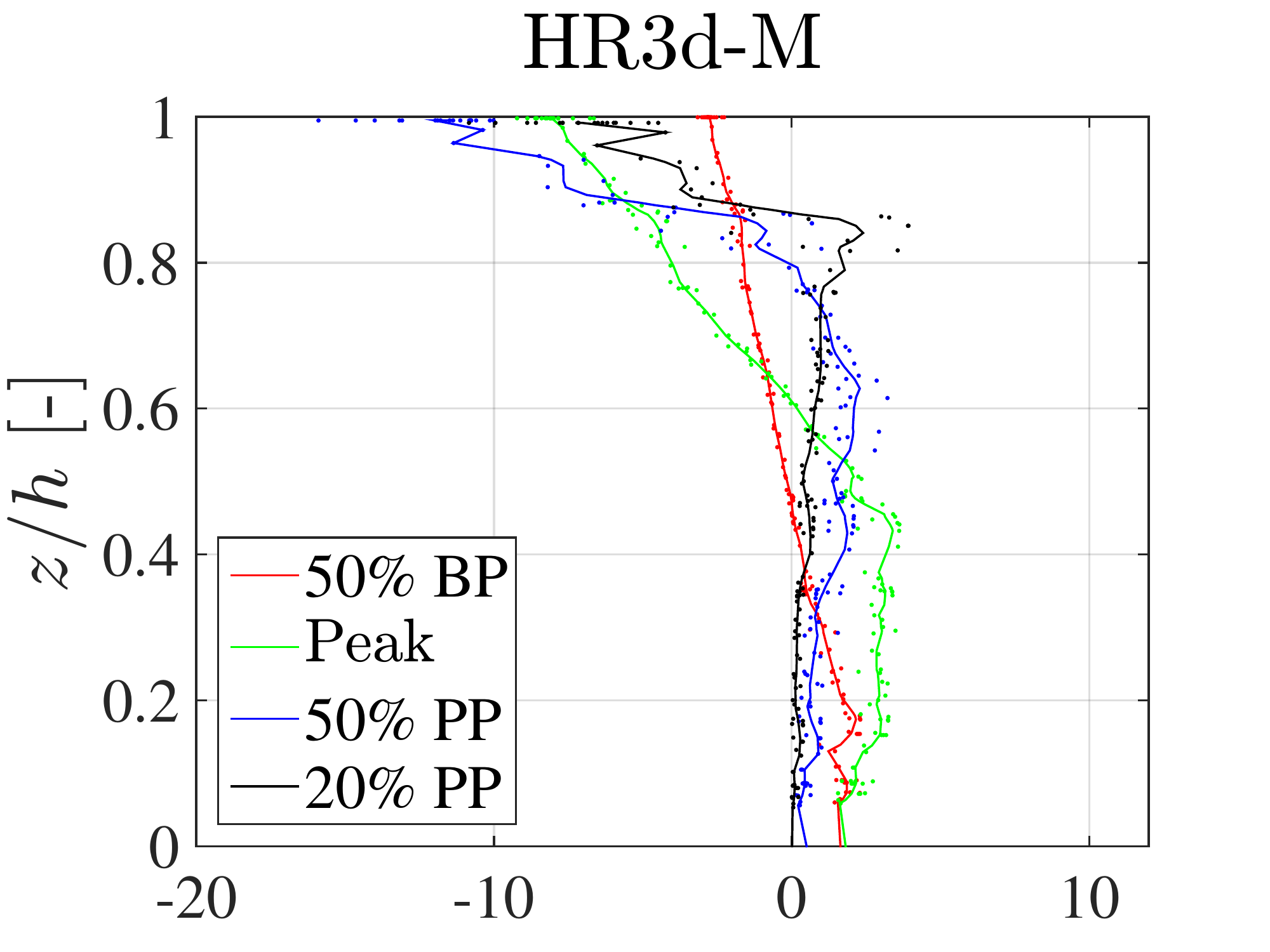} 
    \includegraphics[height=1.35in]{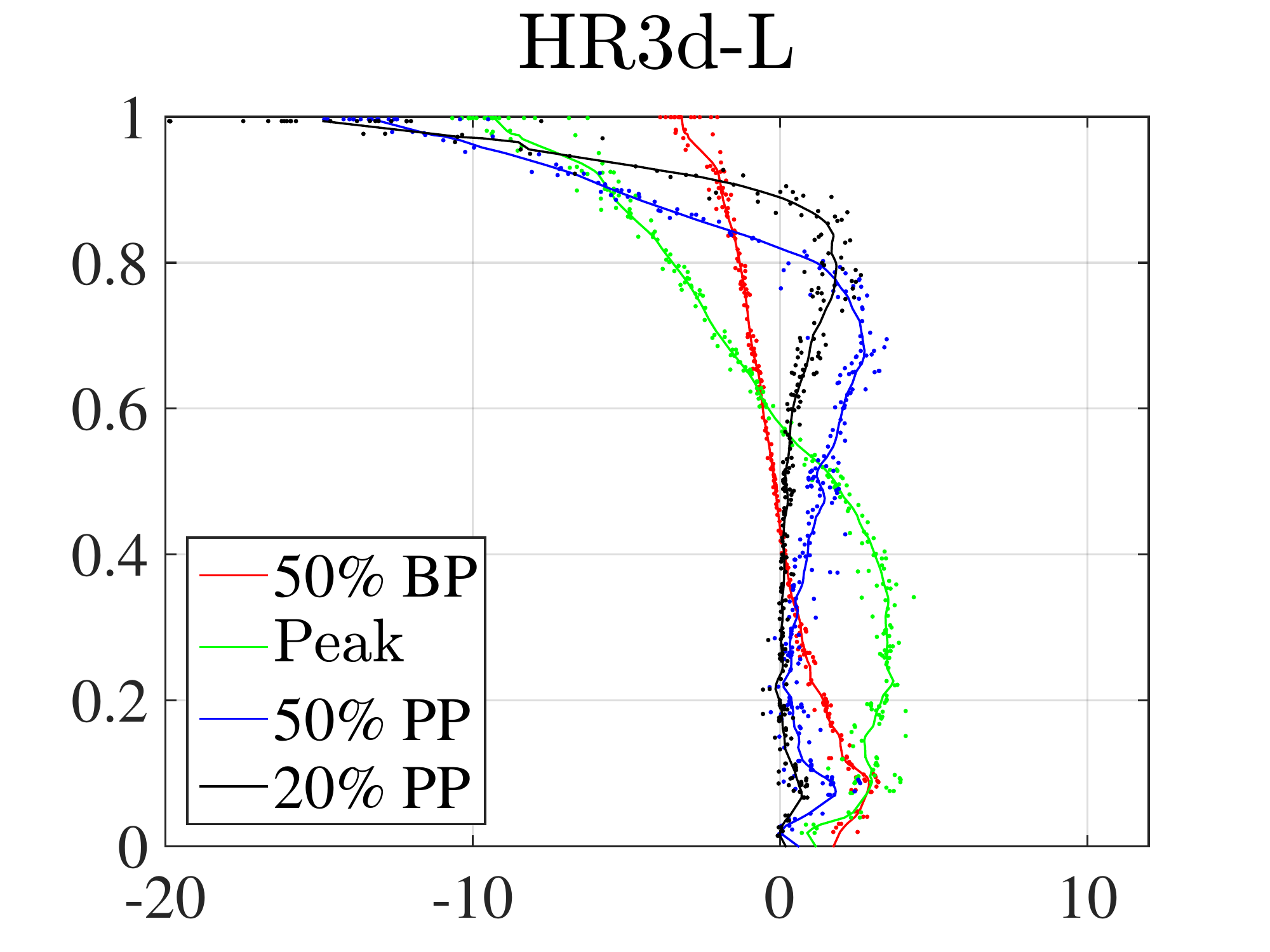} 
        \includegraphics[height=1.35in]{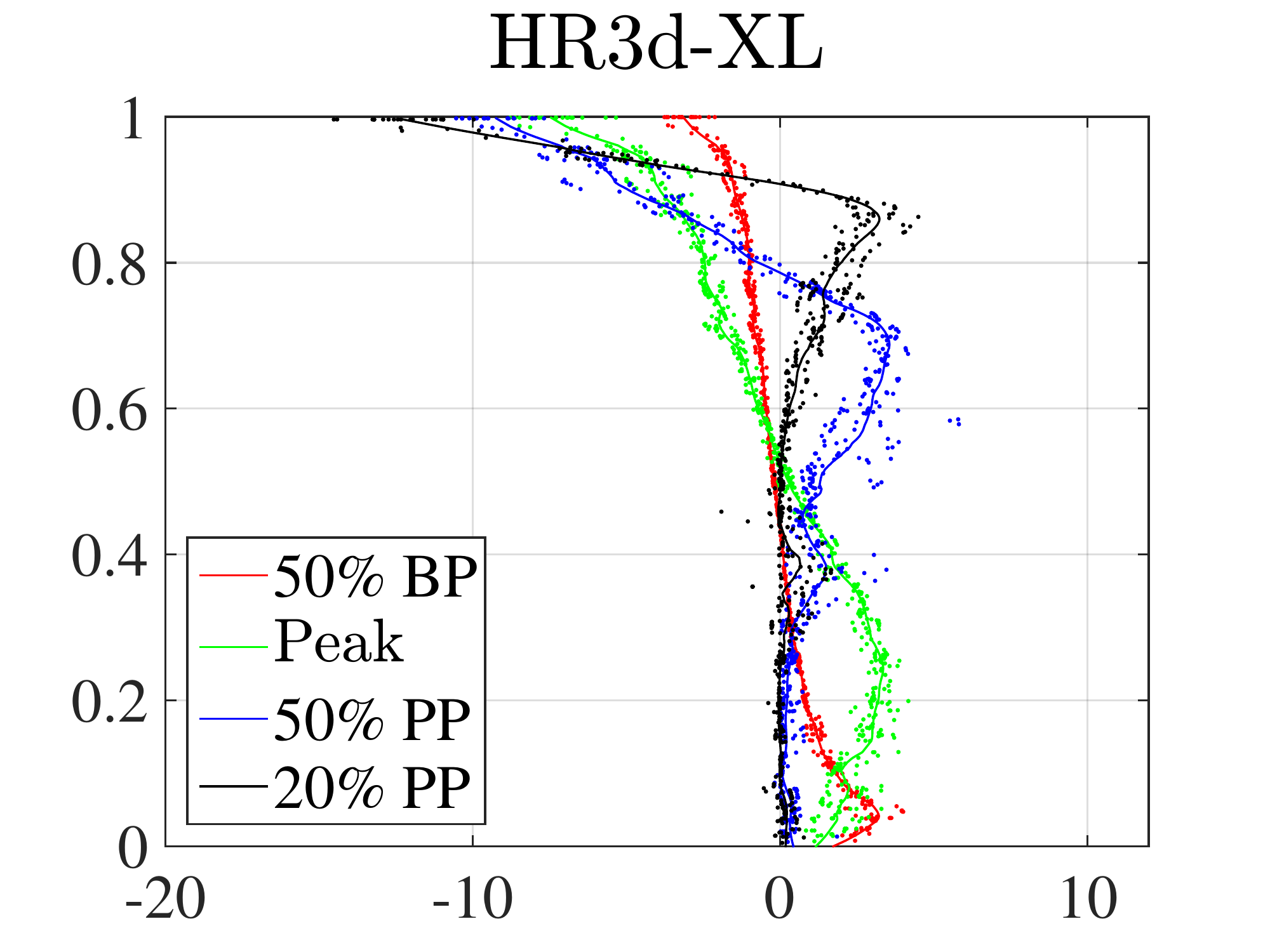} 
            \includegraphics[height=1.35in]{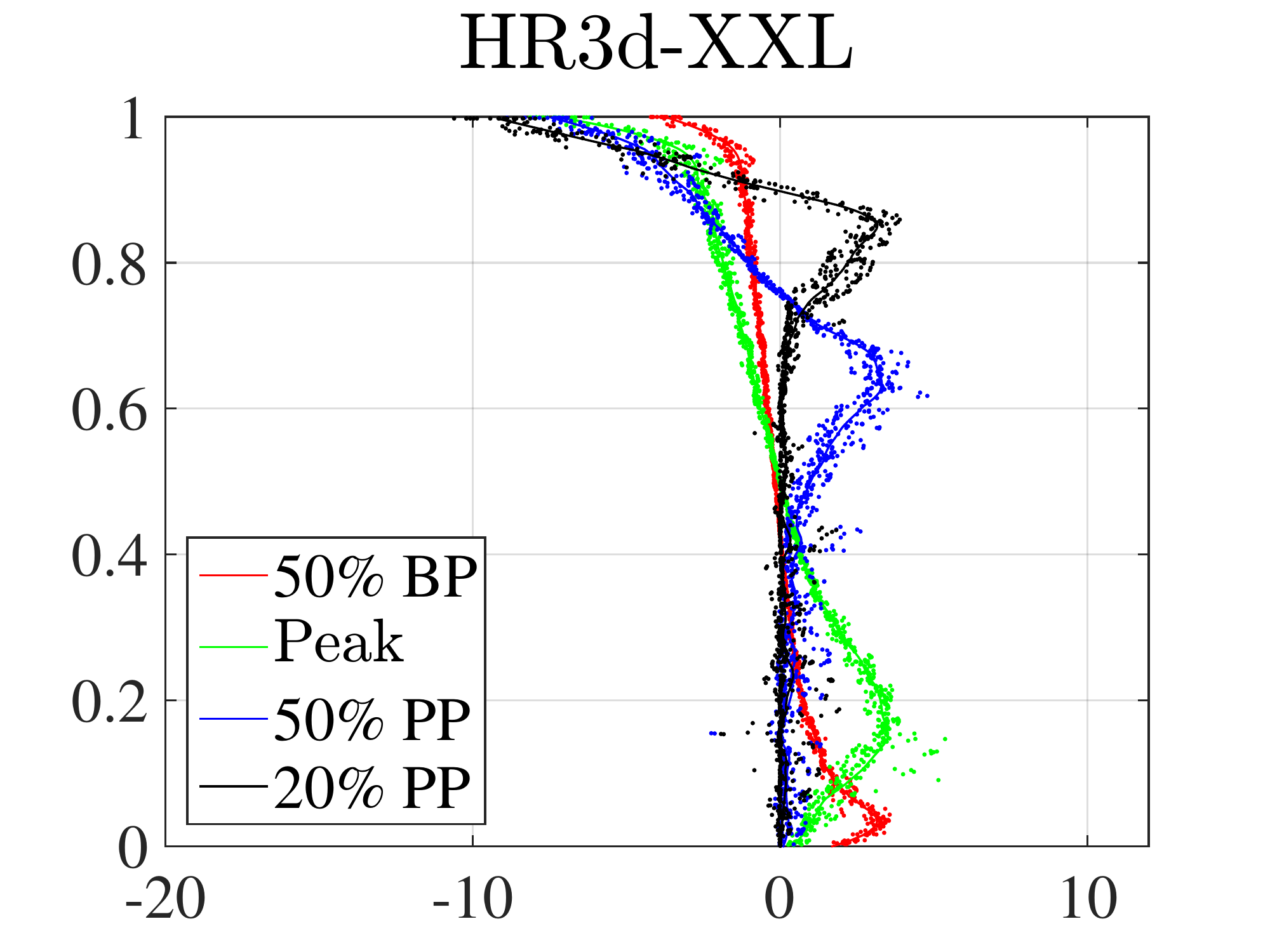} \\
  \includegraphics[height=1.35in]{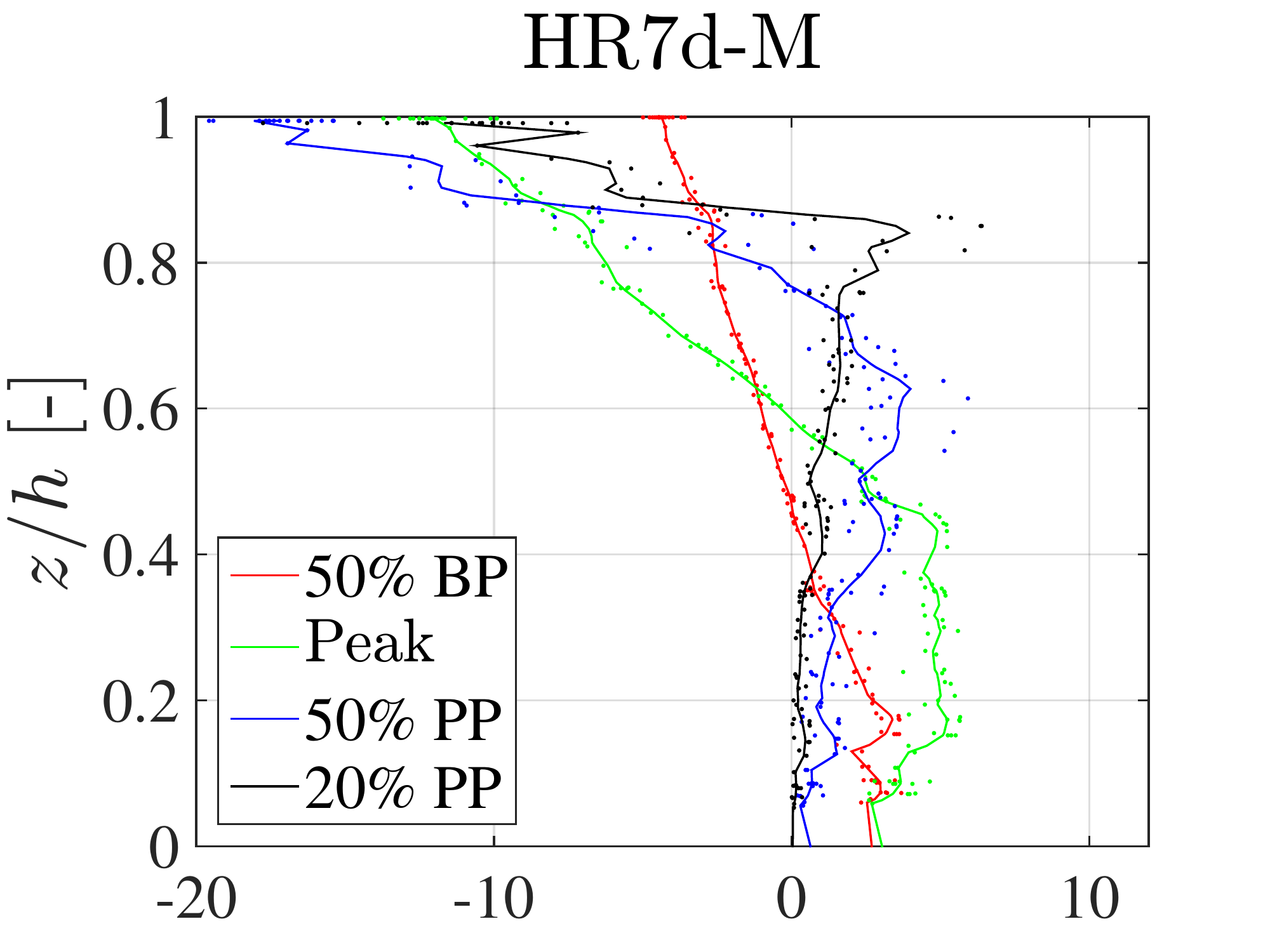} 
    \includegraphics[height=1.35in]{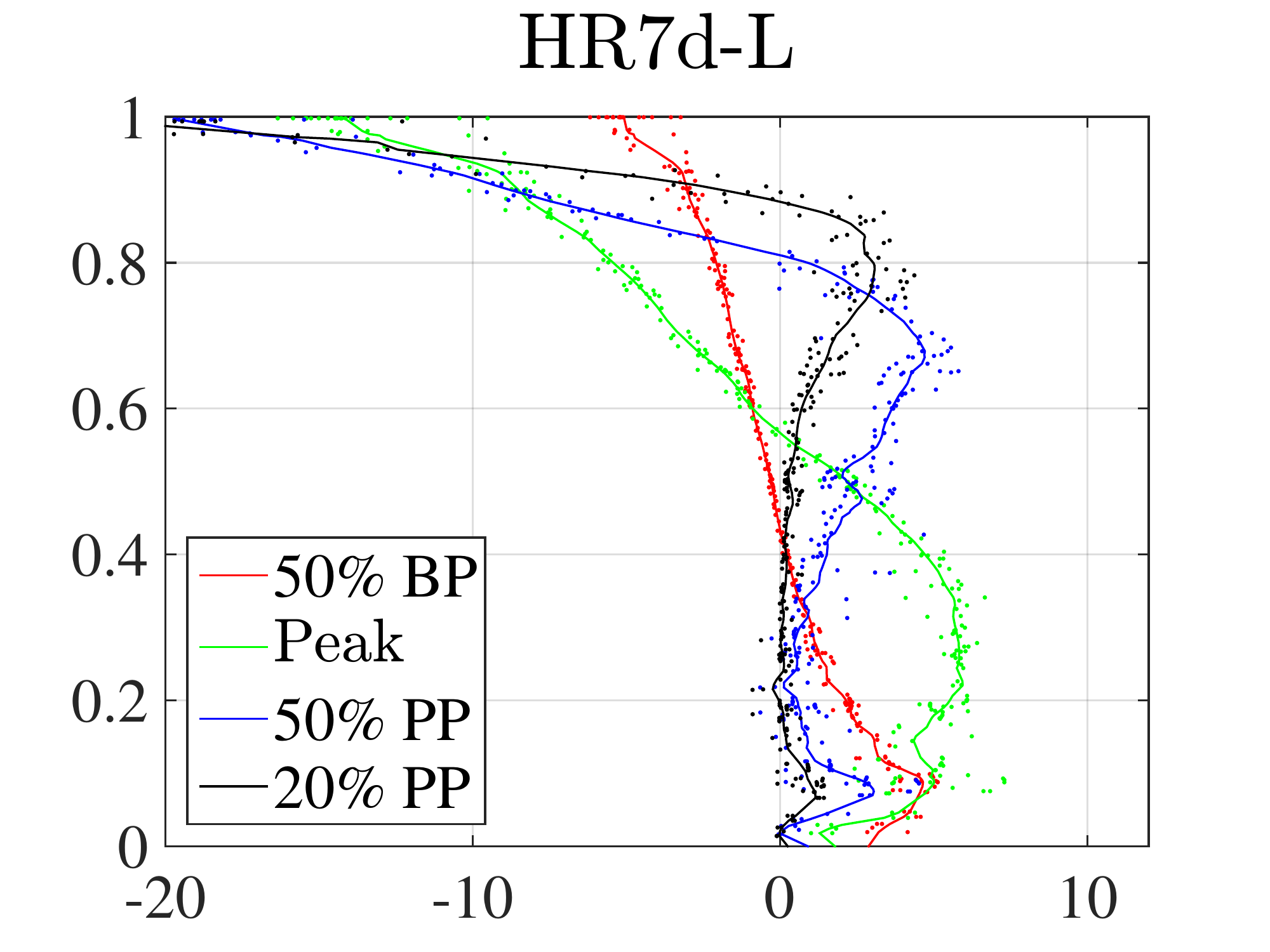} 
        \includegraphics[height=1.35in]{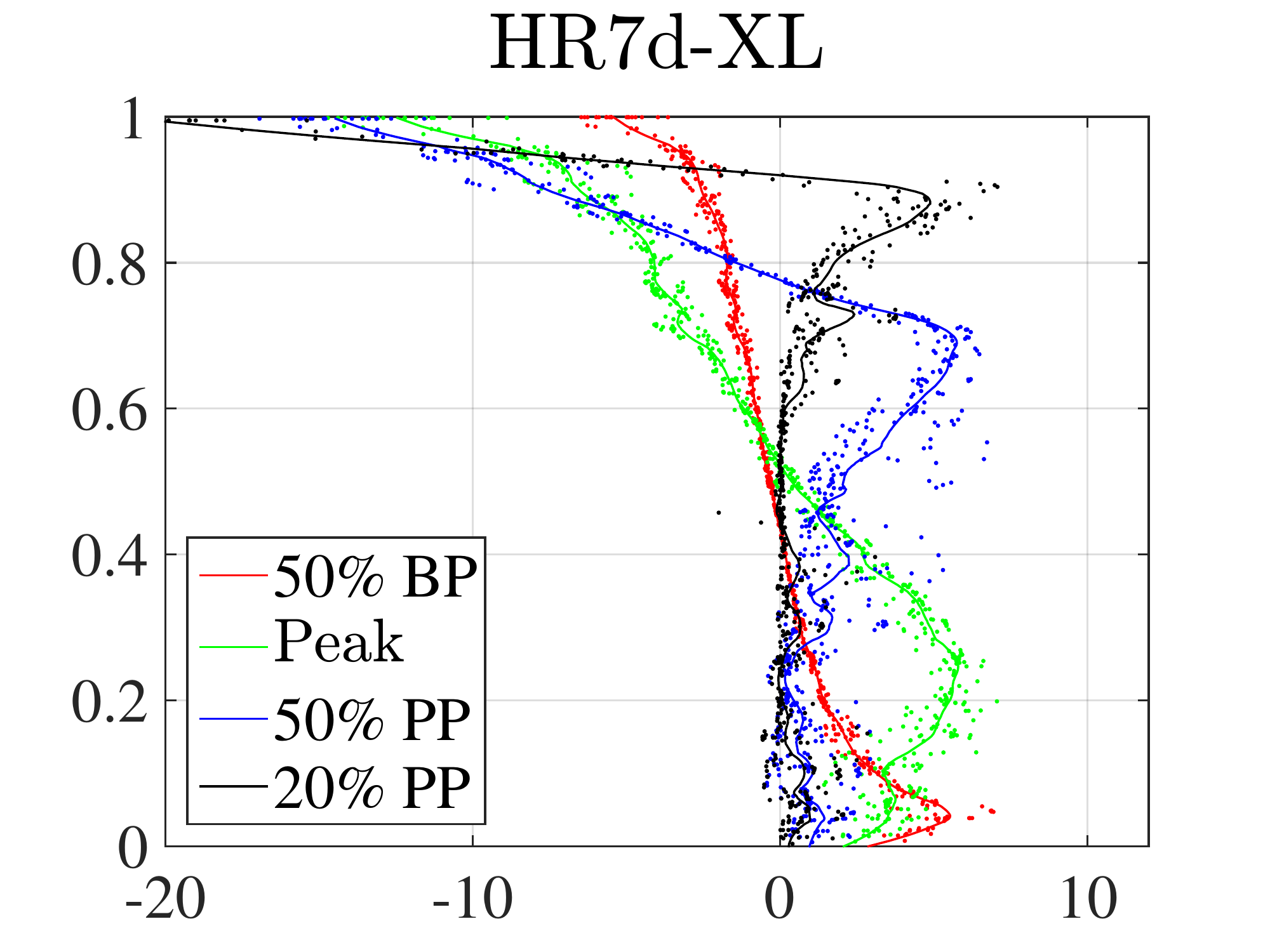} 
            \includegraphics[height=1.35in]{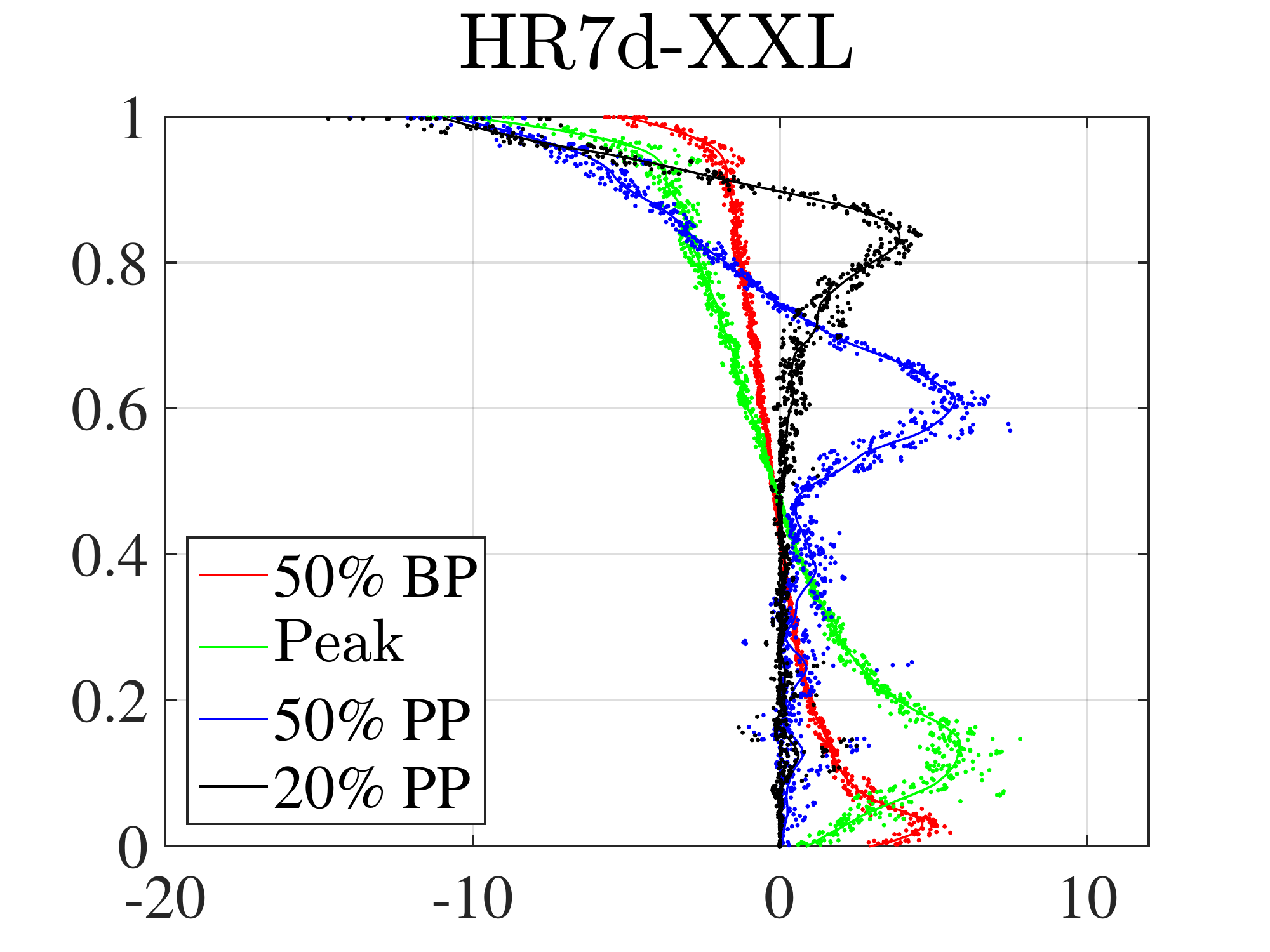} \\
  \includegraphics[height=1.35in]{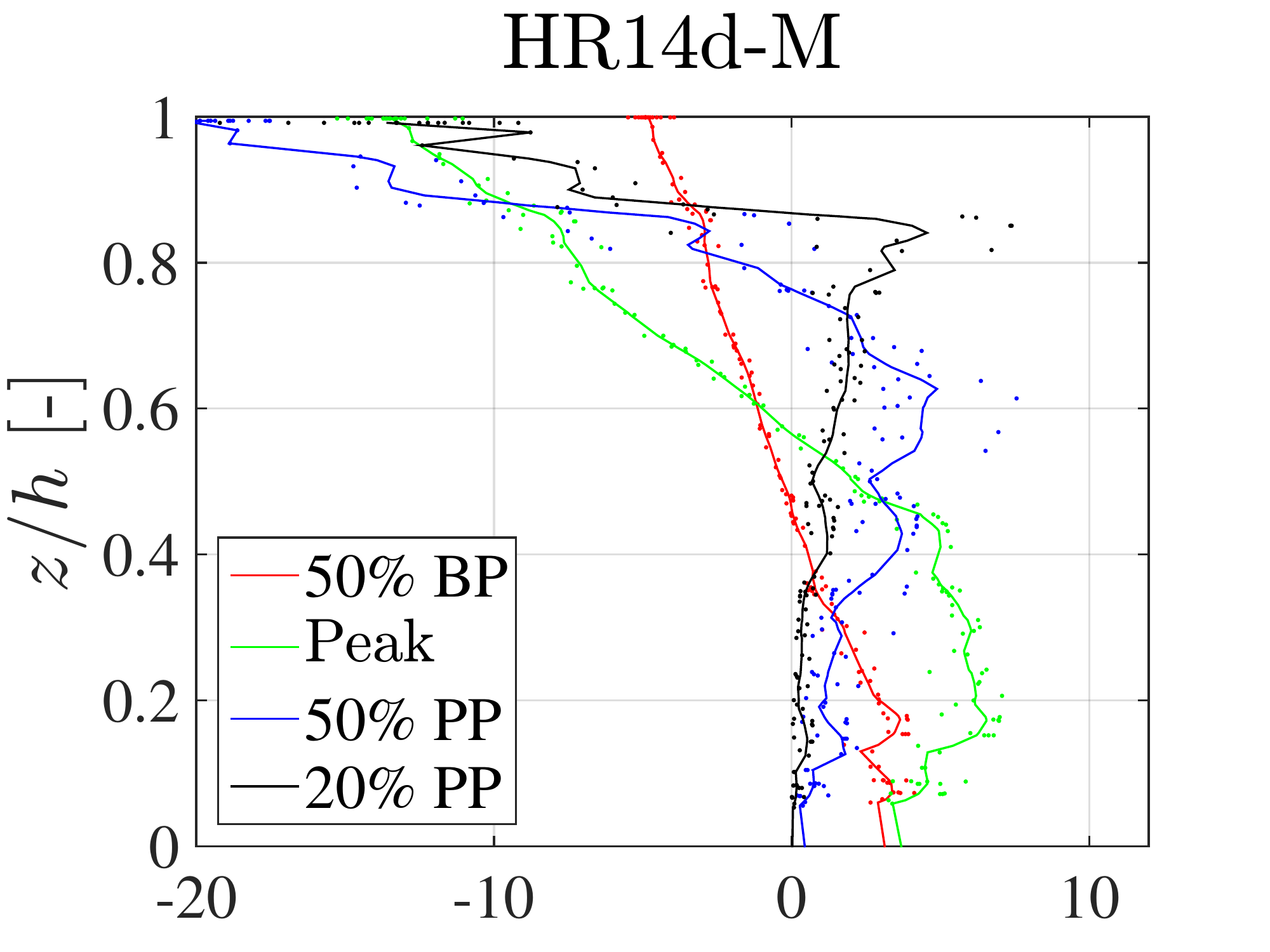} 
    \includegraphics[height=1.35in]{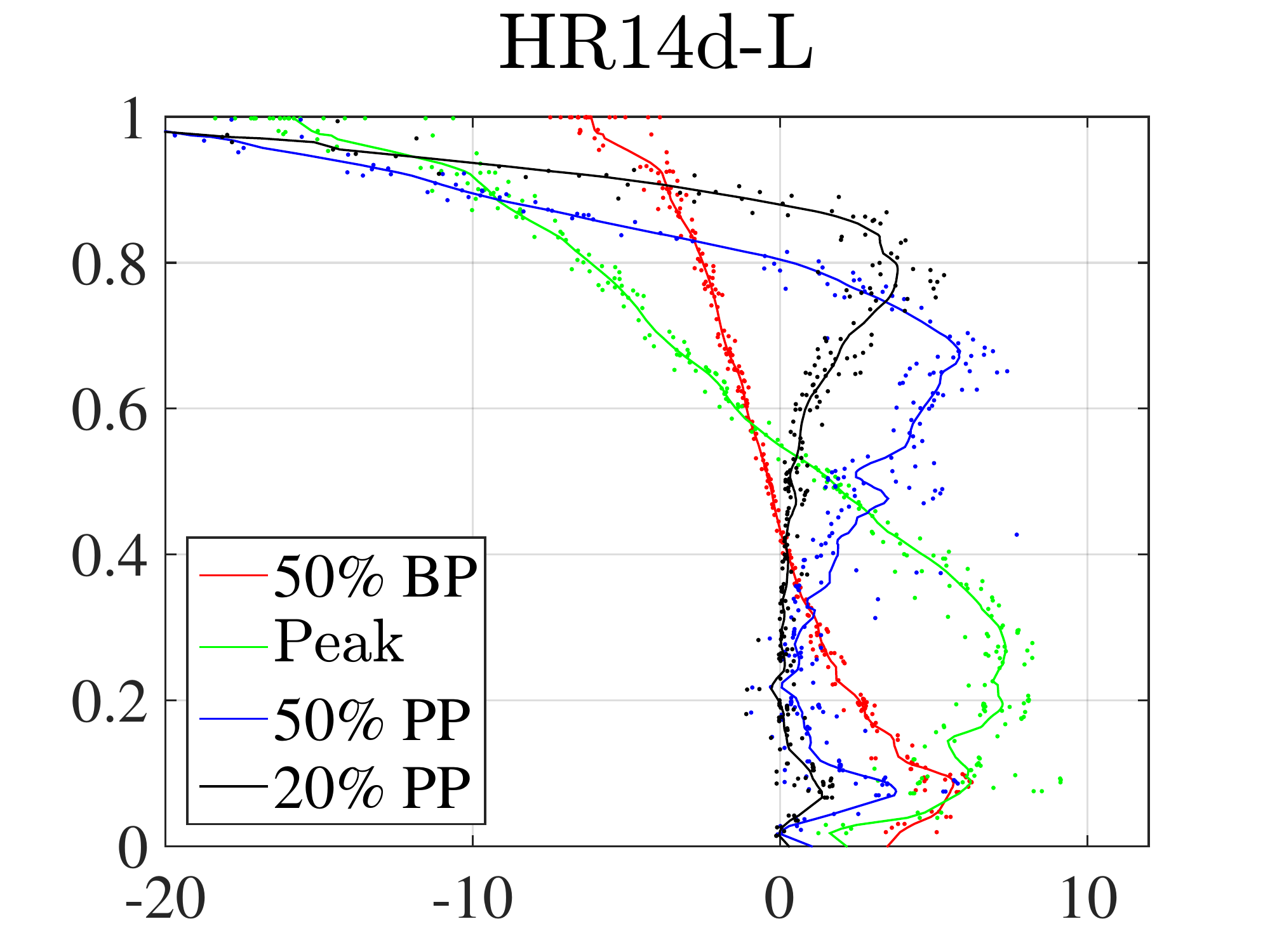} 
        \includegraphics[height=1.35in]{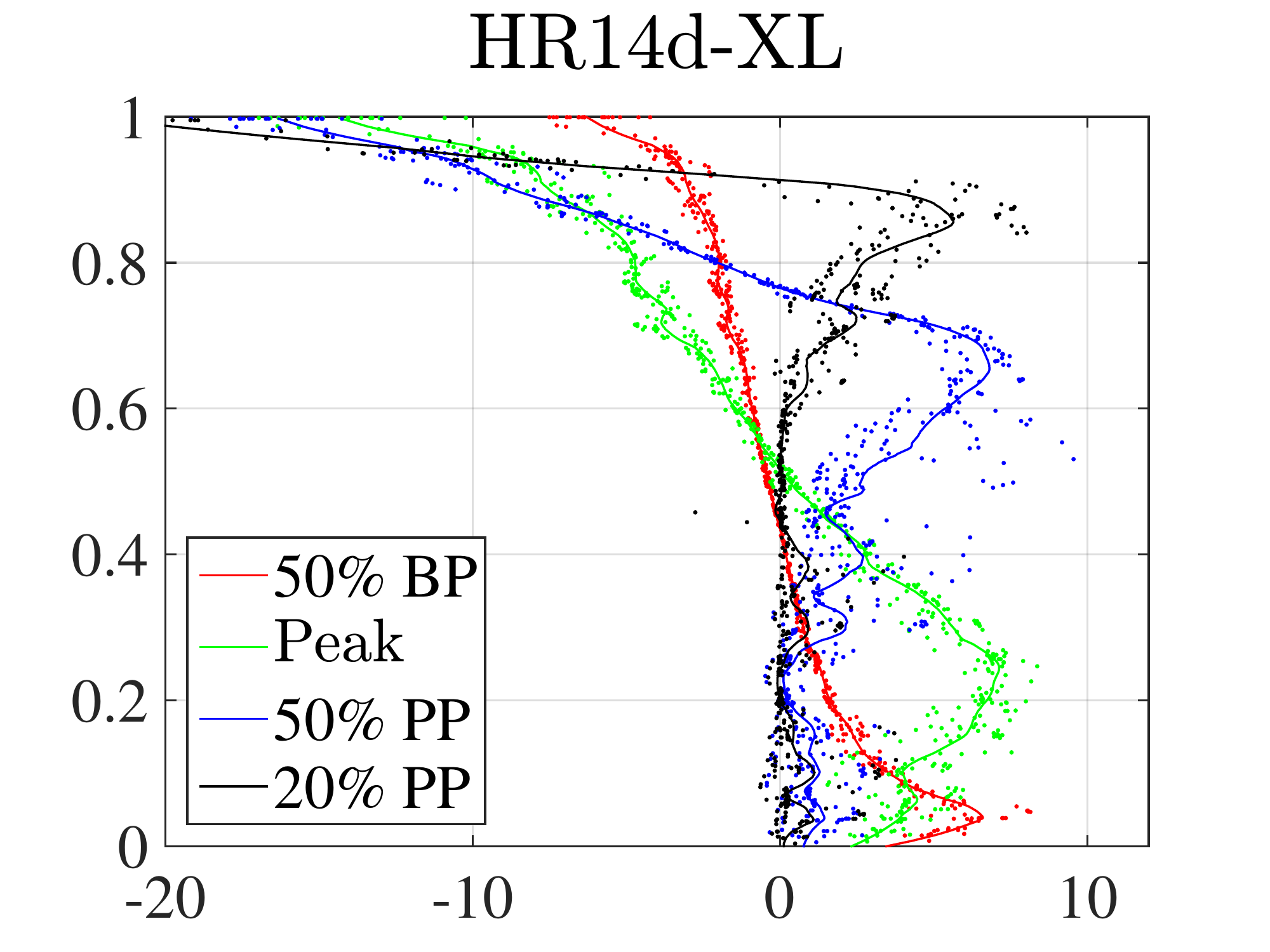} 
            \includegraphics[height=1.35in]{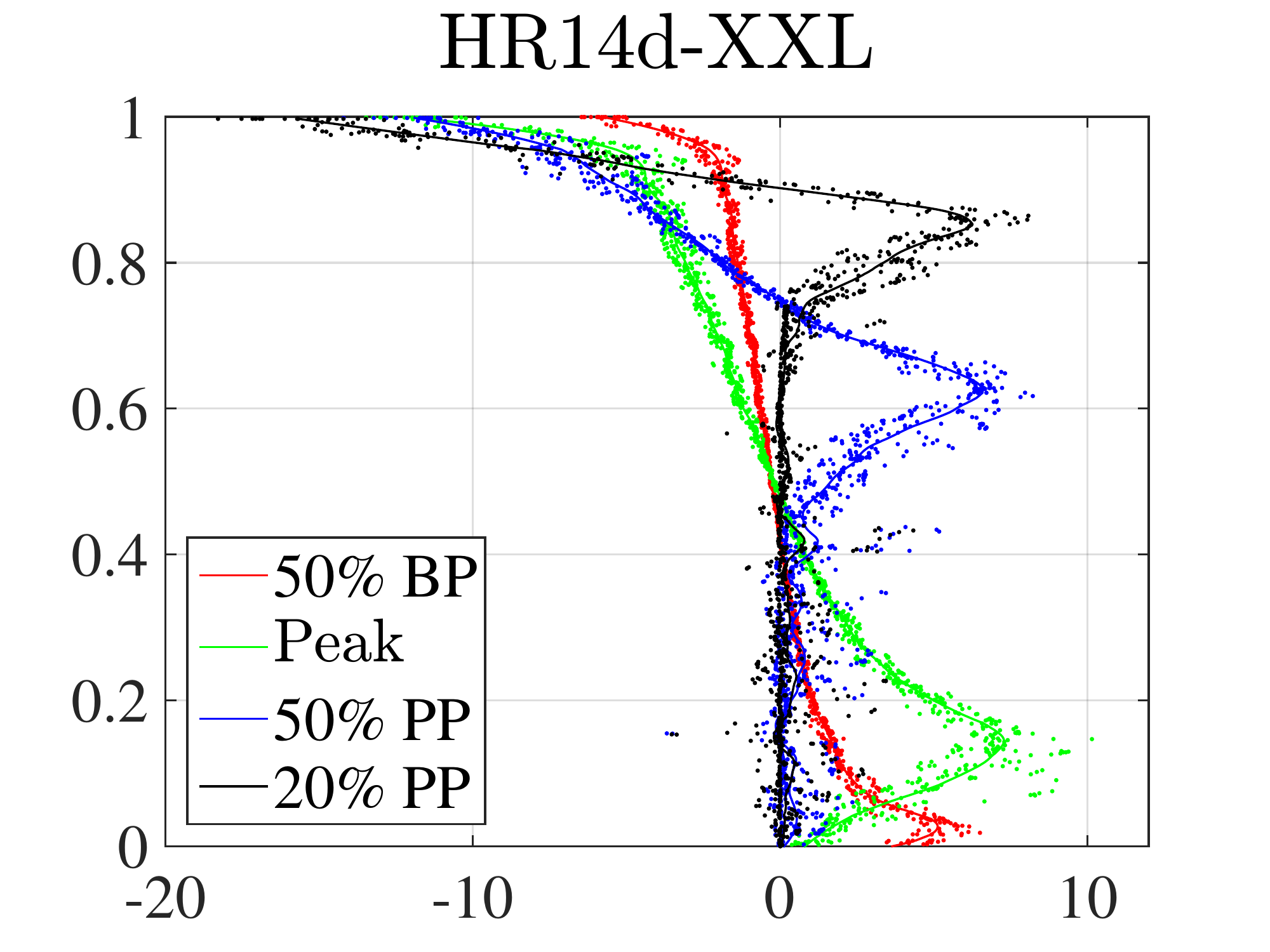} \\
  \includegraphics[height=1.35in]{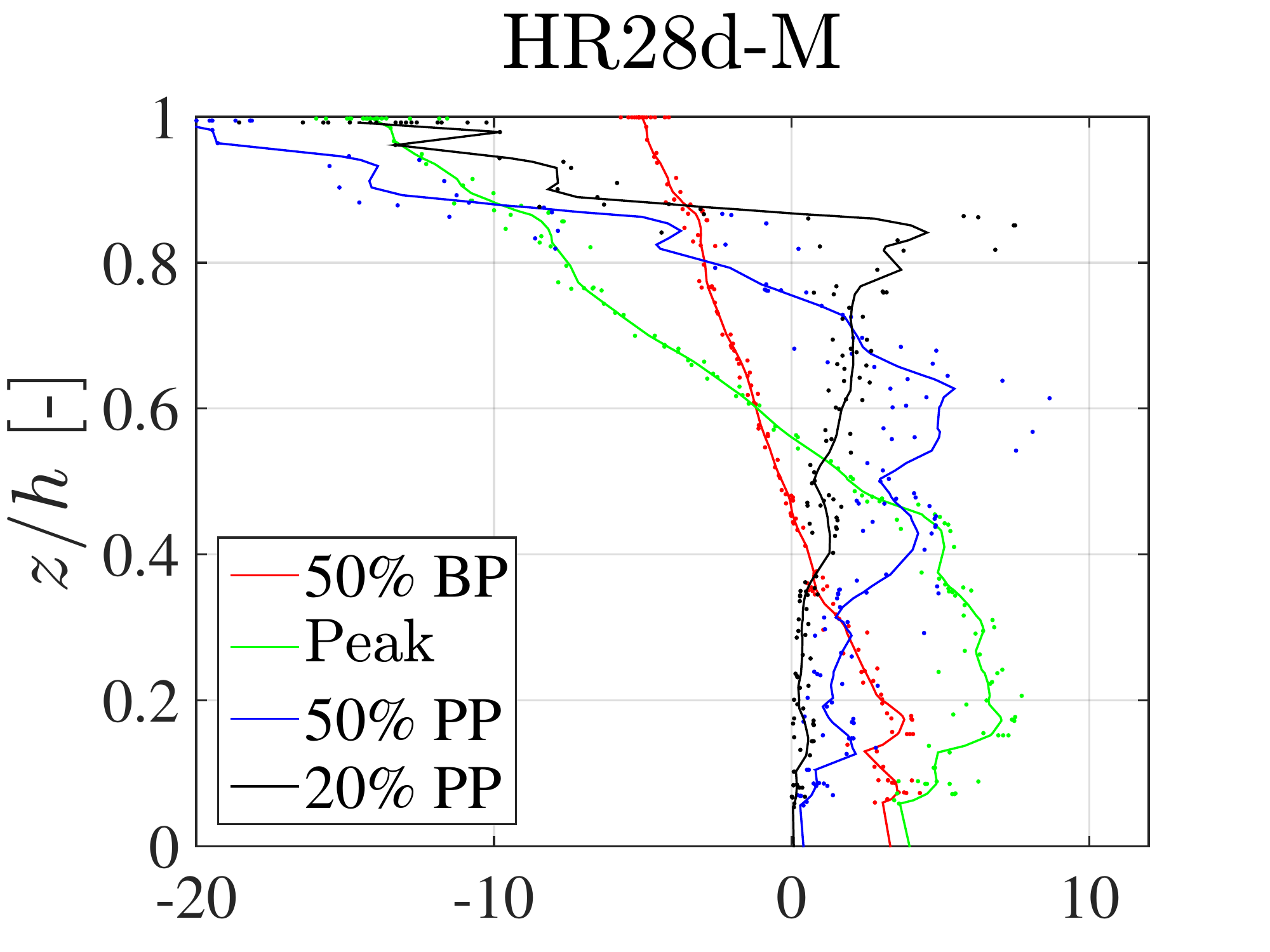} 
    \includegraphics[height=1.35in]{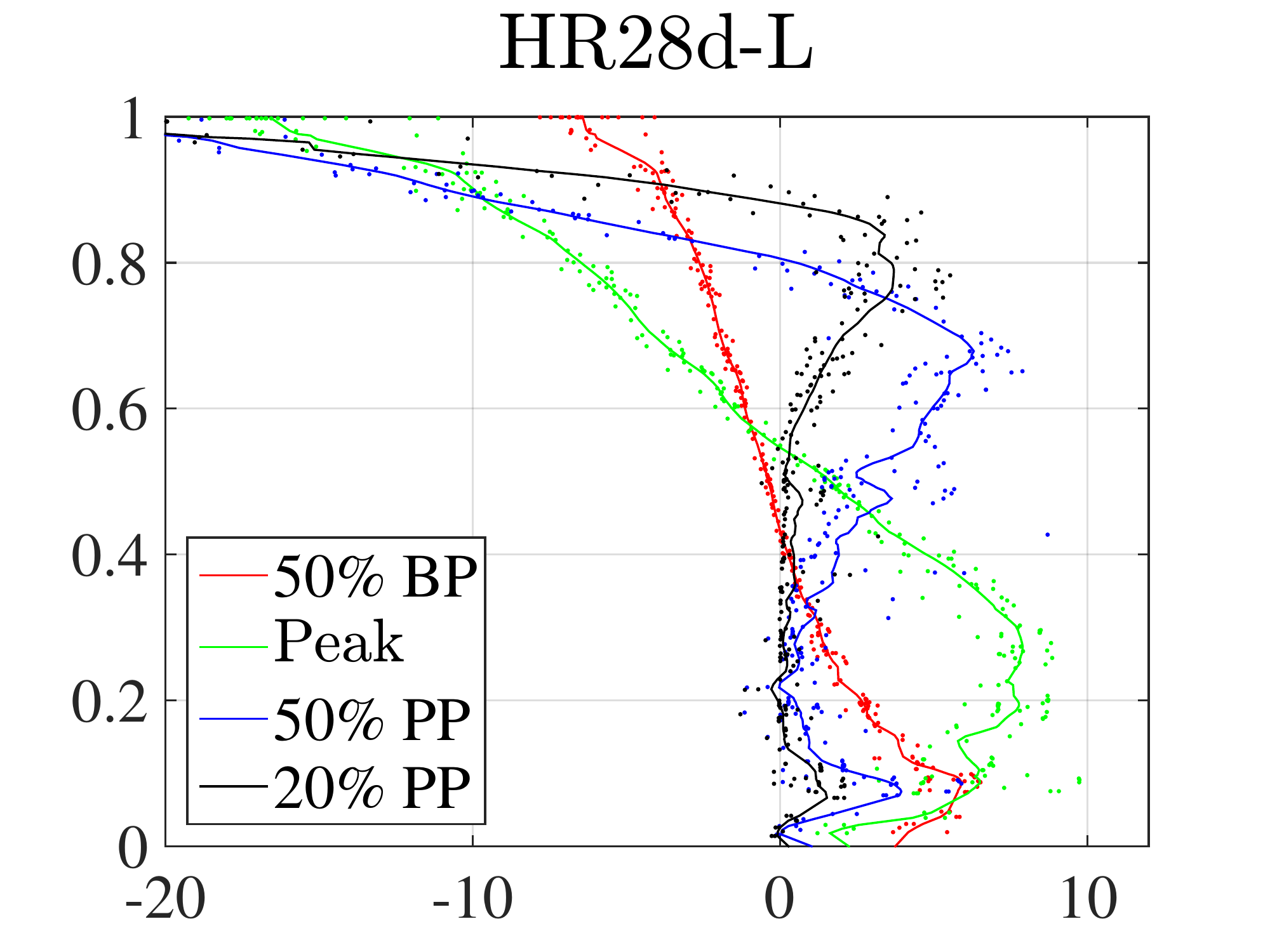} 
        \includegraphics[height=1.35in]{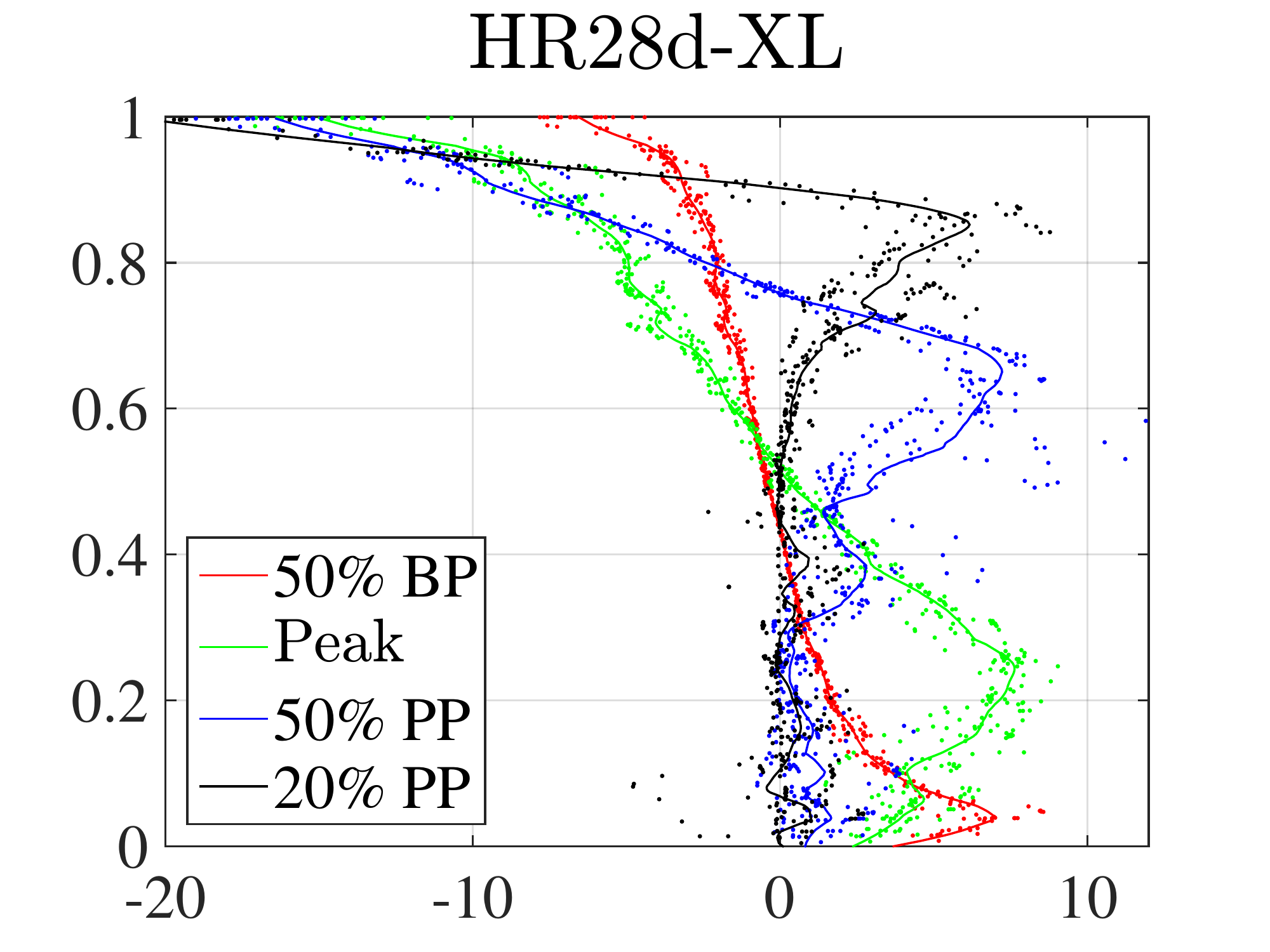} 
            \includegraphics[height=1.35in]{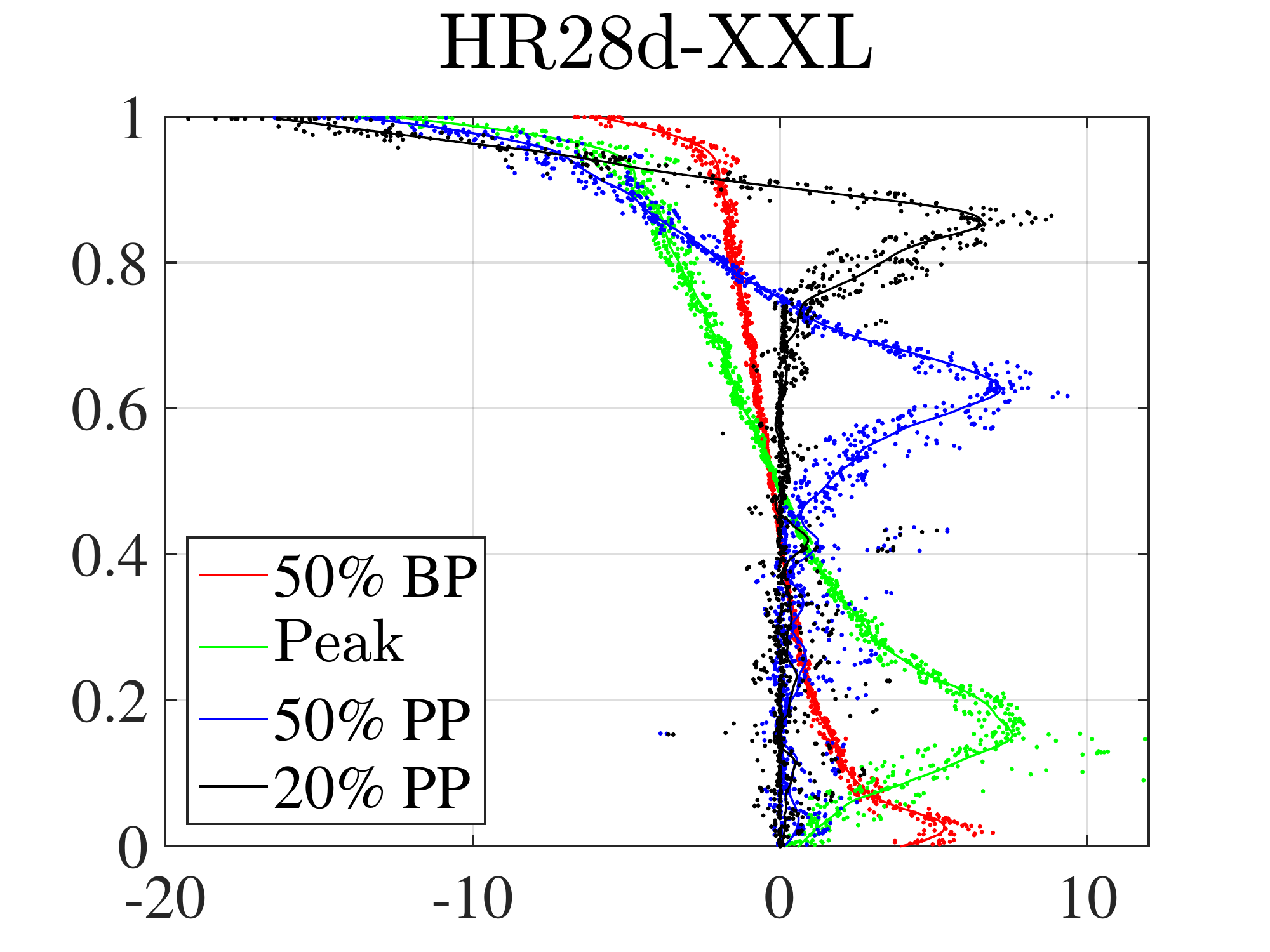} \\
  \includegraphics[height=1.4625in]{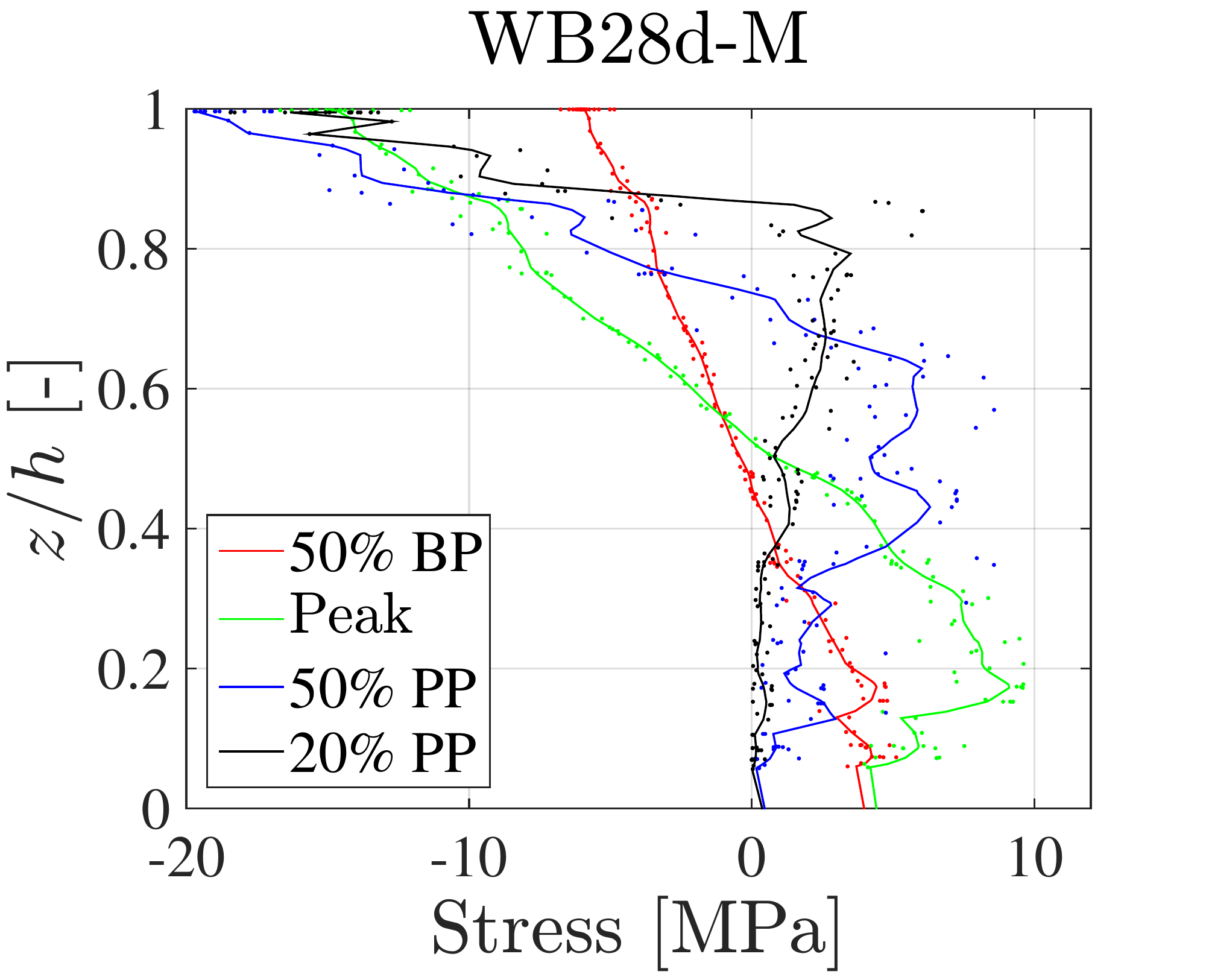} 
    \includegraphics[height=1.4625in]{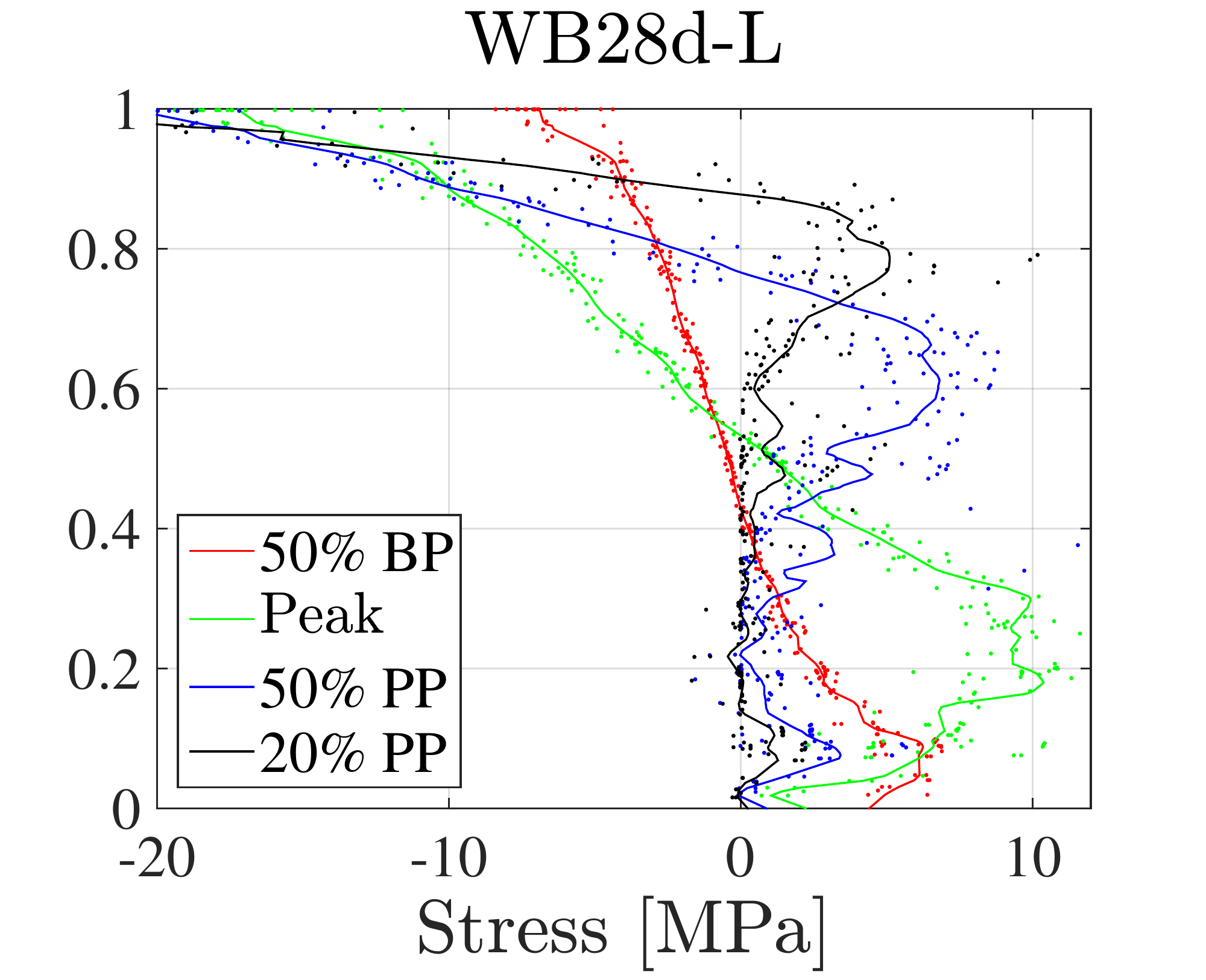} 
        \includegraphics[height=1.4625in]{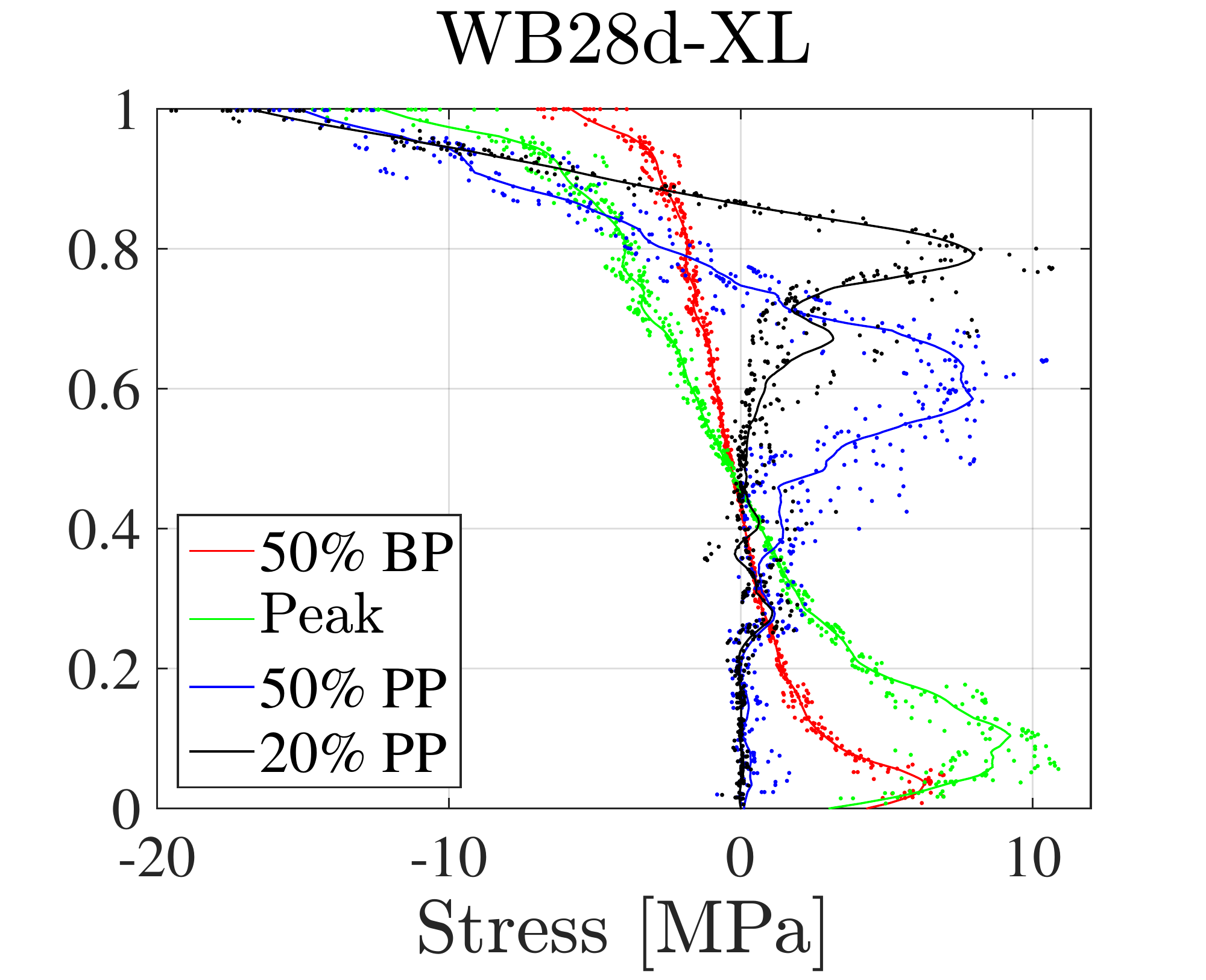} 
            \includegraphics[height=1.4625in]{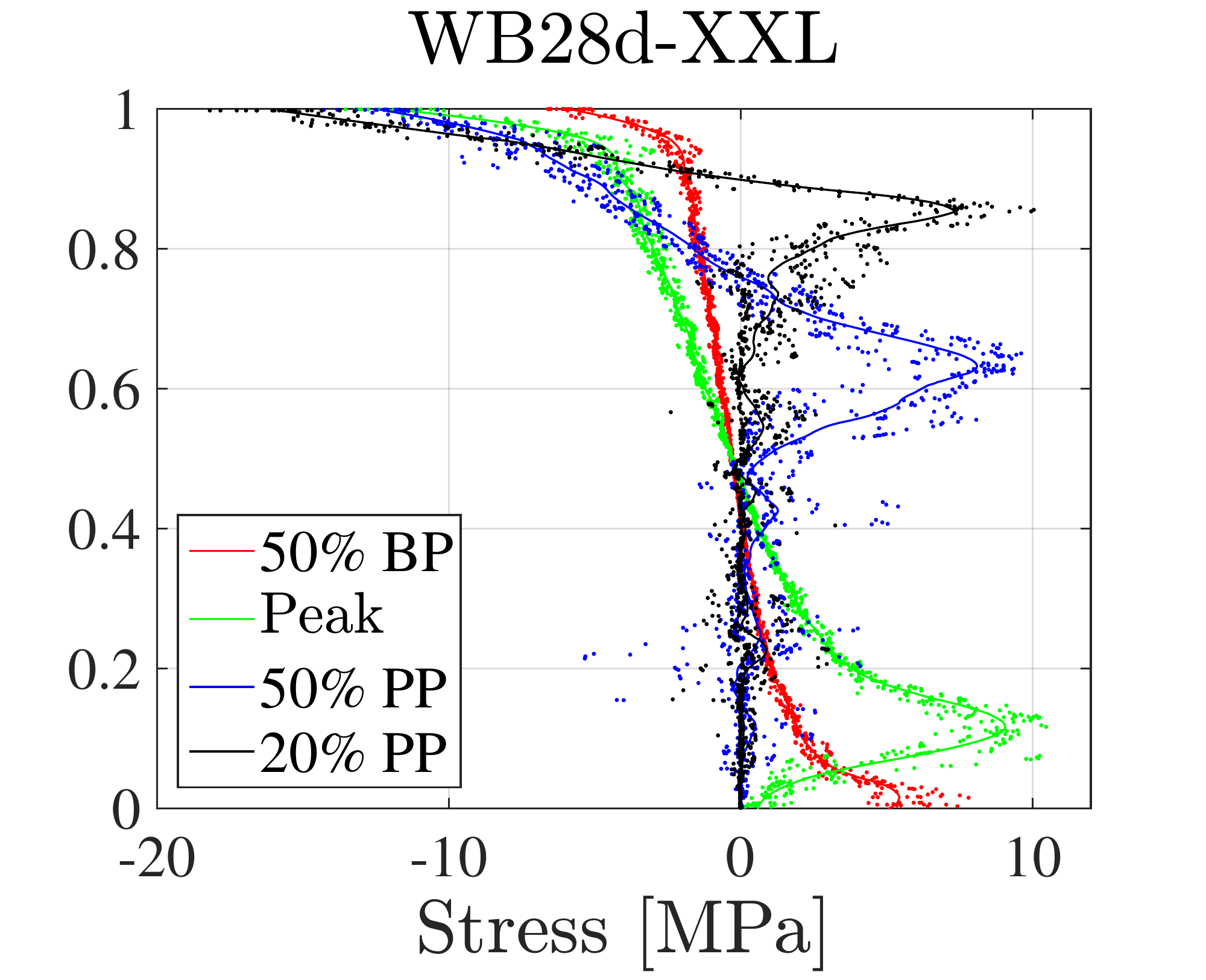} \\
\caption{Size and age dependent stress profiles along relative ligament height $z/h$ as obtained by HTC-LDPM simulations; from left to right: size M, L, XL, XXL; top to bottom: age HR3d, HR7d, HR14d, HR28d, WB28d}    
\label{StressTensor}        
\efi 

\bfi [ht]
\centering
  \includegraphics[height=1.35in]{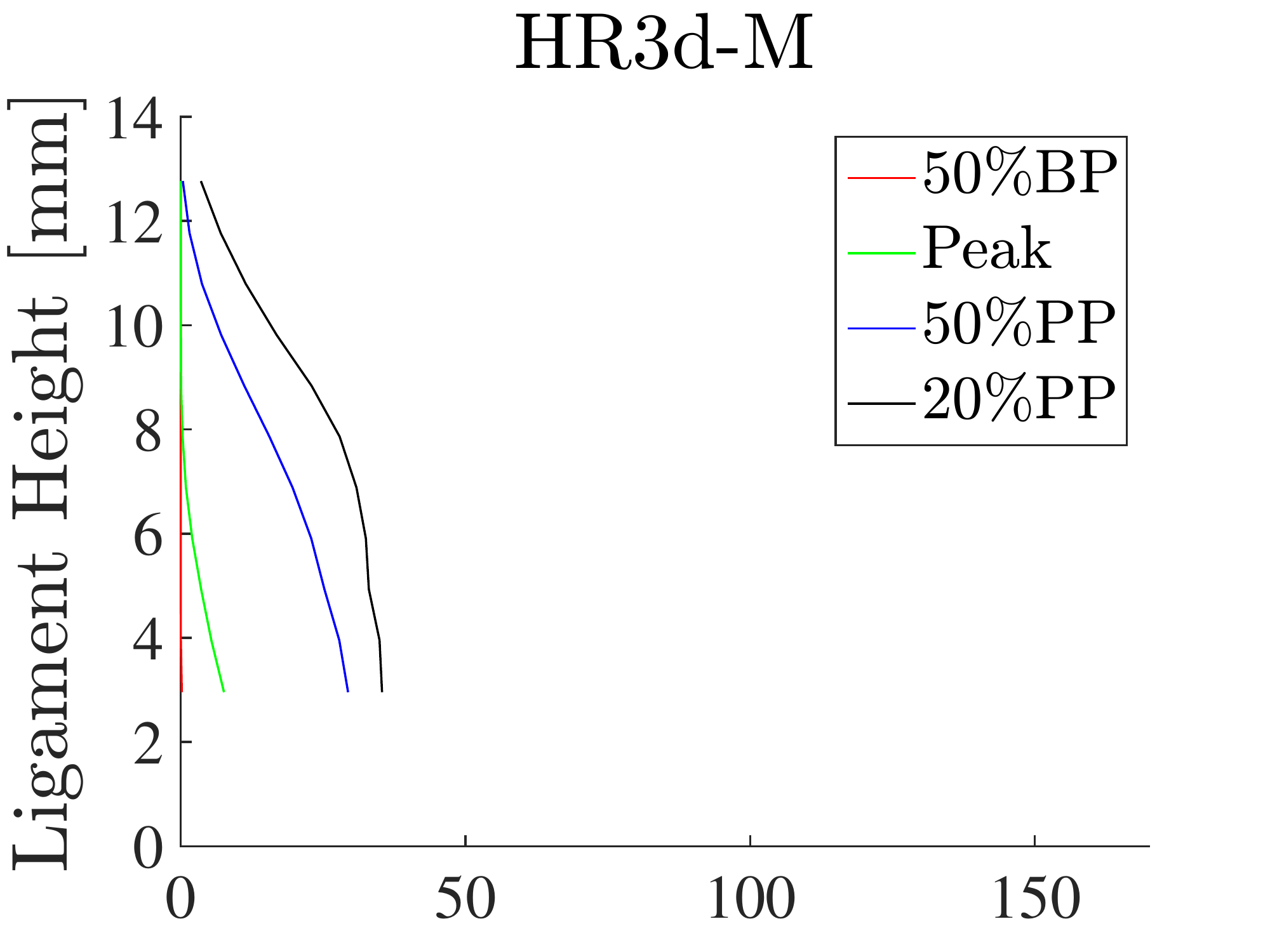} 
    \includegraphics[height=1.35in]{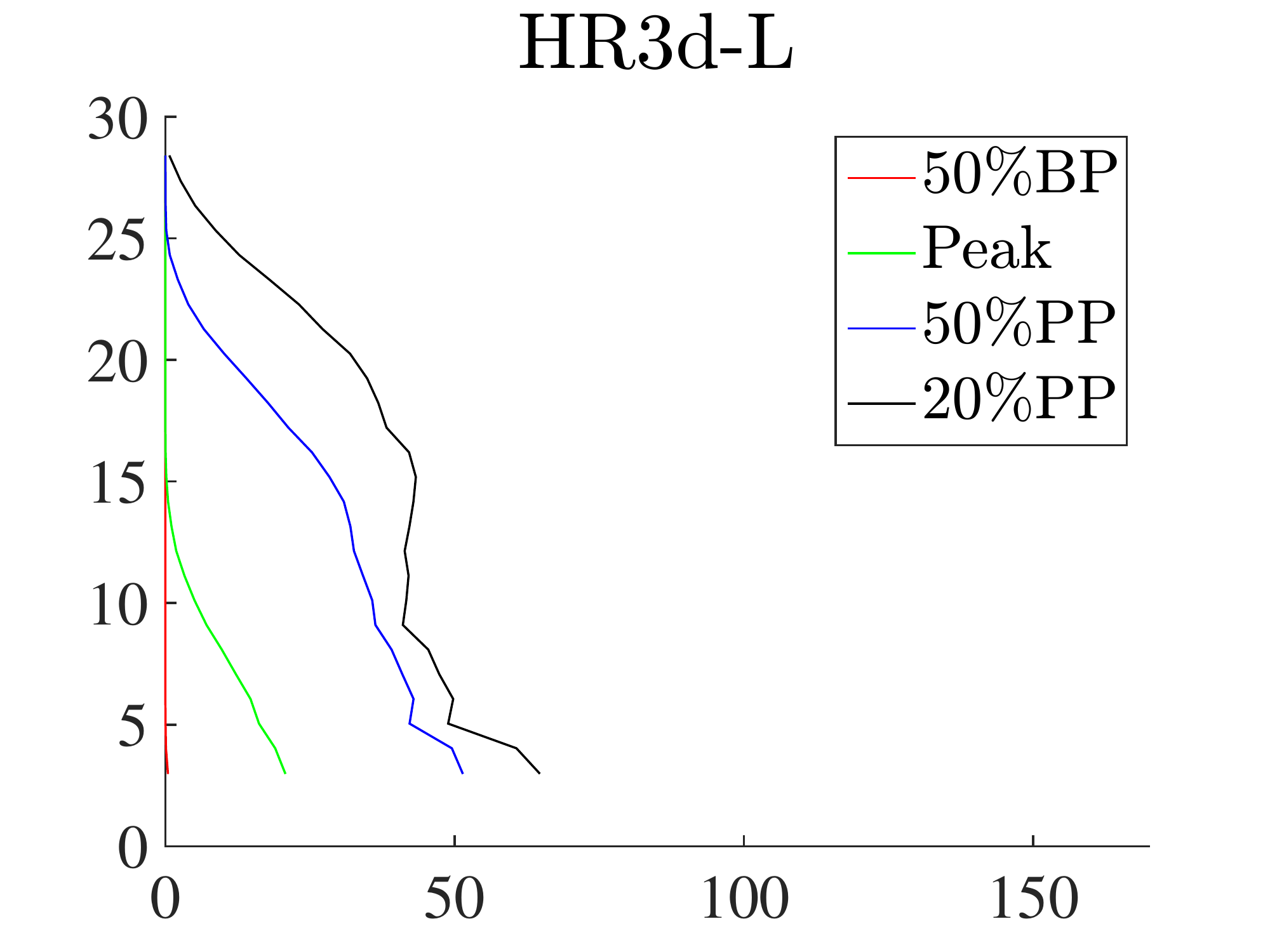} 
        \includegraphics[height=1.35in]{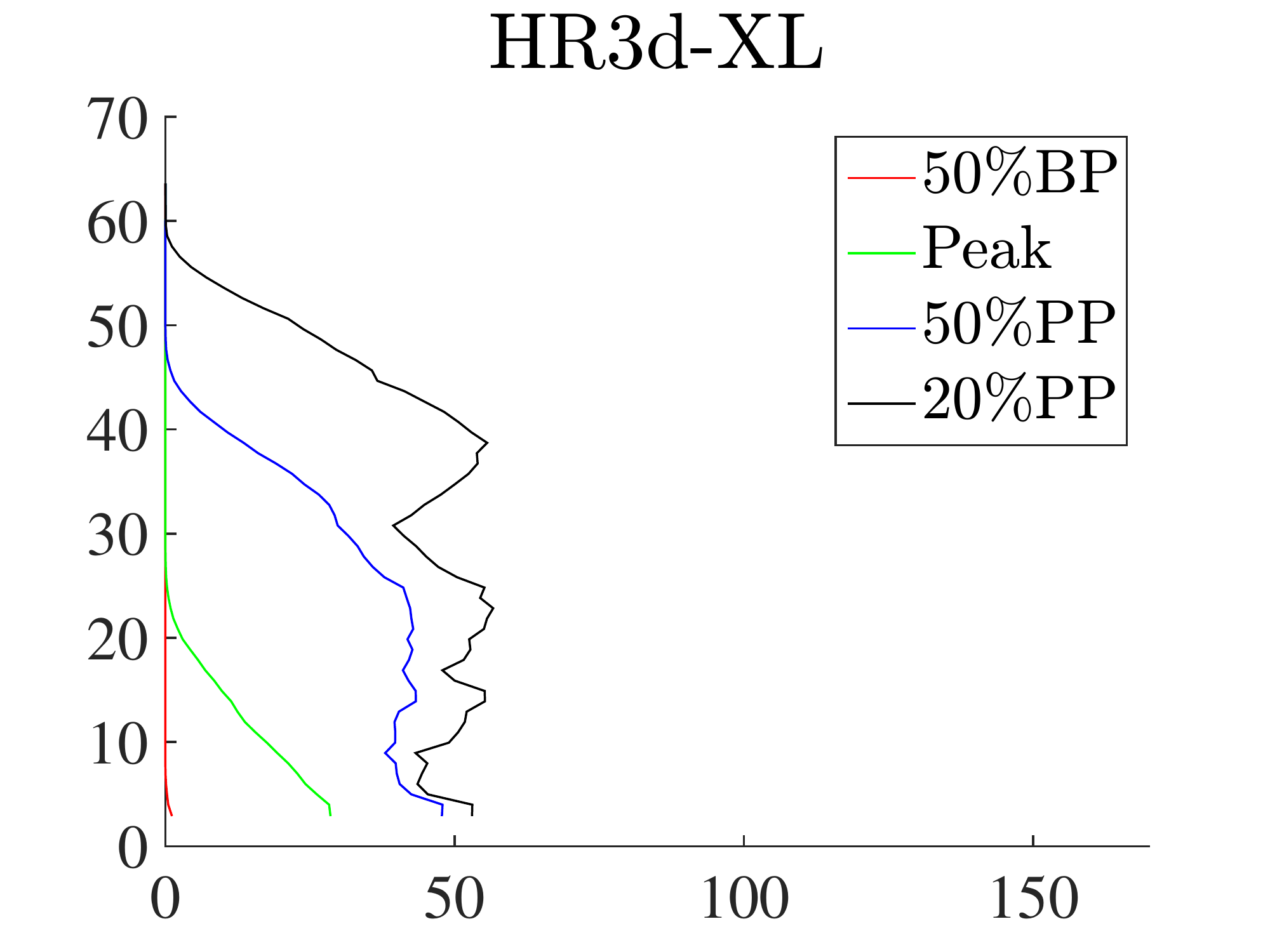} 
            \includegraphics[height=1.35in]{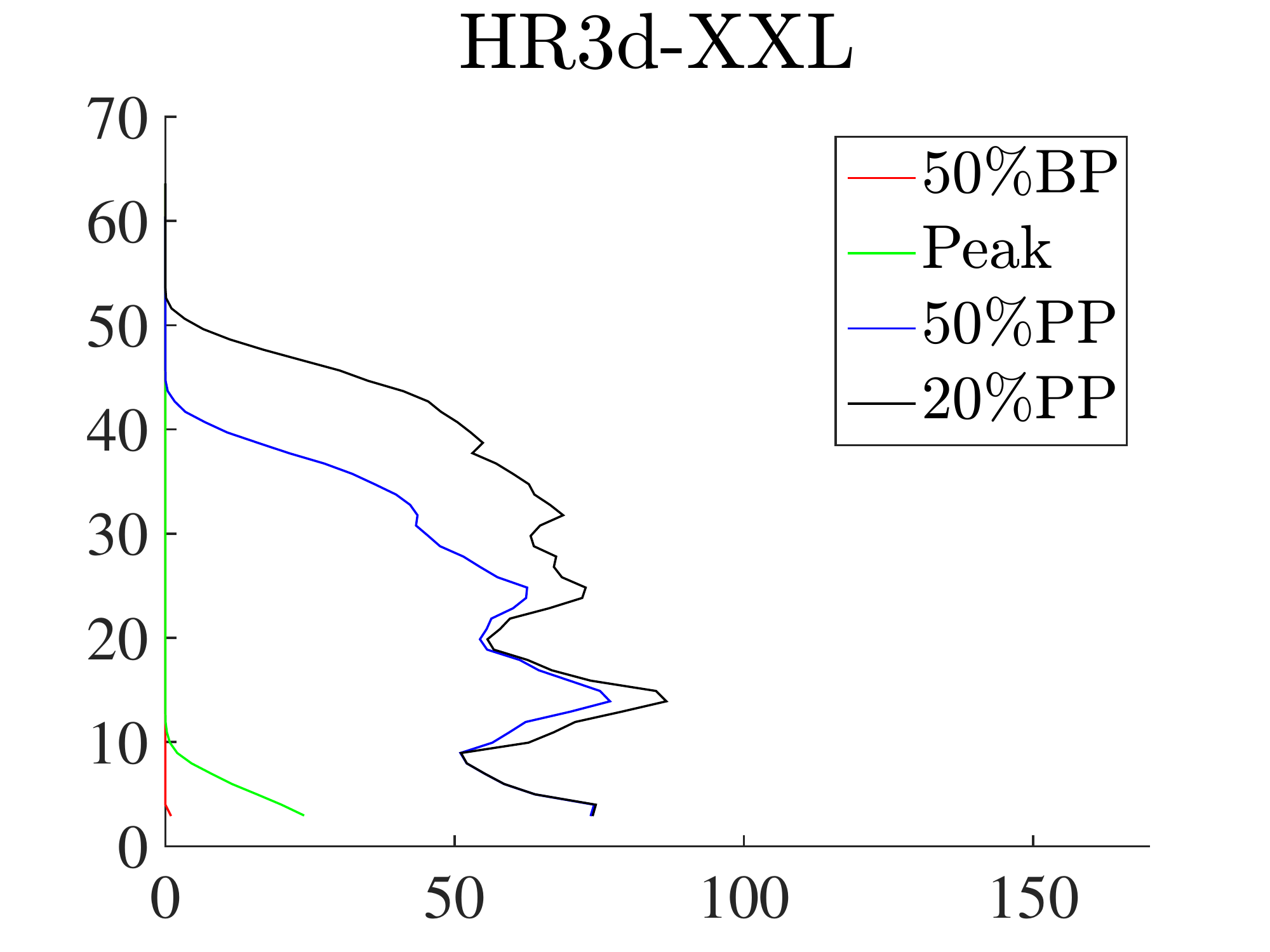} \\
  \includegraphics[height=1.35in]{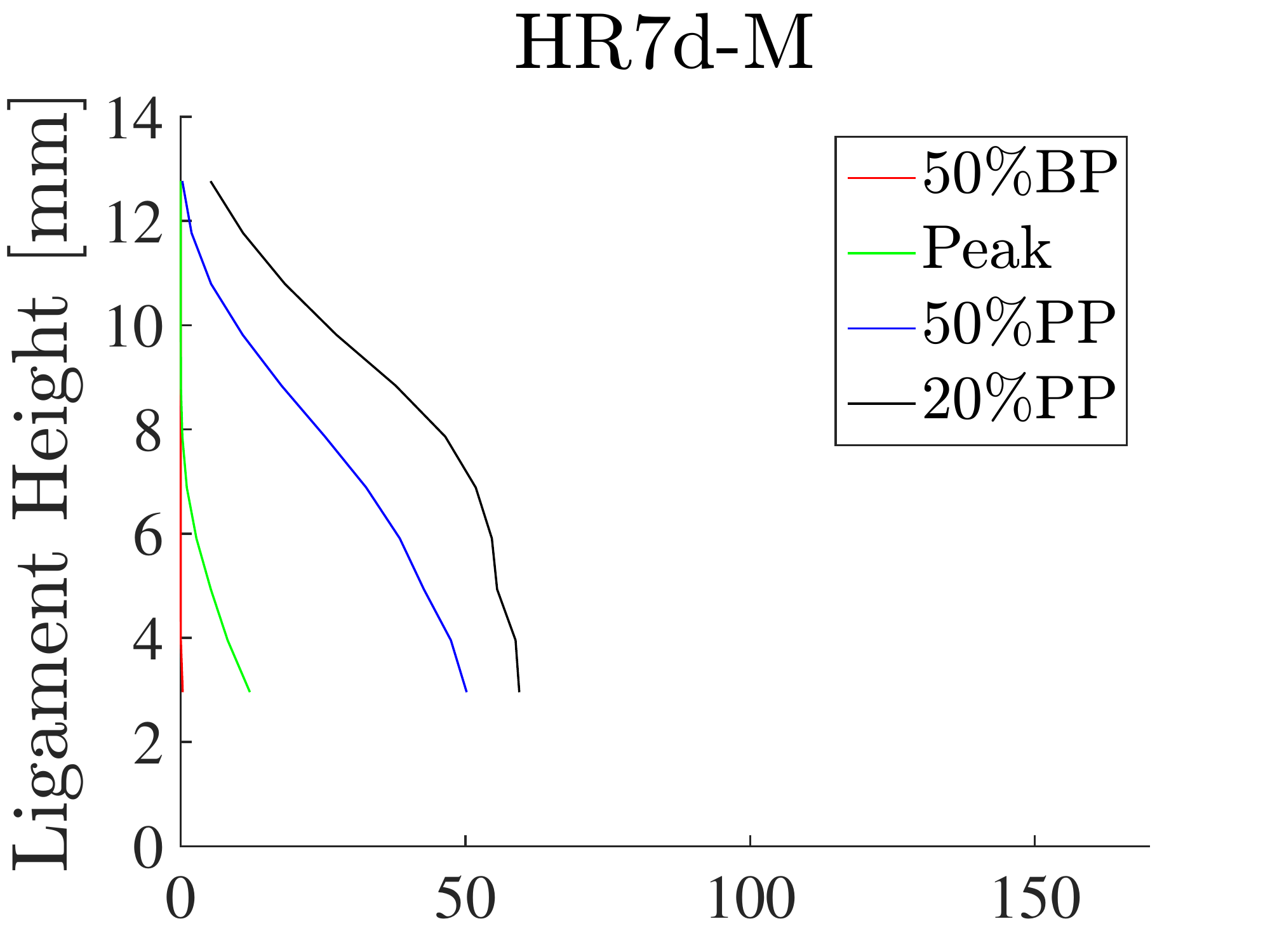} 
    \includegraphics[height=1.35in]{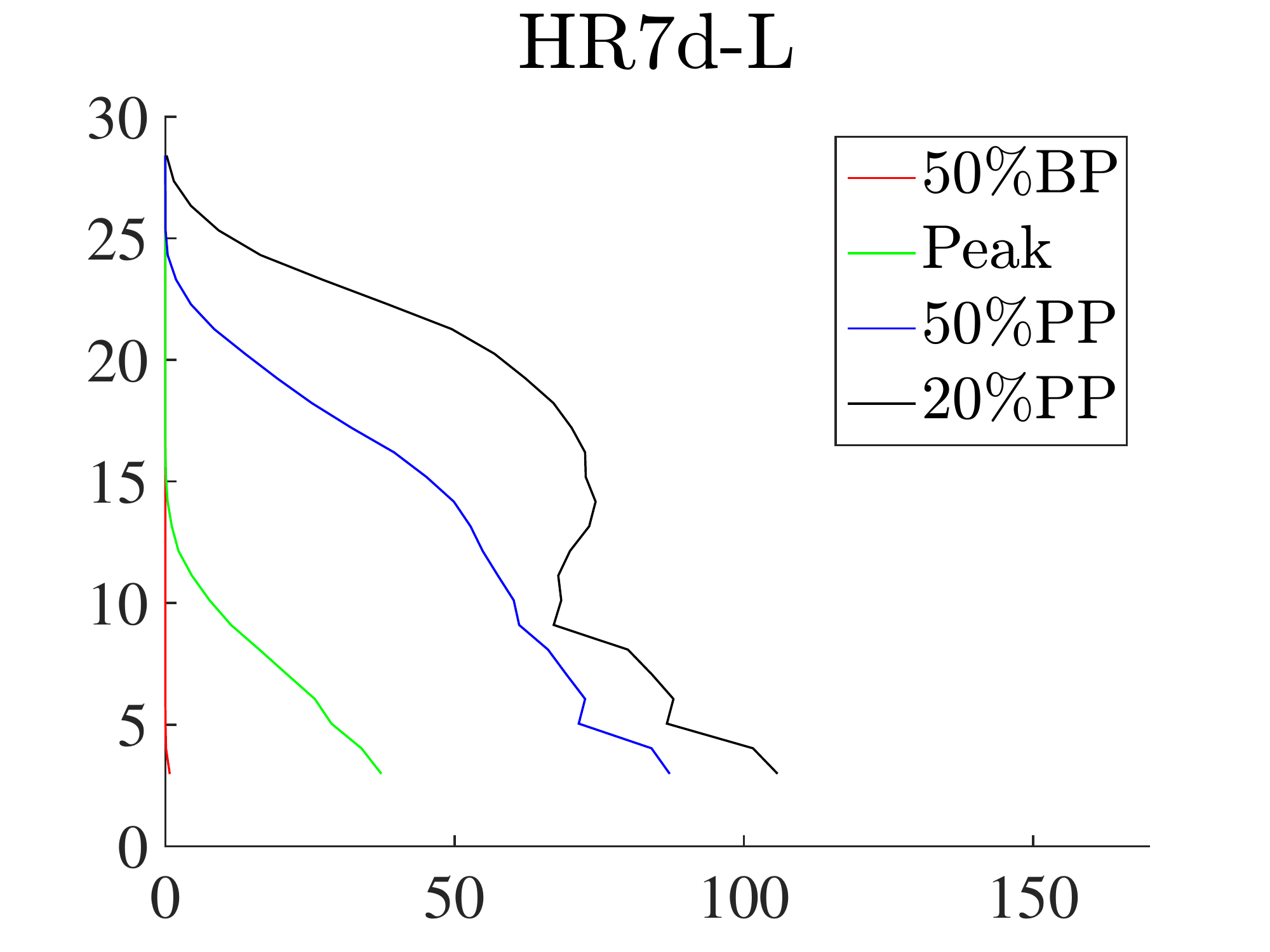} 
        \includegraphics[height=1.35in]{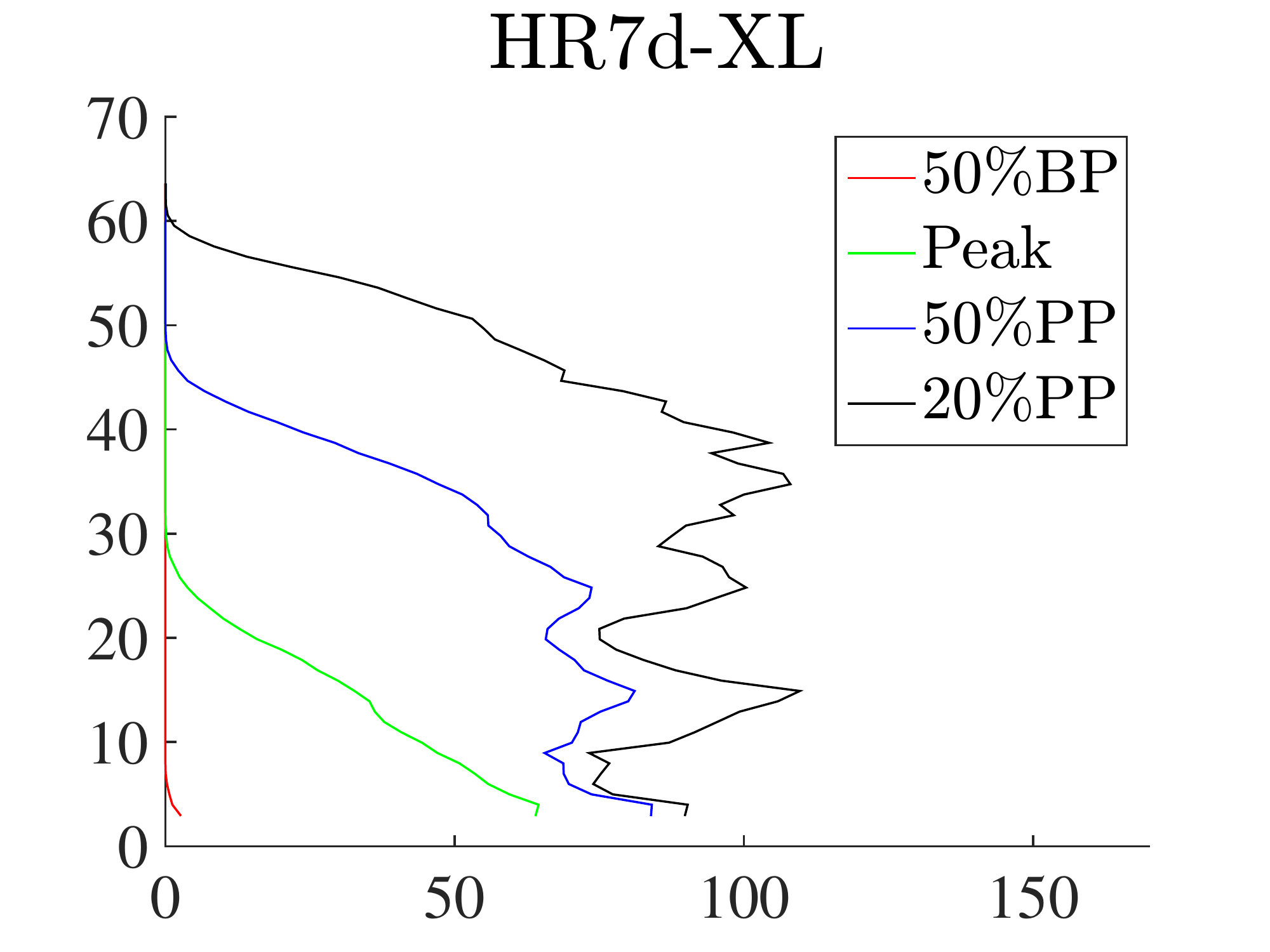} 
            \includegraphics[height=1.35in]{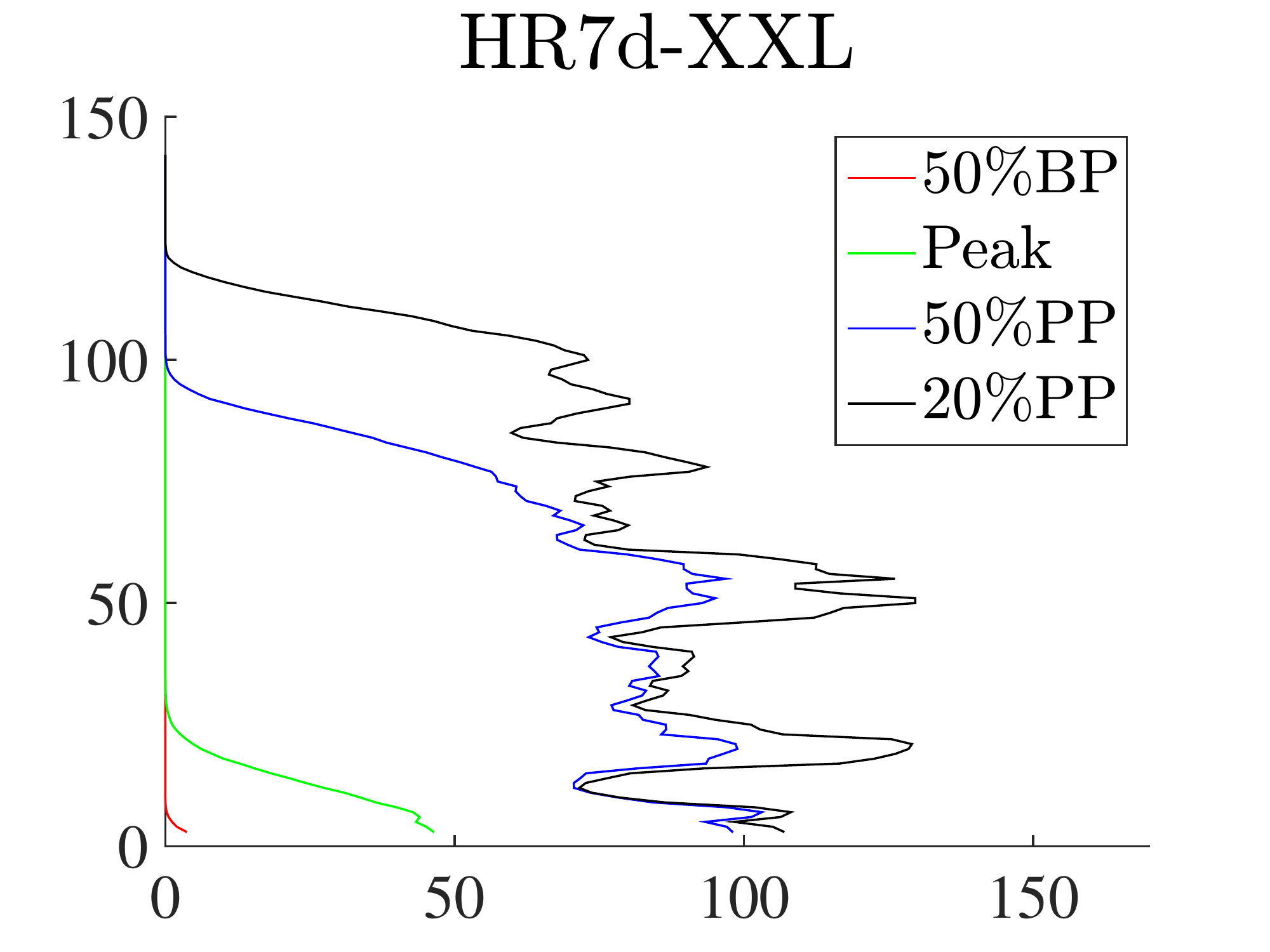} \\
  \includegraphics[height=1.35in]{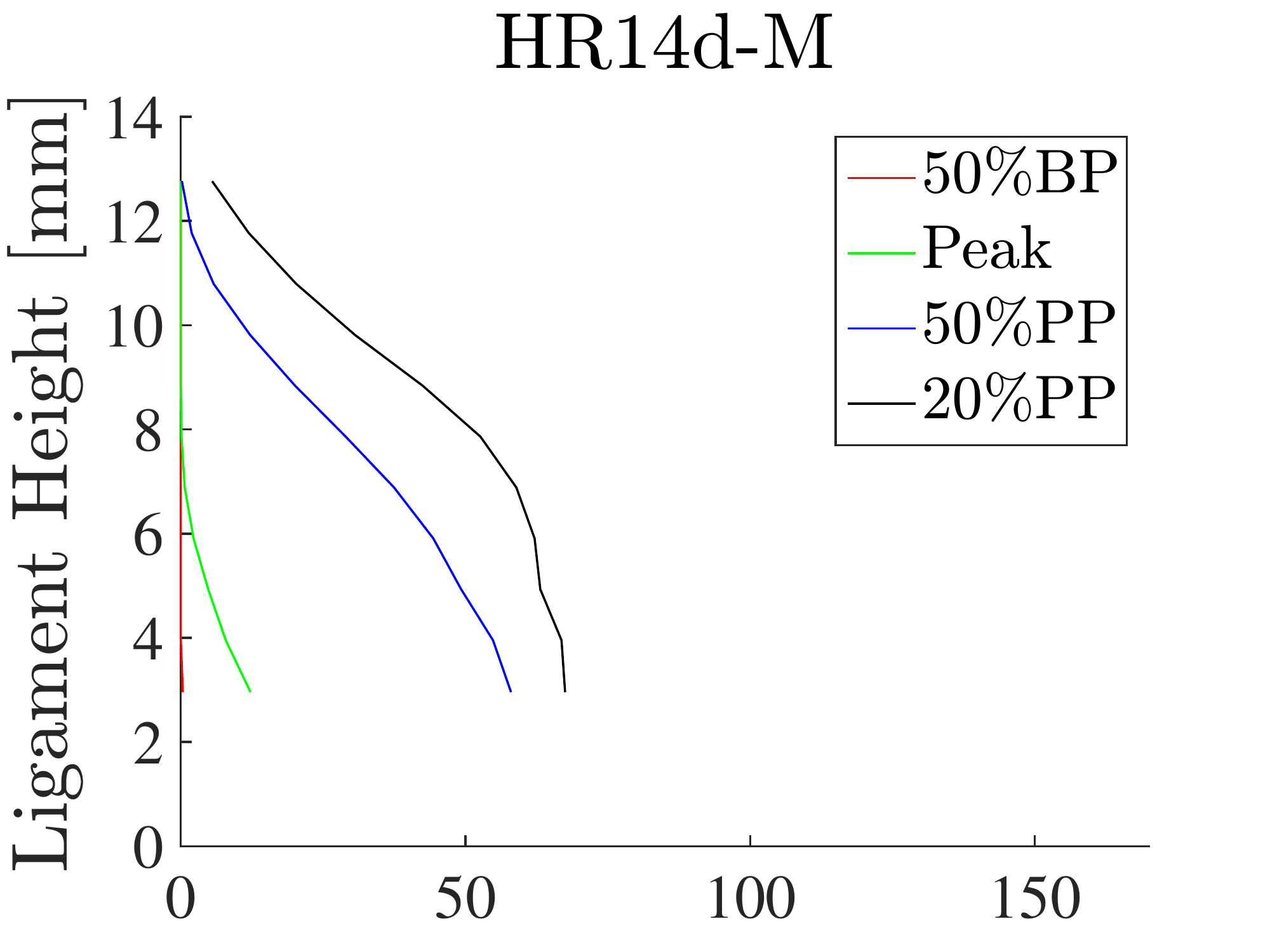} 
    \includegraphics[height=1.35in]{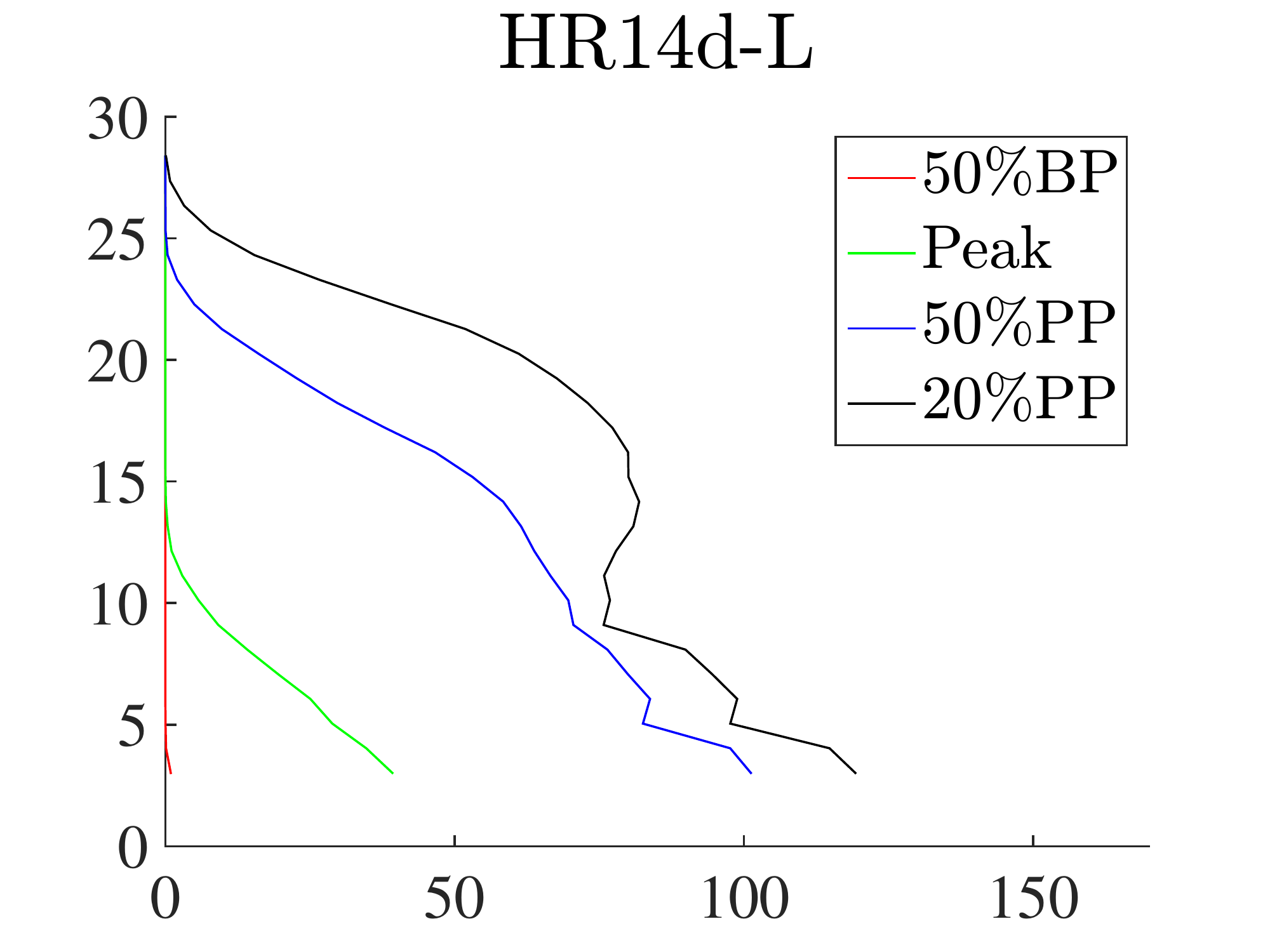} 
        \includegraphics[height=1.35in]{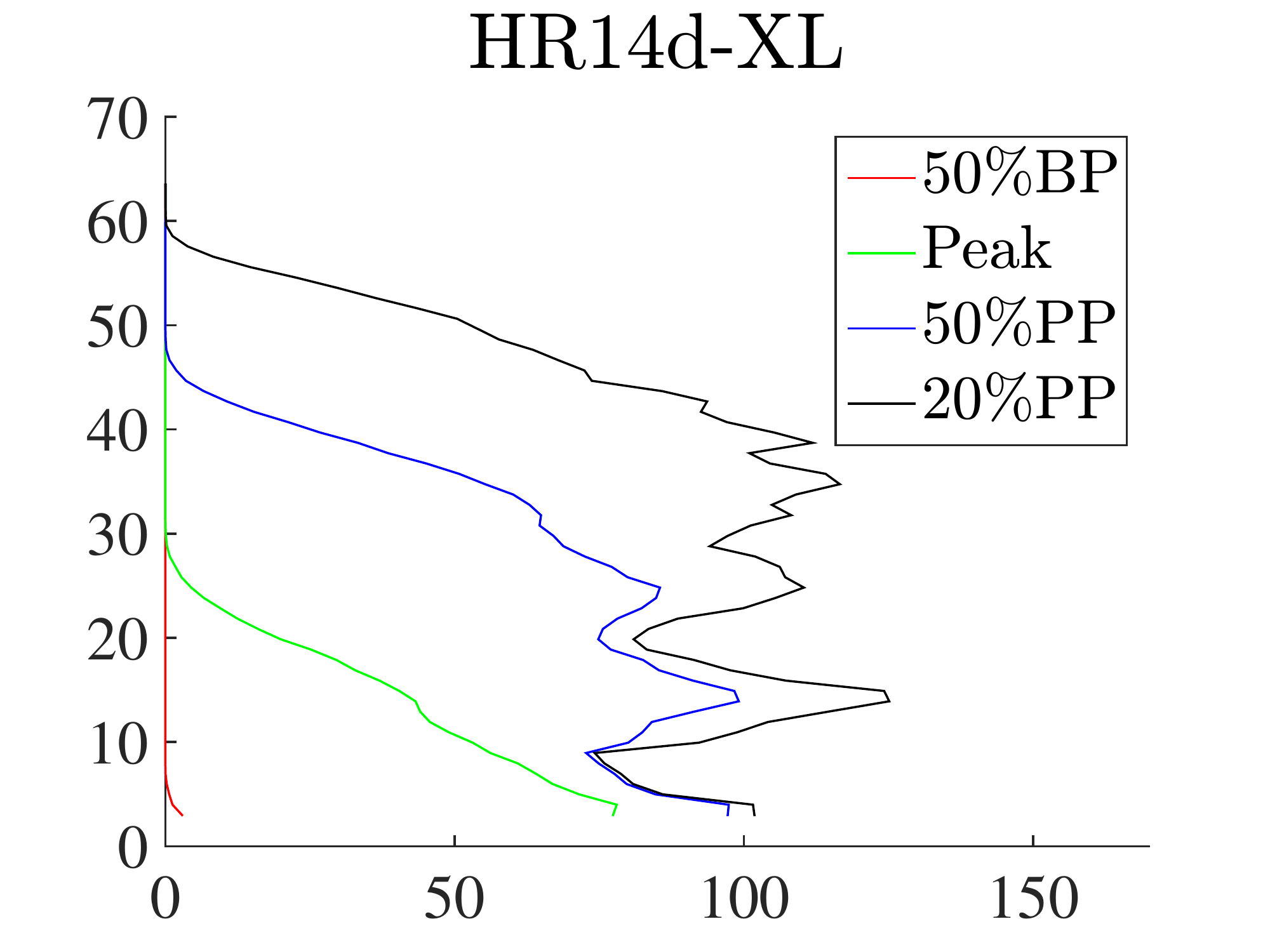} 
            \includegraphics[height=1.35in]{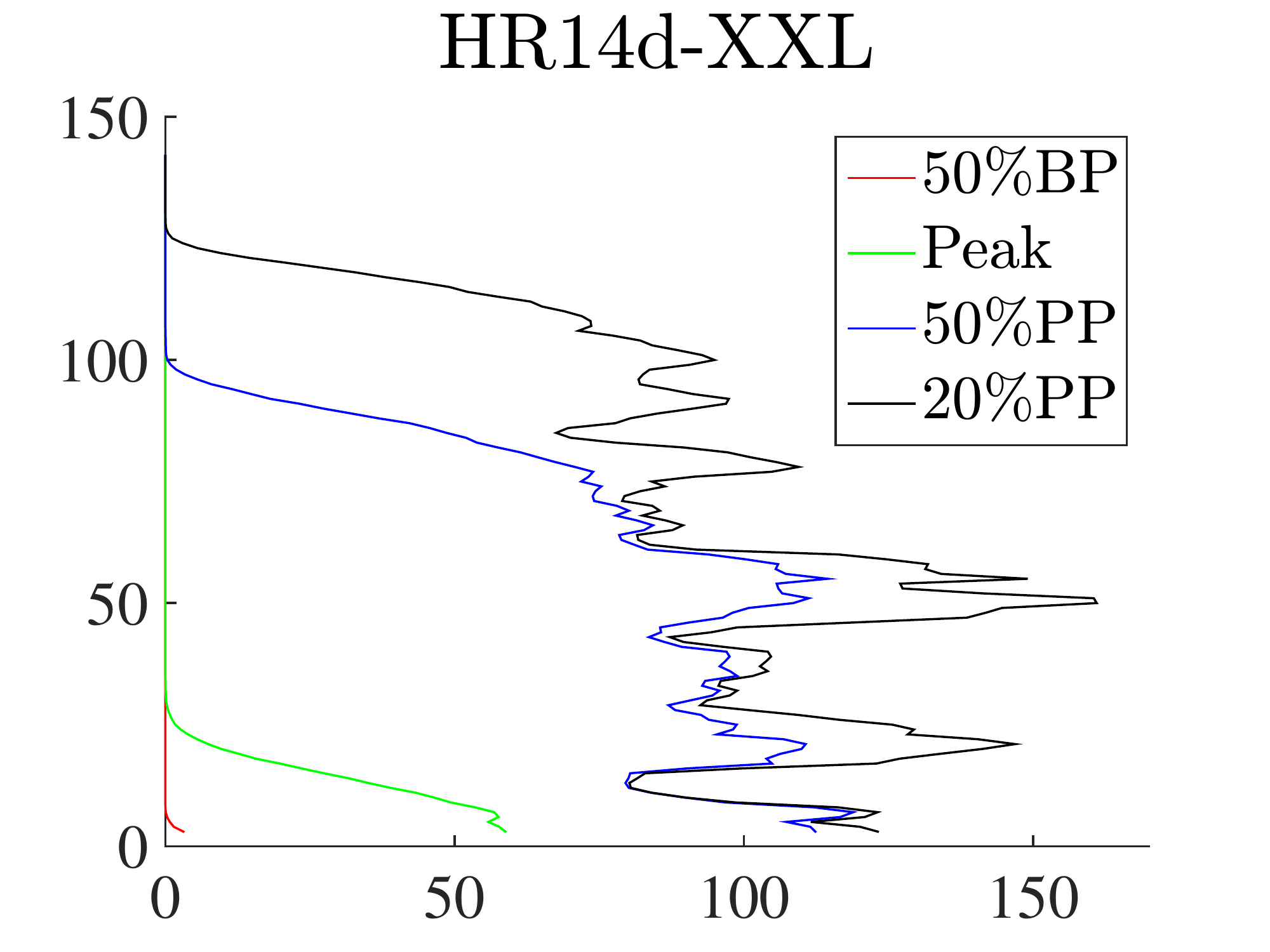} \\
  \includegraphics[height=1.35in]{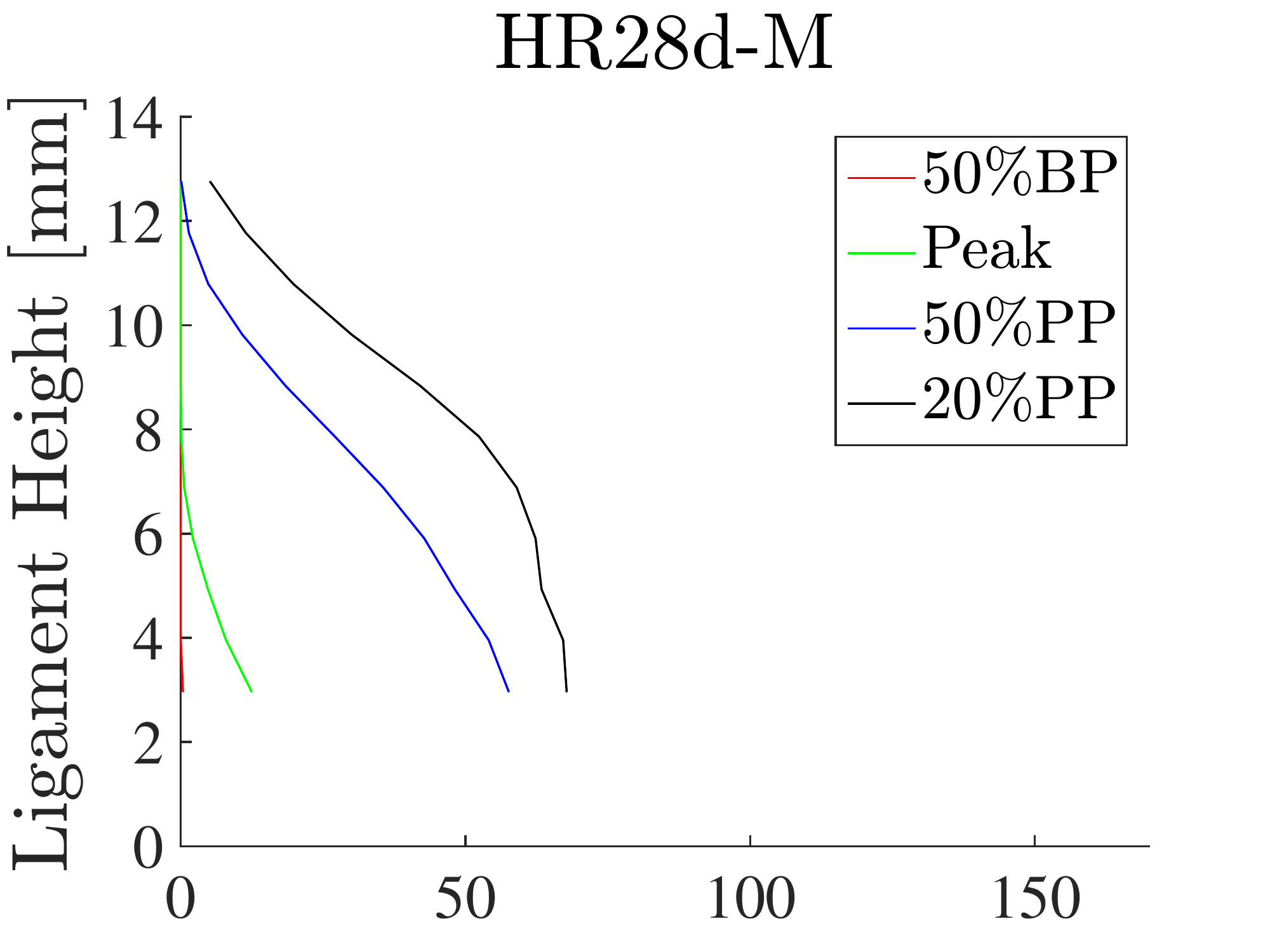} 
    \includegraphics[height=1.35in]{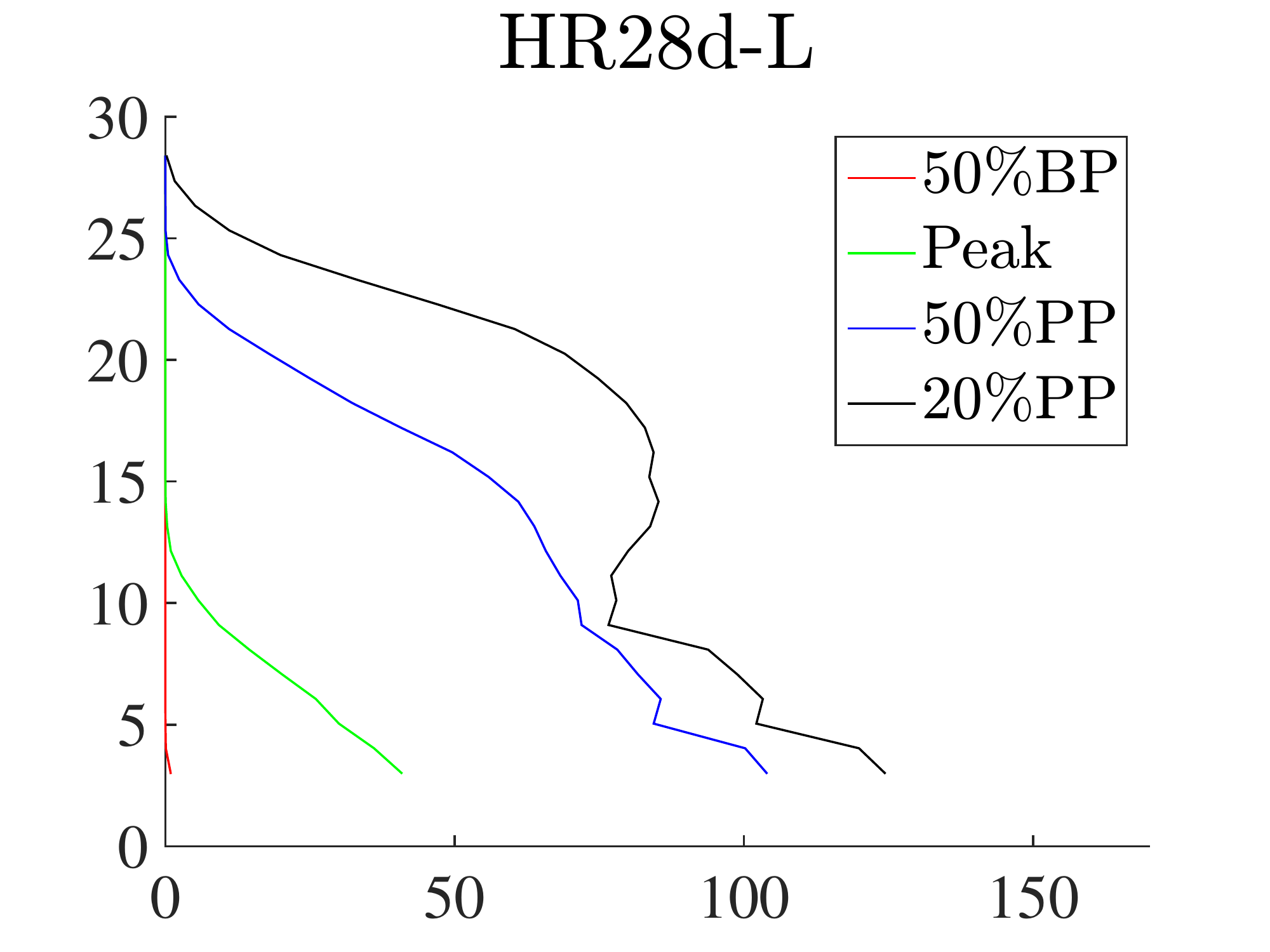} 
        \includegraphics[height=1.35in]{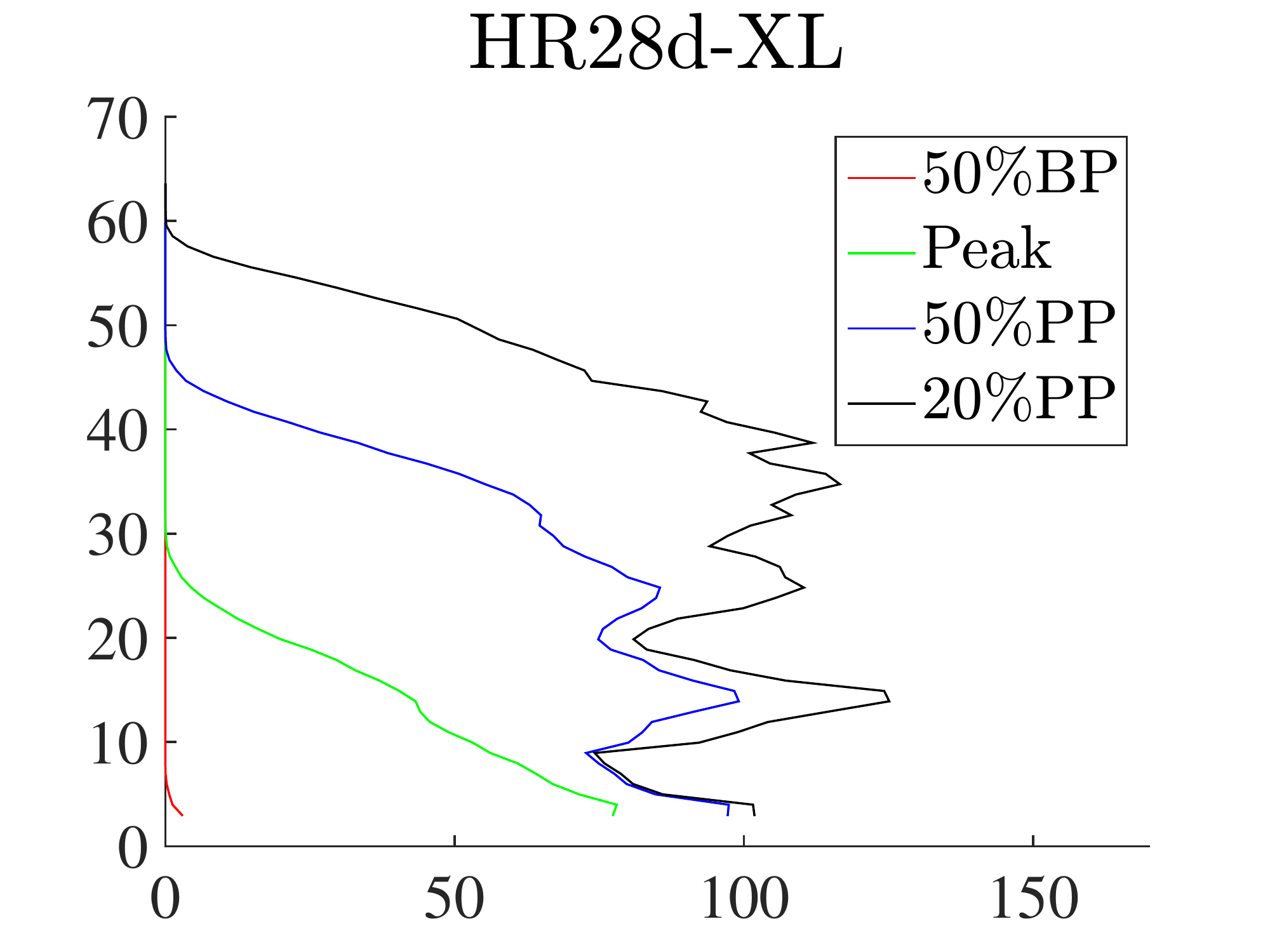} 
            \includegraphics[height=1.35in]{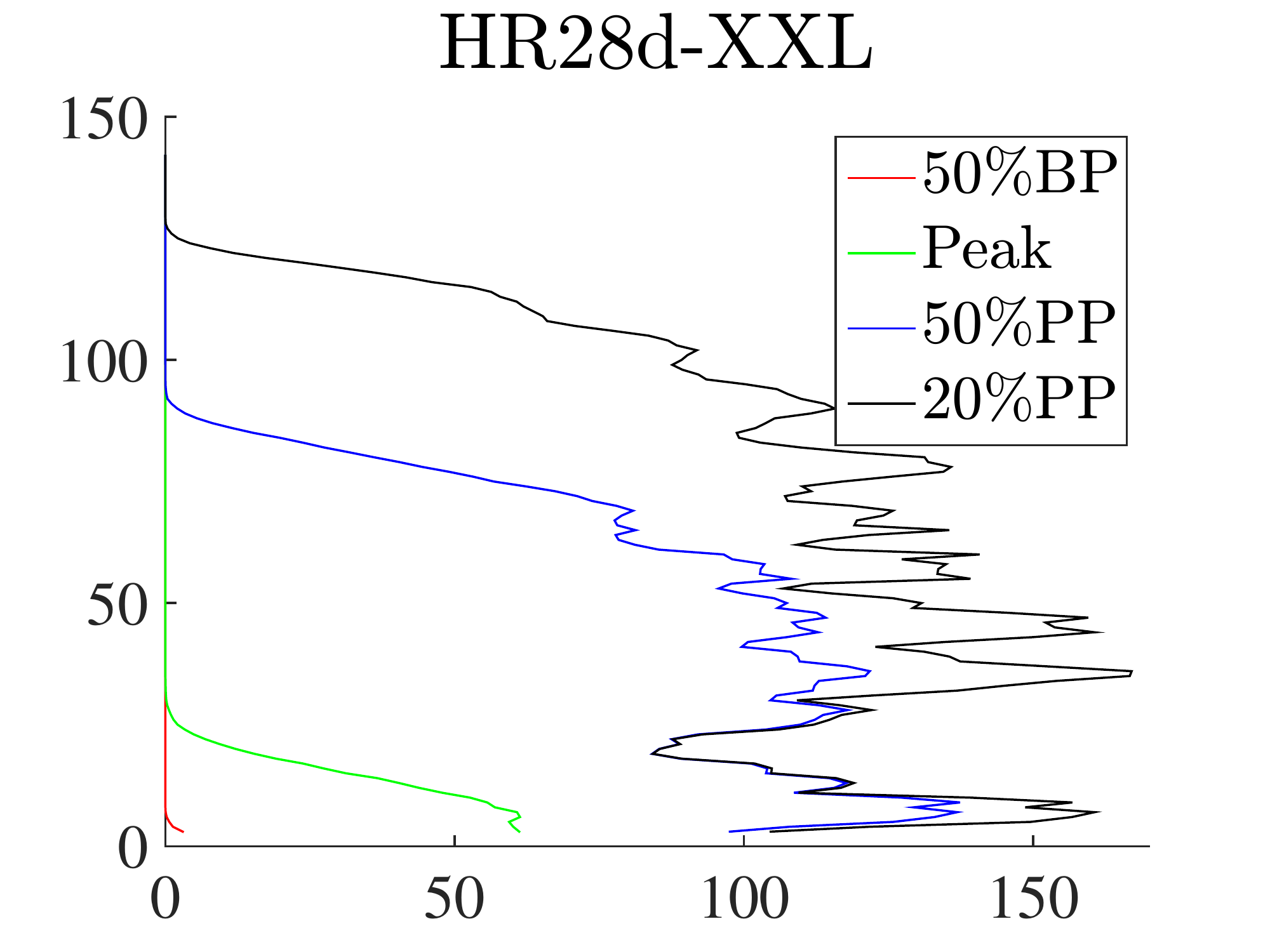} \\
  \includegraphics[height=1.4625in]{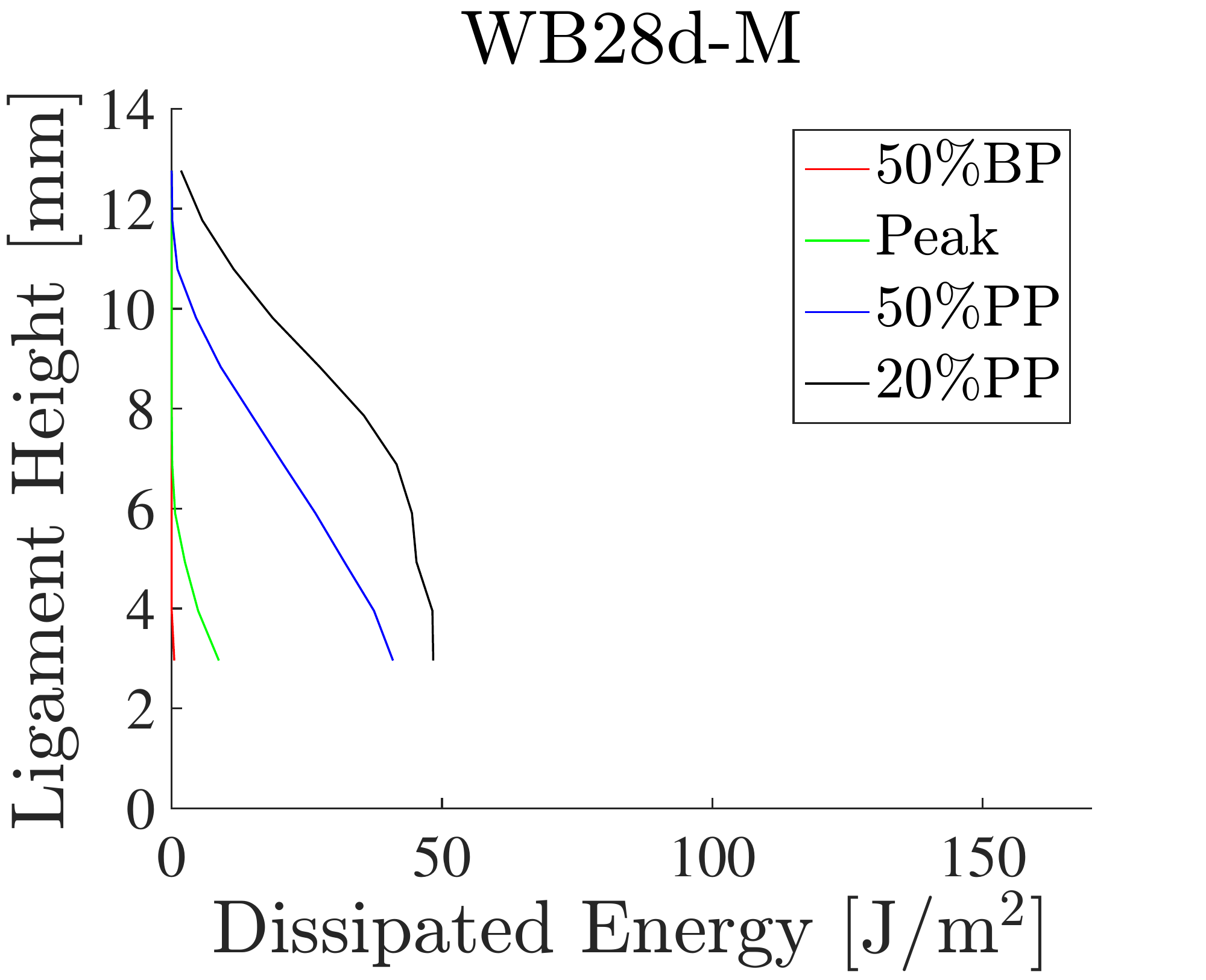} 
    \includegraphics[height=1.4625in]{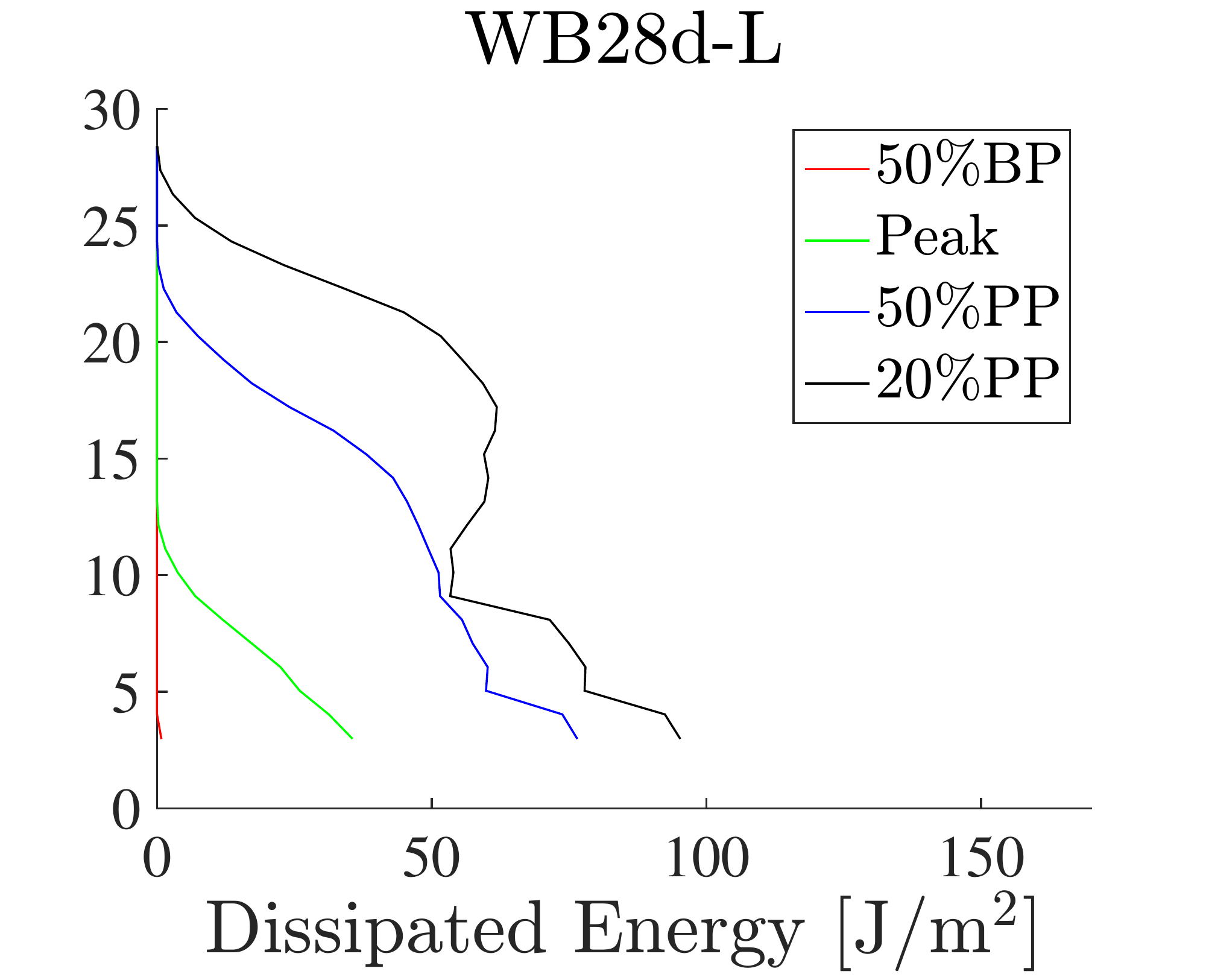} 
        \includegraphics[height=1.4625in]{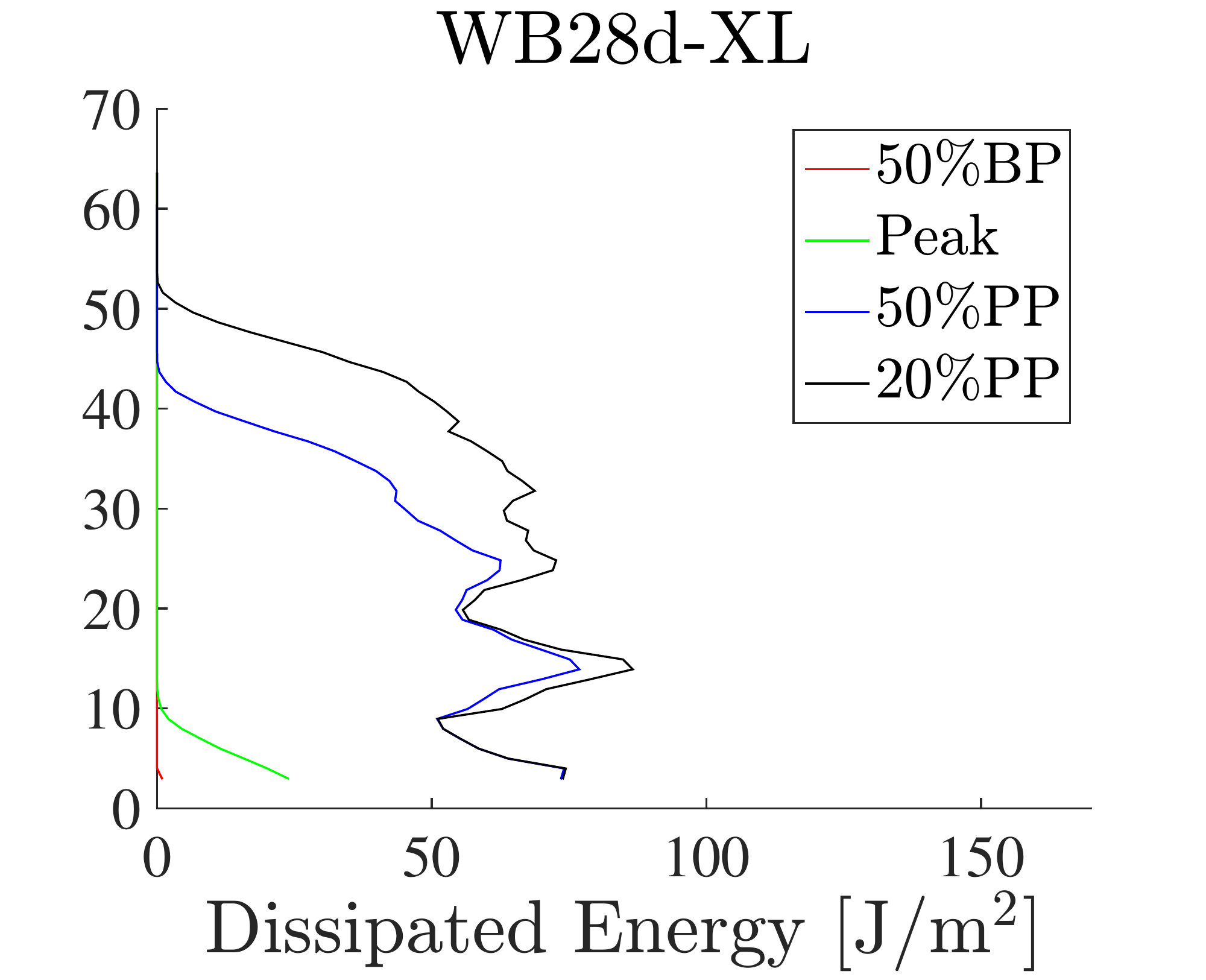} 
            \includegraphics[height=1.4625in]{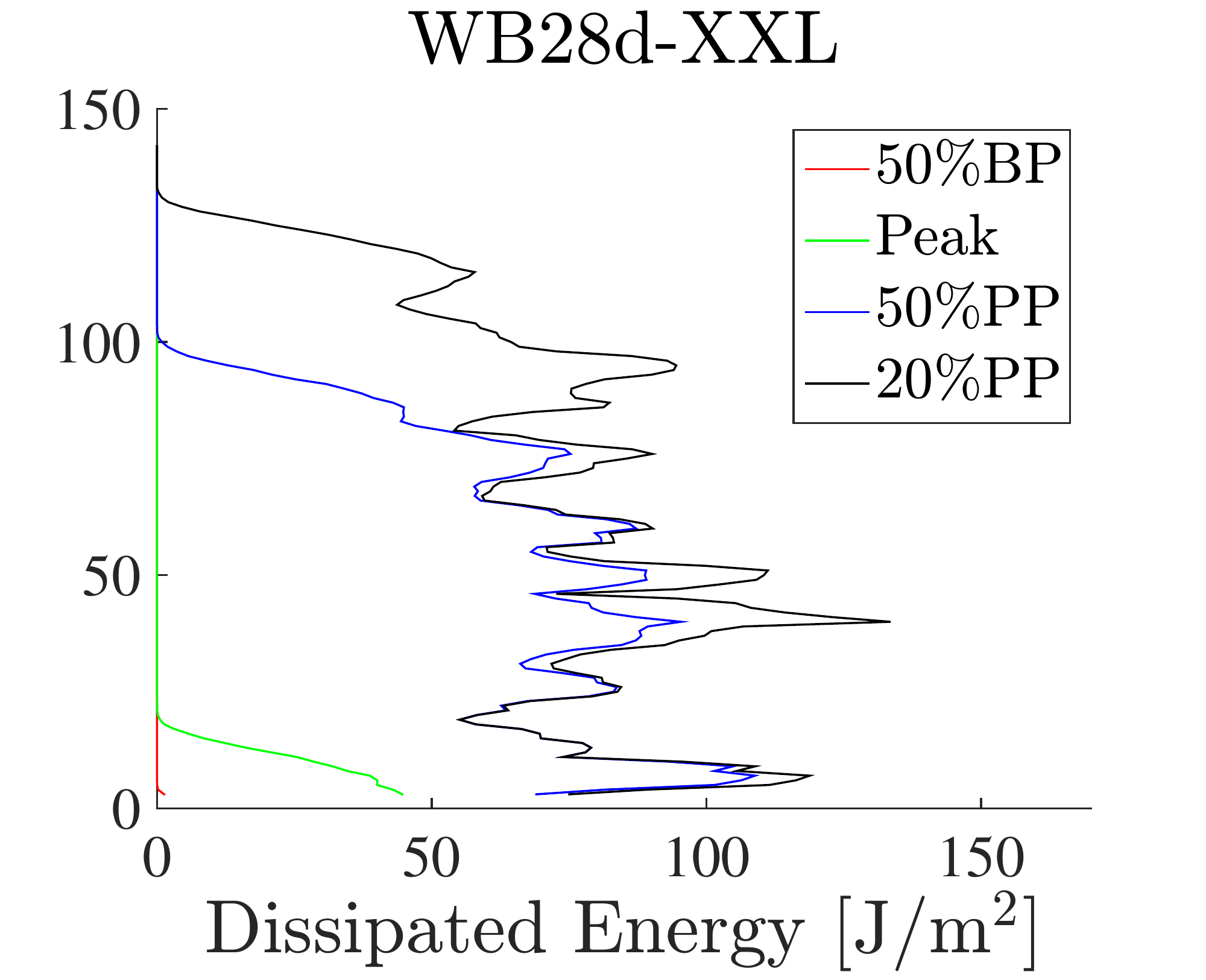} \\
\caption{Size and age dependent dissipated energy profiles along relative ligament height $z/h$ as obtained by HTC-LDPM simulations; from left to right: size M, L, XL, XXL; top to bottom: age HR3d, HR7d, HR14d, HR28d, WB28d}
\label{disenergy}
\efi

\section{Fracture Characteristics at Early Age}

There are a limited number of studies in the literature investigating how fracture energy changes with concrete age. Schutter and Taerwe \cite{Schutter} found, for regular concrete with different types of cement in TPB tests, that the fracture energy increases from day 1 to 28 days of age. Ostergaard \cite{Ostergaard} also concluded from wedge splitting tests (WST), aged 1 $\sim$ 28 days and utilizing a high performance concrete, that the fracture energy increases with age. However, Reis \cite{Reis} conducted TPB tests on PET polymer concrete and concluded that fracture energy decreases from day 1 to day~14. Similarly, Gettu \cite{Gettu} also found that fracture energy decreases from day 4 to 232 days of age, utilizing a high-strength silica fume concrete with mix proportions 
%of cement:sand:gravel:microsilica:water as 1:1.25:1.8:0.2:0.3, which is 
close to the UHPC used in this study. Furthermore, Kim et al. \cite{Kim} observed from WSTs an increasing trend of fracture energy for low strength concrete, however, actually increasing and then decreasing fracture energy for normal strength and high strength concretes from day 1 to day 28. 

Through the available experimental fracture data in the literature the conclusion can be drawn that the fracture energy does not necessarily increase with concrete age or maturity. Furthermore, Wan et al. \cite{WanUHPCI} observed from the TPB fracture tests for the investigated UHPC that the fracture energy first increases then decreases if plotted against increasing aging degree. More specifically, the fracture energy initially increases as the cement hydration and silica fume reaction are ongoing and slowly approach their asymptotic values under humidity room curing conditions. This trend reverses as the material further matures and approaches the ultimate properties under hot water bath curing. Note that the theoretical asymptotes of cement hydration and silica fume reaction vary under different curing conditions, which, for 100\% humidity room curing, are lower than the values for hot water bath curing.

This apparent inconsistency in the published experimental evidence can be explained analytically as well as by the computational early age framework which was developed by Wan et. al \cite{WanUHPCI}. According to Hillerborg \cite{Hillerborg1976} the fracture energy is proportional to the square of critical stress and inversely proportional to the Young's modulus. The proportionality constant is termed tensile characteristic length $\ell_{ch}$.
Since both tensile strength and Young's modulus are independent material properties, Hillerborg's equation implies that either fracture energy or tensile characteristic length is a dependent property. So far fracture energy is considered a material property and more widely used than $\ell_{ch}$, because it is accessible by experiments and intuition. However, if $\ell_{ch}$ were the independent material property one can easily show that, depending on the aging rates of Young's modulus and tensile strength, fracture energy may have a non-monotonic relationship with aging degree as seen in the presented experimental data and in the literature.  

The aging framework as formulated in \cite{WanUHPCI} defines the mesoscale normal modulus $E_0(\lambda) = E^{\infty}_0 * \lambda$ dependent on asymptotic normal modulus $E^{\infty}_0$ and aging degree $\lambda$. The age dependent mesoscale tensile strength is obtained by introducing an aging exponent $n_a$ as $\sigma_t(\lambda) = \sigma_t^{\infty} * \lambda^{n_a}$. The aging fracture properties are defined in terms of a mesoscale tensile characteristic length $\ell_t(\lambda) = \ell^{\infty}_{t} (k_a(1-\lambda) + 1)$ which is linearly decreasing with aging degree. The rate of brittleness increase is given by an aging factor $k_a$. Consequently, the mesoscale fracture energy $G_t=\ell_t f^{'2}_t/2E_0$ is a dependent quantity which can be derived as a function of aging degree in the following form:
\beq
G_t = \lambda^{2n_a} \frac{\ell^{\infty}_{t}\sigma_{t}^{\infty 2}}{2E^{\infty}_0}\left(\frac{k_a+1}{\lambda}-k_a\right) 
 \hspace{10mm}  \propto  \hspace{3mm} \lambda^{2n_a}\left(\frac{k_a+1}{\lambda}-k_a\right) 
\eeq
where $n_a$ and $k_a$ are positive constants. Mathematically, $G_t$ can increase or decrease depending on $n_a$ and $k_a$. The mesoscale aging functions have been derived and validated in the authors' previous work \cite{WanUHPCI} based on extensive experimental evidence. Fig.~\ref{fe} presents the results of the predictive aging size effect study which has been validated by experiments of three sizes for aging degree $\lambda=$0.954. The tensile characteristic length, $\ell_1$, and the tensile strength, $f'_t$, obtained from CSEC and SEL data fitting of simulated size effect data, as well as mesoscale $\ell_t$ and $\sigma_t$ directly from the LDPM simulations are plotted in Fig.~\ref{fe}a\&b against aging degree. The corresponding initial fracture energy, $G_f$, from CSEC and SEL analysis, as well as the mesoscale fracture energy, $G_t$, from HTC-LDPM simulations are plotted in Fig.~\ref{fe}c. The corresponding ages and aging degrees are, from left to right, HR3 (0.558), HR7 (0.706), HR14 (0.776), HR28 (0.810), WB14/28 days (0.954). While the initial fracture energy $G_f$ from SEL is close to that of CSEC, the mesoscale $G_t$ is about one third of the CSEC value. The difference is due to different mechanism of energy dissipation. The mesoscale fracture energy $G_t$ is defined as pure mode~$I$ fracture energy on single facets, while dissipation in shear is explicitly captured by friction laws. However, $G_f$, as obtained by fitting the two size effect laws (CSEC \& SEL) to the simulated peak strength values, or by performing an inverse cohesive crack analysis of the simulated load vs. crack opening data, includes contributions of energy dissipation from both tension and shear inside the fracture process zone. 

Fig.~\ref{fe}d presents a typical softening curve in a generic stripe of a beam after 28 days of water bath curing ($\lambda=$0.954). More specifically, the local stress vs. crack-opening law for a 4~mm high strip located 4-8~mm above the notch tip is plotted together with its optimum fit. The local cohesive stress $\sigma_{ch}$ is obtained from LDPM as the average of the horizontal stress components for the facets directly above the notch and within the given stripe at each time step during the simulated test. The local crack opening $w$ is calculated by equation $w(g_d) = \int_0^{g_d} dg / \sigma_{ch}(g)$, where $dg_d$ is the increment of the dissipated energy computed as the work done by the equivalent cohesive stress for an increment of the crack opening, $dg_d=\sigma_{ch}(g_d)dw$ \cite{Cusatis2007}. The optimum fit of such softening curves can be obtained by assuming a cohesive crack law with an initial plateau followed by a smooth curve consisting of the sum of a straight line and an exponential softening function. The equation of such a curve is $f(w)=f^\prime_t$ for $w\leq w_0$, $f(w)=0$ for $w\geq w_u$, and $f(w)=f^\prime_t[c_1-c_2 (w-w_0)/(w_{ch})+(1-c_1)\exp(-(w-w_0)/w_{ch})]$ for $w_0$\textless$w$\textless$w_u$ \cite{Cusatis2007}. The initial fracture energy describes only the first part of the softening curve (up to a stress drop of 1/3 - 1/2 of the tensile strength) and is defined as the area under its initial tangent. It is calculated as $G_f = f^{\prime 2}_t / (2|f^\prime(0)|)$, where $f^\prime$(0) is the derivative of the softening curve at crack opening equal to zero, or in other words, when $\sigma_{ch} = f_t^\prime$. By plotting and analyzing the softening curves along the ligament for sizes L, XL, \& XXL, the average local initial fracture energy is obtained to be $G_f$ = 44.5 J/m$^2$ and the average local total fracture energy $G_F$~=~93.0~J/m$^2$~$\cong$~2.1$G_f$. Fracture energies along the ligament for size L and XXL on age WB28d are presented as bar plots in Fig.~\ref{lfe}a\&b, respectively. Consistent with the constant LDPM particle size also the height of the stripes was kept constant. As shown, the local initial fracture energy is roughly constant along the ligament, while the total energy slightly increases then decreases along the ligament, a behavior previously discussed by Cusatis and co-workers \cite{Cusatis2007}. The drop of the total fracture energy is due to the presence of the specimen's outer surface that, constraining the propagation of the meso-cracks, induces a reduction of the number of the fractured lattice elements and consequently the width of the fracture process zone. Moreover, the top 20-30\% of the ligament is subject to compressive stresses until late post-peak, thus very low or no energy dissipation is recorded. The local initial fracture energy, $G_f$, is higher than the LDPM mesoscale fracture energy, $G_t=21.1$J/m$^2$, because $G_f$ includes also the energy dissipation due to shear as discussed earlier. The local total fracture energy, $G_F$, on the other hand, agrees with the value obtained by the work of fracture method $G_F\cong 100.6$J/m$^2$ applied to the simulated load deflection data. For the relation between different fracture energies it can be concluded that LDPM $G_t$ $\leq$ local $G_f$ $\leq$ SEL $G_f$ $\leq$($\cong$) $CSEC$ $G_f$ $<$ $G_F$. However, despite the different magnitudes, all fracture energies have the same trend in terms of aging. As shown in Fig.~\ref{fe}c, initial fracture energies from HTC-LDPM, CSEC and SEL all increase and then decrease in terms of aging degree. 

While fracture energy does not have a monotonous relation with concrete maturity expressed in terms of aging degree, the tensile characteristic length does. The mesoscale parameters of the numerical framework as well as the inverse analyses by CSEC \& SEL show a monotonously and even linearly decreasing trend. In all cases also the tensile strength increases monotonously with age \cite{WanUHPCI}. Based on the presented findings it is evident that the tensile characteristic length is the true independent material property together with tensile strength and Young's modulus, and, consequently, fracture energy is dependent on the three and should not be directly utilized as a material property, at least for time-dependent problems. 

Moreover, by evaluating the ``magnitude of size effect'' dependent on the maturity of concrete, it is found that older concretes tend to exhibit a more pronounced size effect. In Fig.~\ref{sea}, a 3-D contour of nominal stress as a function of aging degree and logarithmically normalized size is presented. The specimen span and height in the size effect study are designed to have scaling factor of $\sqrt{5}$, thus the normalized sizes for M, L, XL, XXL have values of $1$, $\sqrt{5}$, $5$, and $5\sqrt{5}$, respectively. As shown, the strength difference with respect to size increases as concrete ages. 

\bfi[ht]
\centering
(a)     \includegraphics[height=2.2in,valign=t]{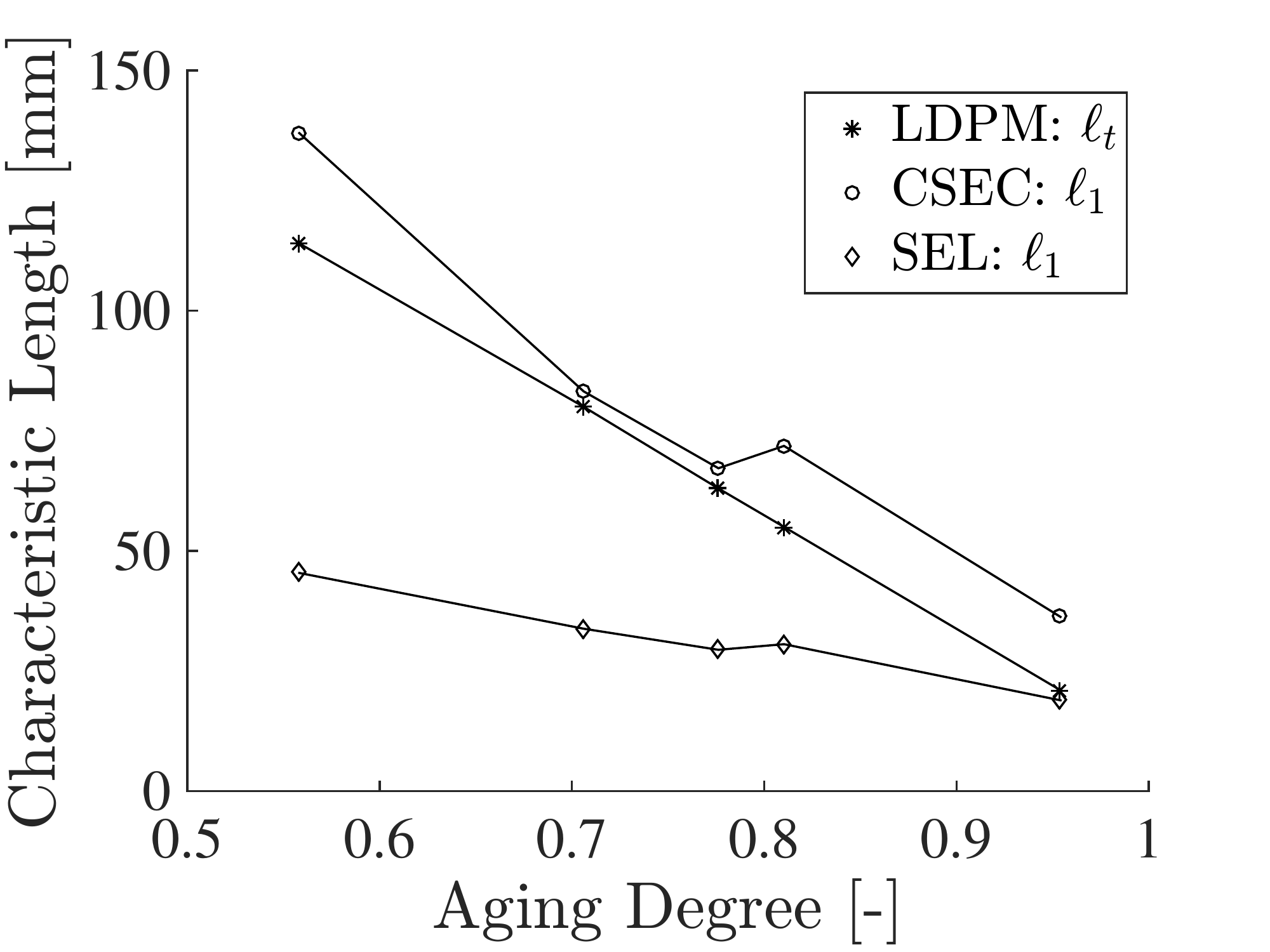} 
(b)    \includegraphics[height=2.2in,valign=t]{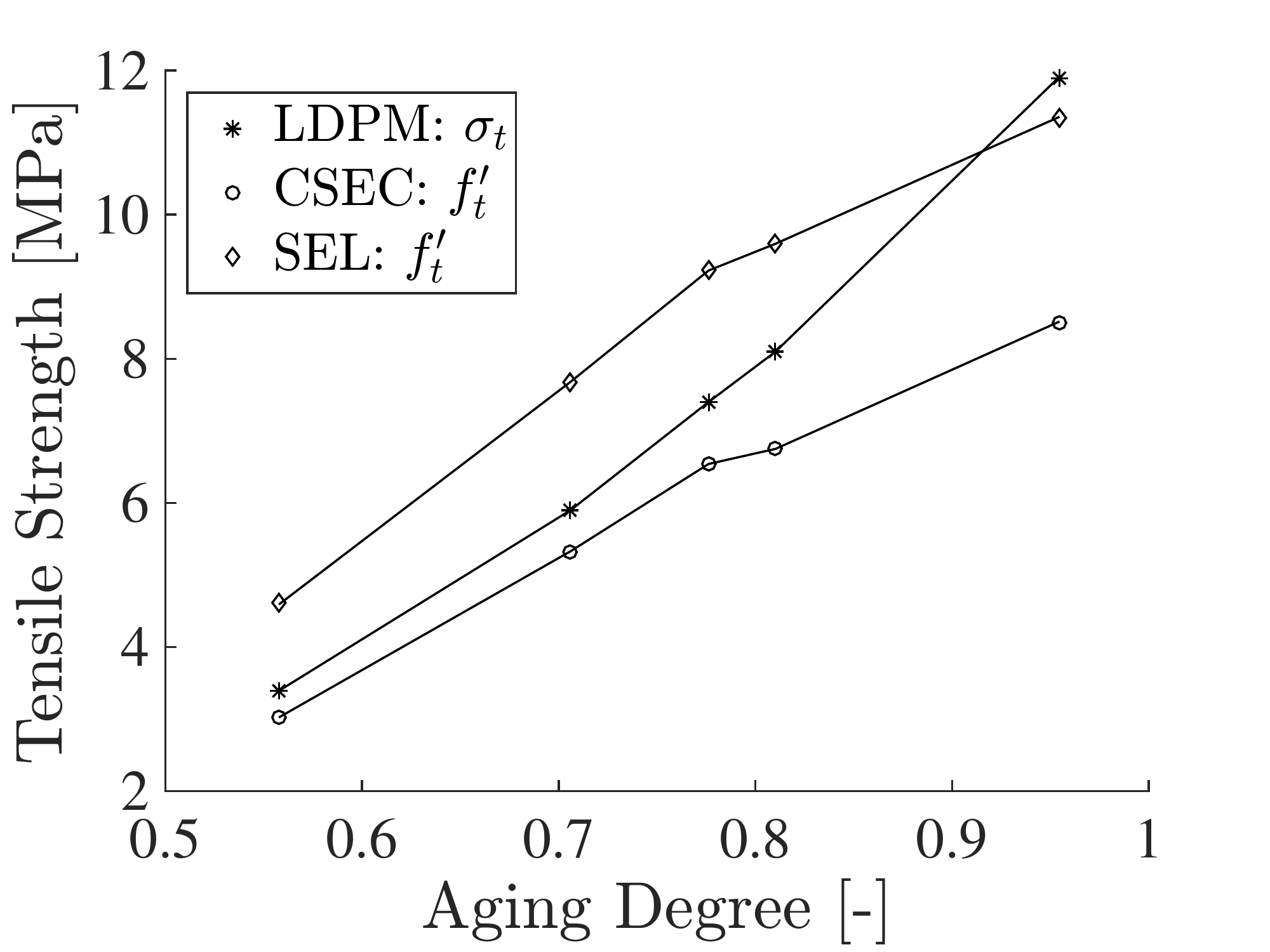} \\
(c)    \includegraphics[height=2.2in,valign=t]{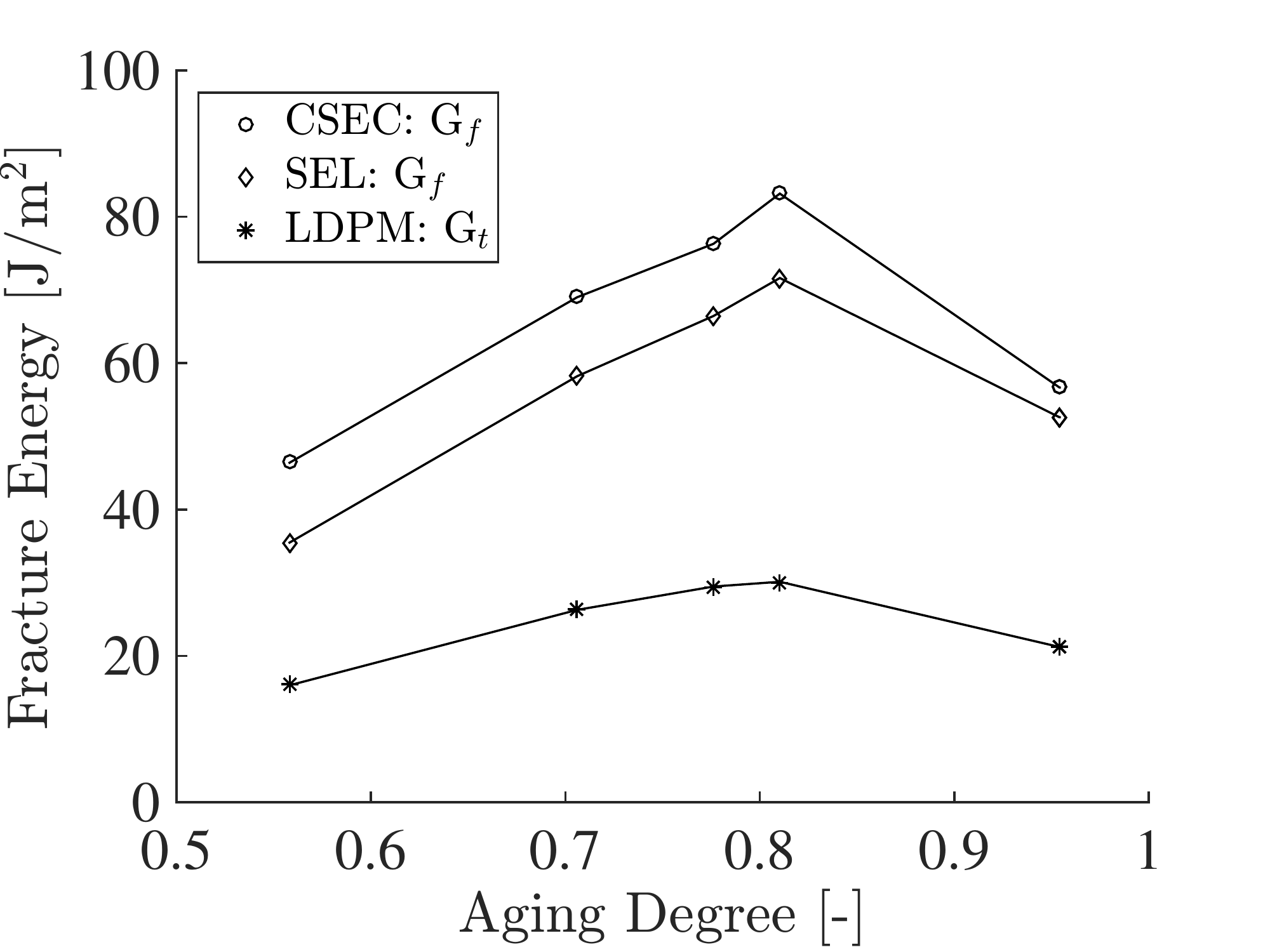} 
(d)    \includegraphics[height=2.2in,valign=t]{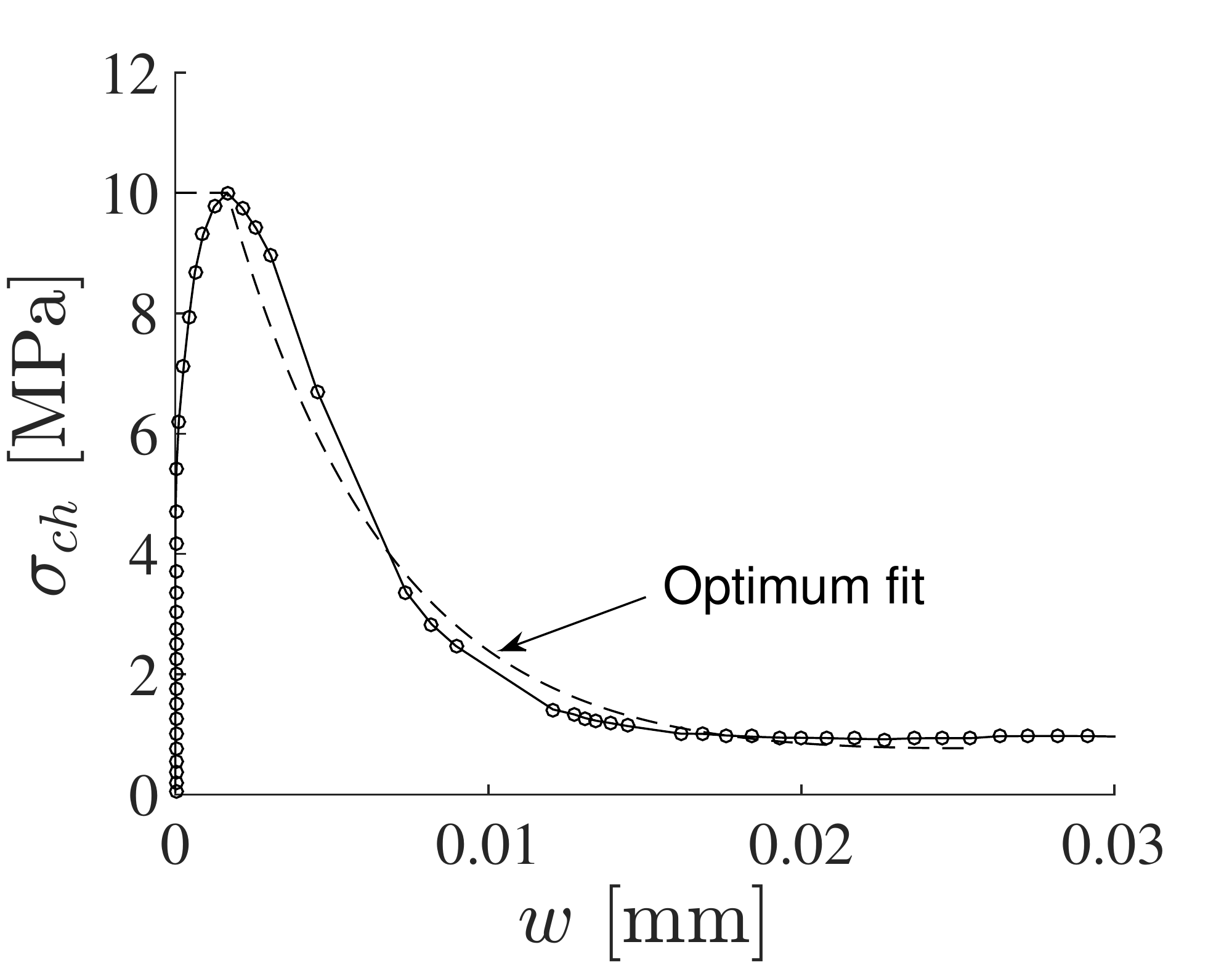} 
   \caption{Fracture characteristics at early age: age dependent (a) characteristic length, (b) tensile strength, (c) fracture energy from HTC-LDPM simulations and CSEC \& SEL analysis, and (d) typical softening curve identified in a generic stripe and its optimum fit for age WB28d }
   \label{fe}
\efi

\bfi[ht]
\centering
(a)   \includegraphics[height=2.2in,valign=t]{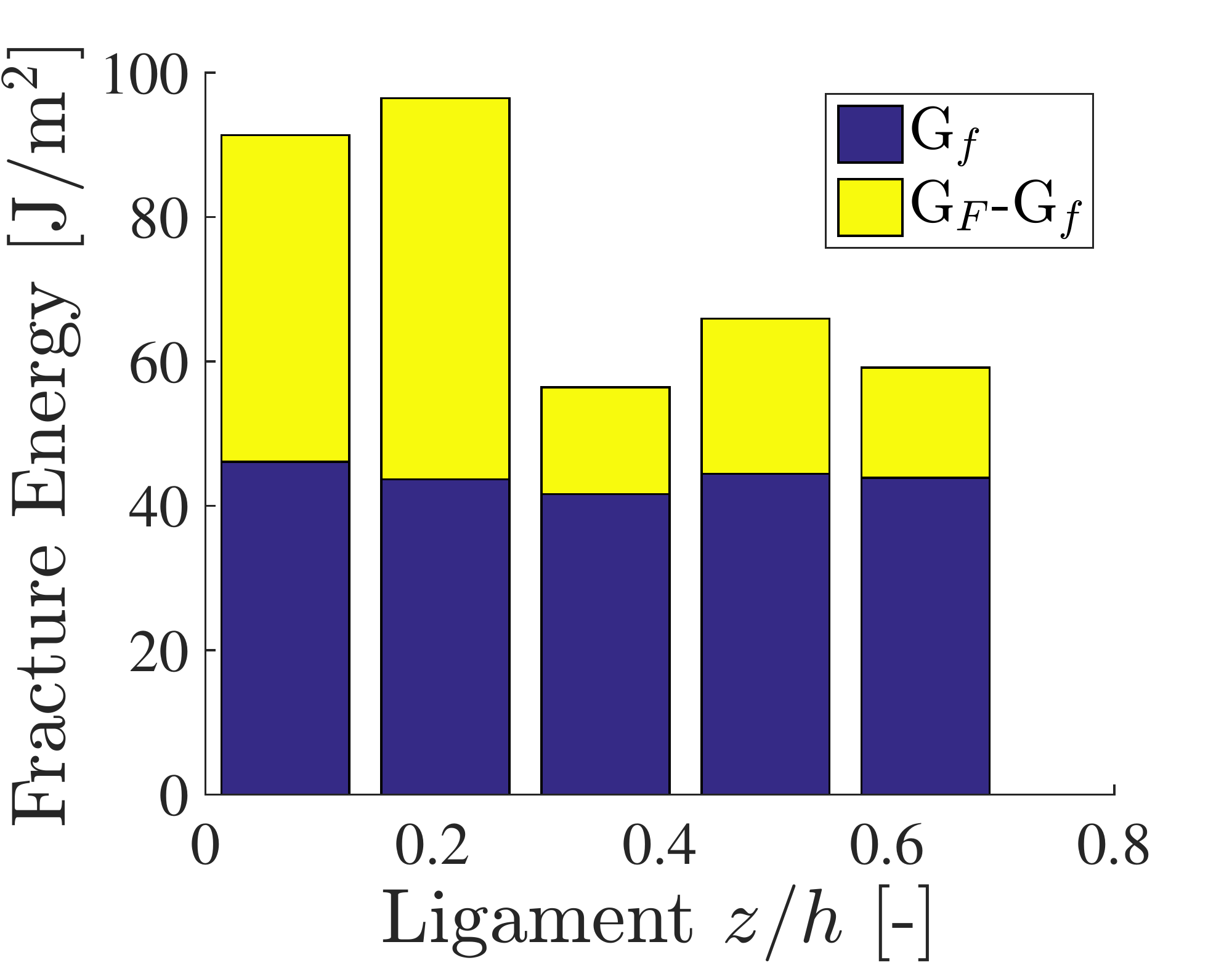} 
(b)   \includegraphics[height=2.2in,valign=t]{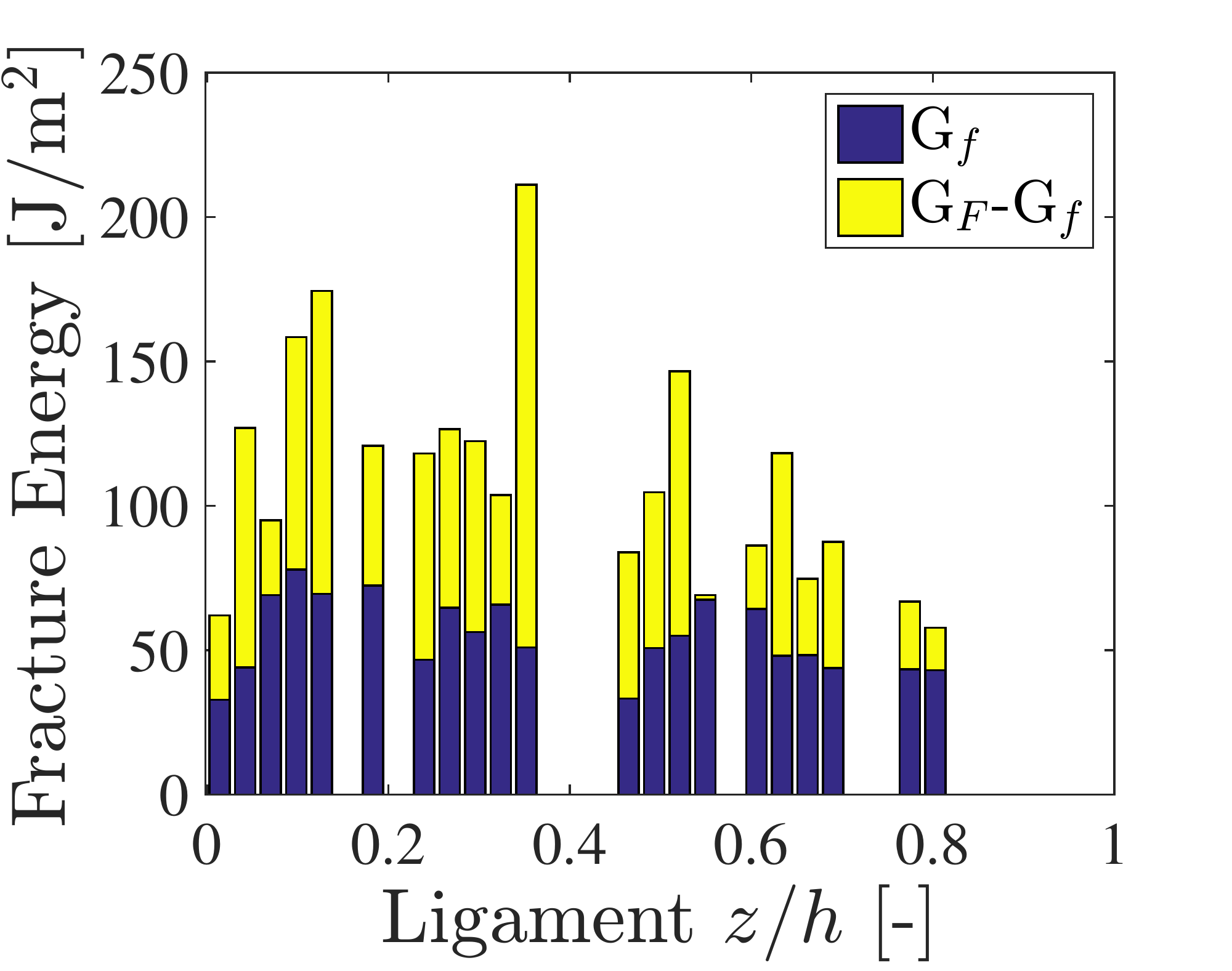} 
\caption{Local fracture energies along the ligament of (a) size L, and (b) size XXL}
 \label{lfe}
\efi

\bfi[ht]
\centering
 \includegraphics[height=3.7in]{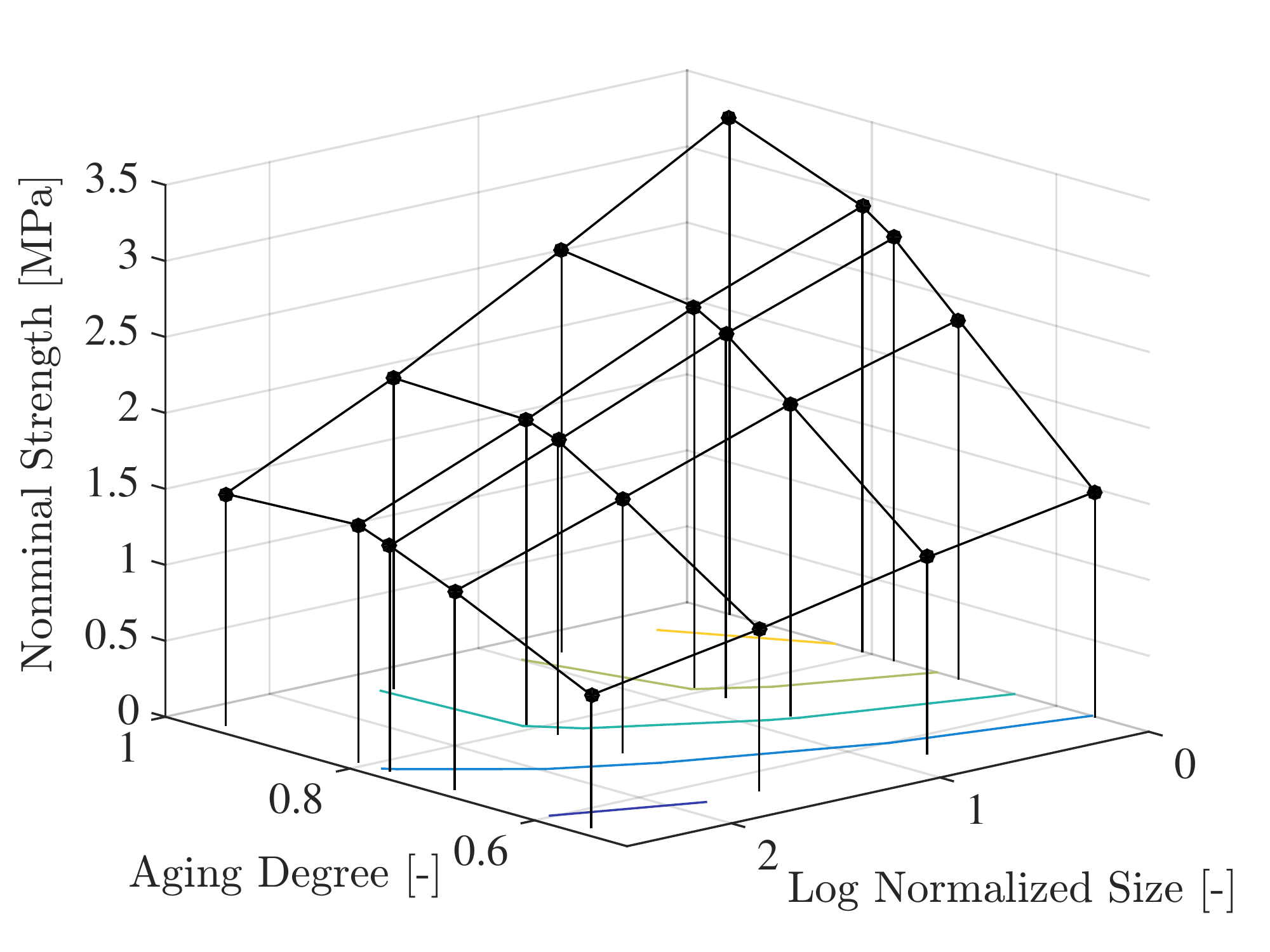} 
 \caption{Age dependent size effect}
 \label{sea}
 \efi
 
\section{Conclusions}

In this contribution the age-dependent fracture characteristics of a selected ultra high performance concrete are presented. The investigation is based on experimental size effect tests of one age and aging tests of one size, all performed in a three point bending configuration. The computational early age framework, introduced in an earlier work by the authors \cite{WanUHPCI}, is employed to augment the experimental size effect data after careful calibration and validation. Both, experimental and simulated size effect data for different ages are analyzed utilizing the cohesive size effect curve (CSEC) and Ba\v{z}ant's classical size effect law. 

According to the experimental studies, the numerical simulations as well as the size effect analyses the following conclusions can be drawn:

\bi
\item The HTC-LDPM model can accurately capture and predict the age dependent fracture characteristics of concrete at early age and beyond. The well calibrated and validated model is able to reproduce the experimentally observed size effect in notched three point bending specimens and provide insights into it's age dependence. 
\item Numerical simulations by HTC-LDPM as well as analytical analyses by CSEC \& SEL show that fracture energy may exhibit a non-monotonous relationship with the maturity of concrete. The reason is that the fracture energy is dependent on modulus, tensile characteristic length, and tensile strength, which may evolve at different rates. Since fracture energy is not an independent material property it is not suitable for the formulation of fracture related aging laws.
\item Contrary to fracture energy the tensile characteristic length changes monotonously in all investigated cases and shows a linear decreasing dependence on aging degree for the studied UHPC.
\item The initial fracture energies identified by SEL, CSEC and local cohesive crack analysis roughly correspond and contain contributions from different dissipative mechanisms on lower scales. Consequently, the pure mode~$I$ mesoscale fracture energy $G_t$ is lower than the macroscopic initial fracture energy $G_f$. Overall, the following relationship is found: LDPM $G_t$ $\leq$ local $G_f$ $\leq$ SEL $G_f$ $\leq$($\cong$) $CSEC$ $G_f$ $<$ $G_F$.
\item For the investigated UHPC the total fracture energy is approximately 2.1 times the initial fracture energy. 
\item Both investigated analytical size effect laws, CSEC and SEL, fit the experimental as well as the numerically generated three point bending size effect data very well. 
\item Consistently for all ages and experimental as well as generated data, the CSEC fits better than SEL. This is qualitatively observed and confirmed by the coefficients of determination. The younger the concrete the larger differences are observed, stemming from the fact that the SEL represents the asymptote of the CSEC for large sizes. 
\item The inverse analysis of size effect data by analytical size effect laws such as CSEC and SEL is highly sensitive to noise in the (experimental) data. By sensitivity analysis it is found that the identification of tensile characteristic length and tensile strength by the CSEC is not stable and can result in unrealistic parameter combinations of $\ell_1$ and $f_t^\prime$ while the corresponding fracture energy, calculated as $G_f=\ell_{1}f^{\prime 2}_t/E$, is quite robust.
\item Size effect of concrete is age-dependent. Older concretes (in the sense of higher aging degrees) exhibit a more pronounced size effect, at least for the investigated UHPC.
\ei

\newpage
\section*{Acknowledgement}\no

The work of the first and last author was supported under National Science Foundation (NSF) Grant CMMI-1237920 to Northwestern University. The work of the first and second author was also supported by the Austrian Federal Ministry of Economy, Family and Youth, and the National Foundation for Research, Technology and Development. The experimental component of this research effort was partially sponsored by the US Army Engineer Research Development Center (ERDC) under Grant W912HZ-12-P-0137. Permission to publish was granted by the director of ERDC geotechnical and structural laboratory. 

\newpage

\section*{Appendix~A - The Computational Framework}
%computational fremework
The proposed hygro-thermo-chemo-mechanical early-age model for cement based concrete consists of two major components: the HTC model and the LDPM with aging material properties. 

\subsection*{\textit{Hygro-Thermo-Chemical (HTC) model}}

The behavior of concrete at early age heavily depends on moisture content and temperature. The overall moisture transport can be described through Fick's law that expresses the flux of water mass per unit time {\bf J} as a function of the spatial gradient of the relative humidity $h$. Assuming that evaporable water $w_e$ is a function of relative humidity $h$, degree of hydration $\alpha_c$, and degree of silica fume reaction $\alpha_s$, one can write $w_e = w_e (h, \alpha_c, \alpha_s)$, which represents an age-dependent sorption/desorption isotherm. Consequently, the moisture mass balance equation reads \cite{DiLuzio2009I}:
$\nabla \cdot (D_h \nabla h) - \partial w_e / \partial h \cdot \partial h / \partial t - \left(\partial w_e / \partial \alpha_c \cdot \dot{\alpha_c} + \partial w_e / \partial \alpha_s \cdot \dot{\alpha_s} + \dot{w_n}\right) = 0$, where $D_h$ is moisture permeability and $w_n$ is nonevaporable water.
The enthalpy balance is also influenced by the chemical reactions occurring at the early age. One can write, at least for temperatures not exceeding 100$^\circ$C \cite{Bazant1996}, 
$\nabla \cdot (k_T \nabla T) - \rho c_T \partial T / \partial t + \dot{\alpha}_ss\tilde{Q}^{\infty}_s + \dot{\alpha}_cc\tilde{Q}^{\infty}_c = 0$,
where $\tilde{Q}^{\infty}_c$ = hydration enthalpy, $\tilde{Q}^{\infty}_s$ = latent heat of silica-fume reaction per unit mass of reacted silica-fume, $\rho$ is the mass density of concrete, $k_T$ is the heat conductivity, and $c_T$ is the isobaric heat capacity of concrete. 

With the assumption that the thermodynamic force is governed by an Arrhenius-type equation and that the viscosity governing the diffusion of water though the layer of cement hydrates is an exponential function of the hydration extent \cite{Ulm1995}, Cervera et al. \cite{Cervera1999} proposed the evolution equation for the hydration degree: $\dot{\alpha}_c = A_c(\alpha_c)e^{-E_{ac}/RT}$, and \mbox{$A_c(\alpha_c) = A_{c1}(
a_{c2}/\alpha^{\infty}_c + \alpha_c)(\alpha^{\infty}_c - \alpha_c) e^{-\eta_c\alpha_c/\alpha^{\infty}_c}$}, where $E_{ac}$ is the hydration activation energy, $R$ is the universal gas constant, and $\eta_c$, $A_{c1}$, and $A_{c2}$ are material parameters. To account for the situation that the hydration process slows down and may even stop if the relative humidity decreases below a certain value, the equation can be rewritten as: $\dot{\alpha}_c = A_c(\alpha_c)\beta_h(h)e^{-E_{ac}/RT}$, where $\beta_h(h) = [1 + (a - ah)^b]^{-1}$. The function $\beta_h(h)$ is an empirical function proposed by Ba\v{z}ant and Prasannan \cite{Bazant1989a} and $a$ \& $b$ are constant model parameters. Similarly, the degree of silica fume (SF) reaction, $\alpha_s$ is introduced \cite{DiLuzio2009I}, $\dot{\alpha}_s = A_s(\alpha_s)e^{-E_{as}/RT}$, and $A_s(\alpha_s) = A_{s1}(
A_{s2} / \alpha^{\infty}_s + \alpha_s)(\alpha^{\infty}_s - \alpha_s)e^{-\eta_s\alpha_s/\alpha^{\infty}_s}$, where $A_s$ is the SF normalized affinity, $E_{as}$ is the activation energy of SF reaction, and $\alpha^{\infty}_s$ is the asymptotic value of the SF reaction degree.

To account for this additional effect from temperature \cite{Cervera2000}, the aging degree $\lambda$ can be used which is formulated as:  
$\dot{\lambda} = \left[ (T_{max} - T)/(T_{max} - T_{ref}) \right] ^{n_\lambda}(B_{\lambda} - 2A_{\lambda}\alpha)$,
where $B_{\lambda} = [1 + A_{\lambda}(\alpha^2_\infty - \alpha^2_0)]/(\alpha_\infty - \alpha_0)$, $n_{\lambda}$ and $A_{\lambda}$ are model parameters obtained from fitting experimental data, and $\alpha$ is the overall degree of reaction defined as \cite{DiLuzio2013}:
$\alpha(t)= 
[ \alpha_c(t)c\tilde{Q}_c^\infty + \alpha_s(t)s\tilde{Q}_s^\infty ] / (c\tilde{Q}_c^\infty + s\tilde{Q}_s^\infty)$.

\subsection*{\textit{Age-dependent Lattice Discrete Particle Model}}

The Lattice Discrete Particle Model (LDPM) is a mesoscale discrete model that simulates the mechanical interaction of coarse aggregate pieces embedded in a cementitious matrix (mortar). In LDPM, rigid body kinematics is used to describe the deformation of the lattice particle system and the displacement jump, $\llbracket \mathbf{u}_{C} \rrbracket$, at the centroid of each facet from which the measures of strain are derived as 
$ e_{N}= (\mathbf{n}^\mathrm{T} \llbracket \mathbf{u}_{C} \rrbracket) / \ell$; 
$ e_{L}= (\mathbf{l}^\mathrm{T} \llbracket \mathbf{u}_{C} \rrbracket) / \ell$; 
$ e_{M}= (\mathbf{m}^\mathrm{T} \llbracket \mathbf{u}_{C} \rrbracket) / \ell$,
where $\ell=$ inter-particle distance; and $\mathbf{n}$, $\mathbf{l}$, and
$\mathbf{m}$, are unit vectors defining a local system of reference attached to each facet. 

Next, a vectorial constitutive law governing the behavior of the material is imposed at the centroid of each facet. In the elastic regime, the normal and shear stresses are proportional to the corresponding strains: $t_{N}= E_N e^*_{N} =E_N (e_{N}-e^0_{N});~ t_{M}= E_T e^*_{M} = E_T (e_{M}-e^0_{M});~ t_{L}= E_T e^*_{L} = E_T (e_{L}-e^0_{L})$, where $E_N=E_0$, $E_T=\alpha E_0$, $E_0=$ effective normal modulus, and $\alpha=$ shear-normal coupling parameter; and $e^0_{N}$, $e^0_{M}$, $e^0_{L}$ are mesoscale eigenstrains that might arise from a variety of phenomena such as, but not limited to, thermal expansion, creep, shrinkage, and chemical reactions, e.g. alkali-silica reaction.

For stresses and strains beyond the elastic limit, the LDPM formulation considers the following nonlinear mesoscale phenomena \cite{Cusatis2011a}: (1) fracture and cohesion; (2) compaction and pore collapse; and (3) friction.

\textbf{Fracture and cohesion due to tension and tension-shear.} For tensile loading ($e^*_N>0$), the fracturing behavior is formulated through an effective strain, $e^* = \sqrt{e_N^{*2}+\alpha (e_M^{*2} + e_L^{*2})}$, and stress, $t = \sqrt{{ t _{N}^2+  (t _{M}^2+t _{L}^2) / \alpha}}$, which define the normal and shear stresses as \mbox{$t _{N}= e_N^*(t / e^*)$}; \mbox{$t _{M}=\alpha e^*_{M}(t / e^*)$}; \mbox{$t _{L}=\alpha e^*_{L}(t / e^*)$}. The effective stress $t$ is incrementally elastic ($\dot{t}=E_0\dot{e}$) and must satisfy the inequality $0\leq t \leq \sigma _{bt} (e, \omega) $ where $\sigma_{bt} = \sigma_0(\omega) \exp \left[-H_0(\omega)  \langle e-e_0(\omega) \rangle / \sigma_0(\omega)\right]$, $\langle x \rangle=\max \{x,0\}$, and $\tan(\omega) =e^* _N / \sqrt{\alpha} e^* _{T}$ = $t_N \sqrt{\alpha} / t_{T}$, and $e_T^*=\sqrt{e_M^{*2} + e_L^{*2}}$. The post peak softening modulus is defined as $H_{0}(\omega)=H_{t}(2\omega/\pi)^{n_{t}}$, where $H_{t}$ is the softening modulus in pure tension ($\omega=\pi/2$) expressed as $H_{t}=2E_0/\left(\ell_t/\ell-1\right)$; $\ell_t=2E_0G_t/\sigma_t^2$; $\ell$ is the length of the tetrahedron edge; and $G_t$ is the mesoscale fracture energy. LDPM provides a smooth transition between pure tension and pure shear ($\omega=0$) with parabolic variation for strength given by $\sigma_{0}(\omega )=\sigma _{t}r_{st}^2\Big(-\sin(\omega) + \sqrt{\sin^2(\omega)+4 \alpha \cos^2(\omega) / r_{st}^2}\Big) / [2 \alpha \cos^2(\omega)]$, where $r_{st} = \sigma_s/\sigma_t$ is the ratio of shear strength to tensile strength. 
% MA
  
\textbf{Compaction and pore collapse from compression.} Normal stresses for compressive loading ($e^*_N<0$) must satisfy the inequality $-\sigma_{bc}(e_D, e_V)\leq t_N \leq 0$, where $\sigma_{bc}$ is a strain-dependent boundary depending on the volumetric strain, $e_V$, and the deviatoric strain, $e_D=e_N-e_V$. The volumetric strain is computed by the volume variation of the Delaunay tetrahedra as $e_V= \Delta V/ 3V_0$ and is assumed to be the same for all facets belonging to a given tetrahedron. Beyond the elastic limit, $-\sigma_{bc}$ models pore collapse as a linear evolution of stress for increasing volumetric strain with stiffness $H_{c}$ for $-e_V \leq e_{c1} = \kappa_{c0} e_{c0}$: $\sigma_{bc} = \sigma_{c0} + \langle-e_V-e_{c0}\rangle H_c(r_{DV})$; $H_c(r_{DV})=H_{c0}/(1 + \kappa_{c2} \left\langle r_{DV} - \kappa_{c1} \right\rangle)$; $\sigma_{c0}$ is the mesoscale compressive yielding stress; $r_{DV}=e_D/e_V$ and $\kappa_{c1}$, $\kappa_{c2}$ are material parameters. Compaction and rehardening occur beyond pore collapse ($-e_V \geq e_{c1}$). In this case one has $\sigma_{bc} = \sigma_{c1}(r_{DV})$ $\exp \left[( -e_{V}-e_{c1} ) H_c(r_{DV})/\sigma_{c1}(r_{DV}) \right]$ and $\sigma_{c1}(r_{DV}) = \sigma_{c0} + (e_{c1}-e_{c0}) H_c(r_{DV})$. 
% MA

\textbf{Friction due to compression-shear.} The incremental shear stresses are computed as  $\dot{t}_M=E_T(\dot{e}^*_M-\dot{e}^{*p}_M)$ and \mbox{$\dot{t}_L=E_T(\dot{e}^*_L-\dot{e}^{*p}_L)$}, where  \mbox{$\dot{e}_M^{*p}=\dot{\xi} \partial \varphi / \partial t_M$}, \mbox{$\dot{e}_L^{*p}=\dot{\xi} \partial \varphi / \partial t_L$}, and $\xi$ is the plastic multiplier with loading-unloading conditions  $\varphi \dot{\xi} \leq 0$ and $\dot{\xi} \geq 0$. The plastic potential is defined as \mbox{$\varphi=\sqrt{t_M^2+t_L^2} - \sigma_{bs}(t_N)$}, where the nonlinear frictional law for the shear strength is assumed to be $\sigma_{bs} = \sigma_s + (\mu_0 - \mu_\infty)\sigma_{N0}[1 - \exp(t_N / \sigma_{N0})] - \mu_\infty t_N$; $\sigma_{N0}$ is the transitional normal stress; $\mu_0$ and $\mu_\infty$ are the initial and final internal friction coefficients.  
% % MA
 
The compatibility and constitutive equations discussed above are completed through the equilibrium equations of each individual particle to arrive at the complete set of governing equations of the LDPM framework. 

The chemo-mechanical coupling between HTC and LDPM framework is achieved by a set of aging functions \cite{WanUHPCI} defining the mesoscale material parameters in terms of aging degree as: 
$E_0 = E^{\infty}_0 \lambda;\:
\sigma_t = \sigma_t^{\infty} \lambda^{n_a} ; \:
\sigma_c = \sigma_c^{\infty} \lambda^{n_a} ; \:
\sigma_{N0} = \sigma_N^{\infty} \lambda^{n_a};\:
\ell_{t} = \ell^{\infty}_{t} (k_a(1-\lambda) + 1)$, where $n_a$ and $k_a$ are positive constants. As seen, the normal modulus, $E_0$, which is related to the elastic modulus, is assumed to have a linear relation with aging degree $\lambda$.  Tensile strength, $\sigma_t$, compressive yielding stress, $\sigma_c$, and transitional stress, $\sigma_{N0}$, on the other hand, are assumed to have power-law type relations with aging degree. Lastly, the tensile characteristic length, $\ell_{t}$, is assumed to be a linear decreasing function with aging degree to simulate the well known brittleness increase with age. All the aging functions are formulated such that the corresponding parameters approach their asymptotic values for $\lambda$ approaching the value of 1. The other LDPM mesoscale parameters, are assumed to be age-independent due to a lack of relevant experimental data on the response in compression under confinement.

\end{document}